\documentclass[11pt]{article}
 \csname
@addtoreset\endcsname{equation}{section}

\def\a{\alpha} \def\ad{\dot{\a}} \def\hA{{\widehat A}}
\def\dB{\dot B}
\def\b{\beta}  \def\bd{\dot{\b}} 
\def\c{\gamma} \def\cd{\dot{\c}}
\def\C{\Gamma}
\def\d{\delta} 
\def\D{\Delta}
\def\e{\epsilon} 

\def\f{\phi}
\def\F{\Phi}
\def\vf{\varphi}
\def\k{\kappa}
\def\l{\lambda}
\def\L{\Lambda}
\def\m{\mu}
\def\n{\nu}
\def\r{\rho}
\def\s{\sigma}
\def\S{\Sigma}
\def\t{\tau}
\def\th{\theta} \def\tb{\bar\theta}

\def\x{\xi}

\def\O{\Omega}
\def\o{\omega}

\def\cL{{\cal L}}

\def\cO{{\cal O}}
\def\cA{{\cal A}}
\def\cM{{\cal M}}
\def\cN{{\cal N}}
\def\cR{{\cal R}}

\def\cV{{\cal V}}
\def\cI{{\cal I}}

\def\cA{{\cal A}}
\def\hcA{\hat\cA}

\def\yb{{\bar y}}
\def\zb{{\bar z}}

\def\cO{{\cal O}}

\def\ra{\rightarrow}

\def\del{\partial}

\let\la=\label
\let\bm=\bibitem
\def\nn{\nonumber}
\newcommand{\eq}[1]{(\ref{#1})}
\newcommand{\ns}[1]{{\normalsize #1}}
\def\tr{{\rm tr}}
\newcommand{\w}[1]{\\[0.#1cm]}
\def\eqs#1#2{(\ref{#1}-\ref{#2})} \def\det{{\rm det\,}}
\def\be{\begin{equation}}
\def\ee{\end{equation}}
\def\bea{\begin{eqnarray}}
\def\eea{\end{eqnarray}}
\def\ba{\begin{array}}
\def\ea{\end{array}}

\def\mx#1#2#3#4{\left#1\begin{array}{#2} #3 \end{array}\right#4}

\def\ft#1#2{{\textstyle{{\scriptstyle #1}
\over {\scriptstyle #2}}}}

\def\ket#1{|#1\rangle}

\def\tfos{
\begin{table}[t]
\begin{center}
{\footnotesize \tabcolsep=1mm
\begin{tabular}{|c|cccccccccccccc|}\hline
& & & & & & & & & & & & &
& \\
{\large${}_{\ell}\backslash s$} & $0$ & \ns{$\ft12$} & $1$ &
\ns{$\ft32$} & $2$ & \ns{$\ft52$} & $3$ & \ns{$\ft72$} & $4$ &
$\ft92$ & $5$ & $\ft{11}2$ &
$6$ & $\cdots$ \\
& & & & & & & & & & & & & & \\ \hline & & & & & & & & & & & & &
& \\
$0$ & $70$ & $56$ & $28$ & $8$ & $1$ &
& & & & & & & & \\
$1$ & $1\!+\!1$ & $8$ & $28$ & $56$ & $70$ & $56$ & $28$ & $8$ &
$1$ &
& & & &  \\
$2$ & & & & & $1$ & $8$ & $28$ & $56$ & $70$
& $56$ & $28$ & $8$ & $1$  &  \\
$3$ & & & & & & & & & $1$ & $8$ & $28$ & $56$ & $70$
& $\cdots$ \\
$4$ & & & & & & & & & & & &
& $1$ & $\cdots$ \\
$\vdots$ & & & & & & & & & & & & & &  \\ \hline
\end{tabular}}
\end{center}
\caption{{\small The $SO(3,2)\times SO(8)$ content of the
symmetric tensor product of two $d=3$, $\cN=8$ singletons. Each
entry refers to the $SO(8)$ content. All $SO(8)$ irreps are
irreducible except $70=35_++35_-$ and all the states have
$E_0=s+1$ except the scalars in one of the $35$-plets at level
$\ell=0$ and one of the scalars at level $\ell=1$.  The
representations have been arranged into a tower of $OSp(8|4)$
supermultiplets labeled by a level index $\ell$. The zeroth level
is the $D=4$, $\cN=8$ supergravity multiplet with $2^8$ degrees of
freedom. The level $\ell\geq1$ multiplets have $2\times 2^8$
degrees of freedom. The spin $s\geq 1$ fields arise in the
$hs(8|4)$ valued master gauge field and the spin $s\leq\ft12$
arise in the quasi-adjoint master zero-form. The minimal bosonic
truncation of the spectrum is obtained by keeping the maximum spin
fields at each level and the (non-pseudo) scalar at level
$\ell=1$.
 }} \la{tfos}
\end{table}
}

\def\tfid{
\begin{table}[h!!!]
\begin{center}
{\footnotesize \tabcolsep=1mm
\begin{tabular}{|c|cccccccc|}\hline
&        &        &         &        &         &
&         &  \\
{\large${}_{|Z|}\backslash s$} & $0$ & \ns{$\ft12$} & $1$ &
\ns{$\ft32$} & $2$    & \ns{$\ft52$} &
$3$ & $\cdots$ \\
&        &        &         &        &         & &
      &   \\
\hline
& & & & & & & &   \\
$0$ & $6$ & $4$ & $1$ & & & & & \\
$\ft12$  & $4$    & $6+1$   & $4$    & $1$     &
&         &  &\\
$1$& $1$ & $4$ & $6$ & $4$ & $1$ & & & \\
$\ft32$  &        & $1$     & $4$    & $6$     & $4$
& $1$     &  &\\
$2$ & & & $1$ & $4$ & $6$ & $4$ & $1$  &
 \\
$\vdots$ & & & & &\vdots & & & \\ \hline
\end{tabular}}
\end{center}
\caption{{\footnotesize The $d=4$, $\cN=4$ singletons. The
quantity $Z$ is the $SU(2,2|4)$ central charge carried by the
supermultiplet. The entries in the Table denote $SU(4)$
representations. Each entry carries an $SO(4)\subset SO(6)\subset
SO(6,2)$ representation $(j_L,0)$, and their complex conjugates
$(0,j_R)$. The states for each value of $|Z|$ form a single
massless irrep of $d=4, {\cN} =4$ Poincar\'e superalgebra, and the
states carry spin $s=j_L$. For all the states $E_0=s+1$, where
$E_0$ is the lowest $AdS_5$ energy. There exists an outer
automorphism group $U(1)_Y$ of $SU(2,2|4)$, and the $U(1)_Y$
charges of $6$, $4$ and $1$ are $0$, $\pm1$ and $\pm2$,
respectively. The $Z=0$ multiplet is the $d=4$, $\cN=4$ SYM
singleton multiplet which has $8+8$ degrees of freedom. All the
other singleton multiplets have $16+16$ degrees of freedom. For
superfield realization of all the singletons listed in this Table,
see Section 3.2. }} \la{tfid}
\end{table}
}

\def\tfis{
\begin{table}[t]
\begin{center}
{\footnotesize \tabcolsep=1mm
\begin{tabular}{|c|cccccccccccccc|}\hline
& & & & & & & & & & & & &
& \\
{\large${}_{\ell}\backslash s$} & $0$ & \ns{$\ft12$} & $1$ &
\ns{$\ft32$} & $2$ & \ns{$\ft52$} & $3$ & \ns{$\ft72$} & $4$ &
$\ft92$ & $5$ & $\ft{11}2$ &
$6$ & $\cdots$ \\
& & & & & & & & & & & & & & \\ \hline & & & & & & & & & & & & &
& \\
$0$ & $42$ & $48$ & $27$ & $8$ & $1$ &
& & & & & & & & \\
$1$ & $1$ & $8$ & $28$ & $56$ & $70$ & $56$ & $28$ & $8$ & $1$ &
& & & &  \\
$2$ & & & & & $1$ & $8$ & $28$ & $56$ & $70$
& $56$ & $28$ & $8$ & $1$  &  \\
$3$ & & & & & & & & & $1$ & $8$ & $28$ & $56$ & $70$
& $\cdots$ \\
$4$ & & & & & & & & & & & &
& $1$ & $\cdots$ \\
$\vdots$ & & & & & & & & & & & & & &  \\ \hline
\end{tabular}}
\end{center}
\caption{{\footnotesize The symmetric tensor product of two $d=4$,
$\cN=4$ SYM singletons arranged into levels $\ell=0,1,2...$ of
$PSU(2,2|4)$ multiplets. The entries denote $USp(8)$
representations: $28=27+1$, $56=48+8$, $70=42+27+1$, which branch
under $SU(4)\times U(1)_Y $ as follows: $1=1_0$,
$8=4_1+\bar{4}_{-1}$, $27=15_0+6_{2}+\bar{6}_{-2}$,
$42=20'_0+10_2+\bar{10}_{-2}+1_4+\bar{1}_{-4}$ and
$48=20_1+\bar{20}_{-1}+4_3+\bar{4}_{-3}$. Each entry also carries
$SO(4)\simeq SU(2)_L\times  SU(2)_R \subset SO(4,2)$ spins
$(j_L,j_R)$. The total spin $s$ is defined as $s=j_L+j_R$ and the
$U(1)_Y$ charge is given by $Y=2(j_R-j_L)$. The level $\ell=0$
multiplet is the $D=5$, $\cN=8$ supergravity multiplet. The level
$\ell=1$ multiplet is the massless Konishi multiplet. The level
$\ell\geq 0$ multiplets have $(4\ell+1)\times 2^8$ degrees of
freedom. The states in the $s\leq \ft12$ sector arise as the
physical states in the master scalar field $\Phi$, as shown in
Table \ref{tfc}. For $s\geq 1$, the states with $Y=0,\pm 1$ arise
in the sector of the master gauge field $A_\mu$ corresponding to
the generators of $hs(2,2|4)$ listed in Table \ref{tfa}. Those
with $Y=\pm 2, \pm 3, \pm 4$ arise in the master scalar field
$\Phi$. With the exception noted in Table \ref{tfc}, these have
dual gauge fields corresponding to the generators of $hs(2,2|4)$
listed in Table \ref{tfb}. The minimal bosonic truncation of the
spectrum is obtained by keeping the maximum spin fields at each
level and the scalar at level $\ell=1$.}} \la{tfis}
\end{table}
}

\def\tfa{
\begin{table}[h!!!]
\begin{center}
{\footnotesize \tabcolsep=1mm
\begin{tabular}{|c|cccccccccccc|}\hline
& & & & & & & & & & & & \\
{\large${}_{\ell}\backslash s$} & $1$ & \ns{$\ft32$} & $2$ &
\ns{$\ft52$} & $3$ & \ns{$\ft72$} & $4$ & $\ft92$ & $5$ &
$\ft{11}2$ &
$6$ & $\cdots$ \\
& & & & & & & & & & & & \\
\hline
& & & & & & & & & & & & \\
$0$ & $15$ & $4$ & $1$ & & & & & & &
& & \\
$1$ & $16'$ & $24$ & $36$ & $24$ & $16'$ & $4$ & $1$
& & & & &  \\
$2$ & & & $1$ & $4$ & $16'$ & $24$ & $36$
& $24$ & $16'$ & $4$ & $1$  &  \\
$3$ & & & & & & & $1$ & $4$ & $16'$ &
$24$ & $36$ & $\cdots$ \\
$4$ & & & & & & & & & & & $1$ &
$\cdots$ \\
$\vdots$ & & & & & & & & & & & &
 \\ \hline
\end{tabular}}
\end{center}
\caption{{\footnotesize The $hs(2,2|4)$ generators with $Y=0,\pm
1$ arranged into levels labeled by
$\ell=\ft14(n_y+n_\yb+n_\th+n_{\bar{\th}}-2)$. The entries are
$SU(4)\times U(1)_Y$ representations as follows: $15=15_0$,
$4=4_1$, $1=1_0$, $16'=15_0+1_0$, $24=20_1+4_1$ and
$36=20'_0+15_0+1_0$, where the $U(1)_Y$ charge is defined by
$Y=n_y-n_\yb$. The $SO(4,1)$ content is given by the highest
weights $m_1\geq m_2\geq \ft12|Y|$ where $m_1=\ft12(n_y+n_\yb)$.
Upon gauging, these generators give rise to spin $s=m_1+1$ gauge
fields which can be used to write a canonical set of covariant
curvature constraints. As a result the gauge fields for $m_2\geq
\ft12|Y|+1$, $s\geq 2$ are auxiliary while those for
$m_2=\ft12|Y|$ contain physical degrees of freedom.}} \la{tfa}
\end{table}
}

\def\tfb{
\begin{table}[h!!!]
\begin{center}
{\footnotesize \tabcolsep=1mm
\begin{tabular}{|c|cccccccccc|}\hline
& & & & & & & & & & \\
{\large${}_{\ell}\backslash s$} & $2$ & \ns{$\ft52$} & $3$ &
\ns{$\ft72$} & $4$ & $\ft92$ & $5$ & $\ft{11}2$ &
$6$ & $\cdots$ \\
& & & & & & & & & & \\ \hline
& & & & & & & & & & \\
$1$ & $16$ & $4$ & $6$ & & & & & & &
\\
$2$ & & & $6$ & $4$ & $16+1$ & $4$ & $6$ & &
&  \\
$3$ & & & & & & & $6$ & $4$ & $16+1$ &
$\cdots$ \\
$\vdots$ & & & & & & & & & &  \\ \hline
\end{tabular}}
\end{center}
\caption{{\footnotesize The $hs(2,2|4)$ generators with
$Y=\pm2,\pm3,\pm4$. The entries are $SU(4)\times U(1)_Y$
representations as follows: $16=10_2+6_2$, $4=4_3$, $6=6_2$ and
$1=1_4$. Further notation is defined in Table \ref{tfa}. These
generators are associated with gauge fields dual to generalized
anti-symmetric tensor fields contained in the scalar master field
$\Phi$; see Table \ref{tfc} for $s\geq 1$.}} \la{tfb}
\end{table}
}

\def\tfc{
\begin{table}[h!!!]
\begin{center}
{\footnotesize \tabcolsep=1mm
\begin{tabular}{|c|cc|cccccccccccc|}\hline
& & & & & & & & & & & & &
& \\
{\large${}_{\ell}\backslash s$} & $0$ & \ns{$\ft12$} & $1$ &
\ns{$\ft32$} & $2$ & \ns{$\ft52$} & $3$ & \ns{$\ft72$} & $4$ &
$\ft92$ & $5$ & $\ft{11}2$ &
$6$ & $\cdots$ \\
& & & & & & & & & & & & & & \\ \hline & & & & & & & & & & & & &
& \\
$0$ & $42$ & $48$ & $\underline{6}$ & & & & &
& & & & & & \\
$1$ & $1$ & $8$ & $\underline{6}$ & $\underline{4}$ &
$16+\underline{1}$ & $4$ & $6$ & & & & & & &
\\
$2$ & & & & & & & $6$ & $4$ & $16+1$ &
$4$ & $6$ & & & \\
$3$ & & & & & & & & & & & $6$ &
$4$ & $16+1$ & $\cdots$ \\
$\vdots$ & & & & & & & & & & & & & &  \\ \hline
\end{tabular}}
\end{center}
\caption{{\footnotesize The physical fields contained in the
master scalar field $\Phi$ arising in the $hs(2,2|4)$ gauge theory
in $D=5$. The entries are the following $SU(4)\times U(1)_Y$
representations for $s<1$:
$42=20'_0+10_2+\bar{10}_{-2}+1_4+\bar{1}_{-4}$,
$48=20_1+\bar{20}_{-1}+4_3+\bar{4}_{-3}$, $8=4_1+\bar{4}_{-1}$ and
$1_0$; for $s\geq 1$: $6_2$, $4_3$, $16=10_2+6_2$ and $1_4$. The
spin $s\geq 1$ sector is realized in the field theory in terms of
two-form potentials and their higher spin generalizations. These
fields obey self-duality in $D=5$ and have dual one-form gauge
fields corresponding to the generators given in Table \ref{tfb},
with the exception of the underlined representations, which have
no one-form duals. Here the form degree refers to the number of
curved indices as opposed to the tangential multi-spinor indices
arising from the $(y,\yb)$-expansion.}} \la{tfc}
\end{table}
}

\def\tsd{
\begin{table}[h!!!]
\begin{center}
{\footnotesize \tabcolsep=1mm
\begin{tabular}{|c|cccccccc|}\hline
& & & & & & & &  \\
{\large${}_{|Z|}\backslash s$} & $0$ & \ns{$\ft12$} & $1$ &
\ns{$\ft32$} & $2$ & \ns{$\ft52$} &
$3$ & $\cdots$ \\
& & & & & & & &   \\ \hline
& & & & & & & &   \\
$0$ & $5$ & $4$ & $1$ & & & & & \\
$\ft12$ & $4$ & $6$ & $4$ & $1$ & & & &  \\
$1$ & $1$ & $4$ & $6$ & $4$ & $1$ & & & \\
$\ft32$ & & $1$ & $4$ & $6$ & $4$ & $1$ & &  \\
$2$ & & & $1$ & $4$ & $6$ & $4$ & $1$  & \\
$\vdots$ & & & & &\vdots & & & \\ \hline
\end{tabular}}
\end{center}
\caption{{\footnotesize The $d=6$, $\cN=(2,0)$ singletons. The
quantity $Z$ denotes the $SU(2)_Z$ spin defined in Section 2.3.
The entries denote $USp(4)_Y \simeq SO(5)\times U(1)_Y$
representations, which are irreducible except $6=5+1$. The
$U(1)_Y$ charges of $1$, $4$, $5$ are $0$, $\pm 1$ and $\pm 2$,
respectively. The $SO(6)$ highest weights $(n_1,n_2,n_3)$
associated with each entry are given by $n_1=n=2=n_3=s$, and the
$AdS_7$ energy by$E_0=s+2$. The level $\ell=0 (Z=0)$ multiplet is
the $d=6$, $\cN=(2,0)$ tensor singleton; see Section (3.3) for
superfield realization of all the singletons shown in the Table,
and composites formed out of the tensor singleton. }} \la{tsd}
\end{table}
}

\def\tss{
\begin{table}[t]
\begin{center}
{\footnotesize \tabcolsep=1mm
\begin{tabular}{|c|cccccccccccccc|}\hline
& & & & & & & & & & & & &
& \\
{\large${}_{\ell}\backslash s$} & $0$ & \ns{$\ft12$} & $1$ &
\ns{$\ft32$} & $2$ & \ns{$\ft52$} & $3$ & \ns{$\ft72$} & $4$ &
$\ft92$ & $5$ & $\ft{11}2$ &
$6$ & $\cdots$ \\
& & & & & & & & & & & & & & \\ \hline & & & & & & & & & & & & &
& \\
$0$ & $14$ & $16$ & $15$ & $4$ & $1$ &
& & & & & & & & \\
$1$ & $1$ & $4$ & $16'$ & $24$ & $36$ & $24$ & $16'$ & $4$ & $1$ &
& & & &  \\
$2$ & & & & & $1$ & $4$ & $16'$ & $24$ & $36$
& $24$ & $16'$ & $4$ & $1$  &  \\
$3$ & & & & & & & & & $1$ & $4$ & $16'$ & $24$ & $36$
& $\cdots$ \\
$4$ & & & & & & & & & & & &
& $1$ & $\cdots$ \\
$\vdots$ & & & & & & & & & & & & & &  \\ \hline
\end{tabular}}
\end{center}
\caption{{\footnotesize The symmetric tensor product of two $d=6$,
$\cN=(2,0)$ tensor singletons arranged into levels $\ell=0,1,2...$
of $OSp(8^*|4)$ multiplets. The entries denote $SO(5)\times
U(1)_Y$ representations as follows: $14=14_0$, $16=16_1$,
$15=10_0+5_2$, $4=4_1$, $16'=10_0+5_2+1_2$, $24=16_1+4_1+4_3$,
$36=14_0+5_0+1_0+10_2+5_2+1_4$. The $SO(6)\subset SO(6,2)$ highest
weights $(n_1,n_2,n_3)$ associated with each entry are given by
$n_1=s$, $n_2=\ft12|Y|$ and $n_3=\ft12 Y$. The level $\ell=0$
multiplet is the $D=7$, $\cN=2$ supergravity multiplet. The level
$\ell\geq 0$ supermultiplets contain $\ft13
(\ell+1)(2\ell+1)(4\ell+3)\times 2^8$ degrees of freedom. The
states with $|Y|\leq 1$, $s\geq 1$ are expected to arise in the
sector of the master gauge field $A_\mu$ corresponding to the
generators given in Table \ref{tsa}. The states with $s\leq
\ft12$, or $|Y|\geq 2$ and $s\geq 1$, which are listed in Table
\ref{tsc}, are expected to arise in a quasi-adjoint master
zero-form $\Phi$. With a few low-lying exceptions which are given
in Table \ref{tsc}, these are generalized Hodge duals of the
$|Y|\geq 2$ sector of the master gauge field $A_\mu$ which
corresponds to the $hs(8^*|4)$ generators listed in Table
\ref{tsb}. The minimal bosonic truncation of the spectrum is
obtained by keeping the maximum spin fields at each level and the
scalar at level $\ell=1$.}} \la{tss}
\end{table}
}

\def\tsa{
\begin{table}[h!!!]
\begin{center}
{\footnotesize \tabcolsep=1mm
\begin{tabular}{|c|cccccccccccc|}\hline
& & & & & & & & & & & & \\
{\large${}_{\ell}\backslash s$} & $1$ & \ns{$\ft32$} & $2$ &
\ns{$\ft52$} & $3$ & \ns{$\ft72$} & $4$ & $\ft92$ & $5$ &
$\ft{11}2$ &
$6$ & $\cdots$ \\
& & & & & & & & & & & & \\
\hline
& & & & & & & & & & & & \\
$0$ & $10$ & $4$ & $1$ & & & & & & &
& & \\
$1$ & $10$ & $20$ & $20'$ & $20$ & $10$ & $4$ & $1$
& & & & &  \\
$2$ & & & $1$ & $4$ & $10$ & $20$ & $20'$
& $20$ & $10$ & $4$ & $1$  &  \\
$3$ & & & & & & & $1$ & $4$ & $10$ &
$20$ & $20'$ & $\cdots$ \\
$4$ & & & & & & & & & & & $1$ &
$\cdots$ \\
$\vdots$ & & & & & & & & & & & &
 \\ \hline
\end{tabular}}
\end{center}
\caption{{\footnotesize The $hs(8^*|4)$ generators with $Y=0,\pm
1$ arranged into levels labelled by
$\ell=\ft14(n_y+n_\yb+n_\th+n_{\bar{\th}}-2)$. The entries are
$SO(5)\times U(1)_Y$ representations as follows: $10=10_0$,
$4=4_1$, $1=1_0$, $20=16_1+4_1$ and $20'=14_0+5_0+1_0$, where the
$U(1)_Y$ charge is defined by $Y=n_y-n_\yb$. The $SO(6,1)$ content
is labelled by highest weights $m_1\geq m_2\geq m_3=\ft12|Y|$
where $m_1=\ft12(n_y+n_\yb)$. Upon gauging, these generators give
rise to spin $s=m_1+1$ gauge fields which can be used to write a
canonical set of covariant curvature constraints. As a result the
gauge fields for $m_2\geq \ft12|Y|+1$, $s\geq 2$ are auxiliary
while those for $m_2=\ft12|Y|$ contain physical degrees of
freedom. }} \la{tsa}
\end{table}
}

\def\tsb{
\begin{table}[h!!!]
\begin{center}
{\footnotesize \tabcolsep=1mm
\begin{tabular}{|c|cccccccccc|}\hline
& & & & & & & & & & \\
{\large${}_{\ell}\backslash s$} & $2$ & \ns{$\ft52$} & $3$ &
\ns{$\ft72$} & $4$ & $\ft92$ & $5$ & $\ft{11}2$ &
$6$ & $\cdots$ \\
& & & & & & & & & & \\ \hline
& & & & & & & & & & \\
$1$ & $15$ & $4$ & $6$ & & & & & & &
\\
$2$ & & & $6$ & $4$ & $16$ & $4$ & $6$ & &
&  \\
$3$ & & & & & & & $6$ & $4$ & $16$ &
$\cdots$ \\
$\vdots$ & & & & & & & & & &  \\ \hline
\end{tabular}}
\end{center}
\caption{{\footnotesize The $hs(8^*|4)$ generators with
$Y=\pm2,\pm3,\pm4$. The entries are $SO(6)\times U(1)_Y$
representations as follows: $15=5_2+10_2$, $4=4_3$, $6=5_2+1_2$
and $16=10_2+5_2+1_4$. These generators are associated with gauge
fields dual to generalized anti-symmetric three-form tensor fields
contained in the scalar master field $\Phi$; see Table \ref{tsc}
for $s\geq 1$. Further notation is defined in Table \ref{tsa}.}}
\la{tsb}
\end{table}
}

\def\tsc{
\begin{table}[h!!!]
\begin{center}
{\footnotesize \tabcolsep=1mm
\begin{tabular}{|c|cc|cccccccccccc|}\hline
& & & & & & & & & & & & &
& \\
{\large${}_{\ell}\backslash s$} & $0$ & \ns{$\ft12$} & $1$ &
\ns{$\ft32$} & $2$ & \ns{$\ft52$} & $3$ & \ns{$\ft72$} & $4$ &
$\ft92$ & $5$ & $\ft{11}2$ &
$6$ & $\cdots$ \\
& & & & & & & & & & & & & & \\ \hline & & & & & & & & & & & & &
& \\
$0$ & $14_0$ & $16_1$ & $\underline{5}_2$ & & & &
& & & & & & & \\
$1$ & $1_0$ & $4_1$ & $\underline{6}_2$ & $\underline{4}_3$ &
$15_2+\underline{1}_4$ & $4_3$ & $6_2$ & & & &
& & & \\
$2$ & & & & & & & $6_2$ & $4_3$ & $15_2+1_4$
& $4_3$ & $6_2$ & & & \\
$3$ & & & & & & & & & & & $6_2$
& $4_3$ & $15_2+1_4$ & $\cdots$ \\
$\vdots$ & & & & & & & & & & & & & &  \\ \hline
\end{tabular}}
\end{center}
\caption{{\footnotesize The physical fields expected to arise in
the master scalar field $\Phi$ in the $hs(8^*|4)$ gauge theory in
$D=7$. The entries are $SO(6)\times U(1)_Y$ representations, where
$6=5+1$ and $15=10+5$. The spin $s\geq 1$ sector is expected to be
realized in $\Phi$ in terms of three-form potentials and their
higher spin generalizations. These fields obey self-duality in
$D=7$ and have dual one-form gauge fields corresponding to the
generators given in Table \ref{tsb}, with the exception of the
underlined representations, which have no one-form duals. Here the
form degree refers to the number of curved indices as opposed to
the tangential multi-spinor indices arising from the
$(y,\yb)$-expansion.}} \la{tsc}
\end{table}
}

\begin{document}

\hfill{CTP-TAMU-08/02}
\\[-20pt]

\hfill{UU-01-10}
\\[-20pt]

\hfill{hep-th/0205131}

\hfill{\today}

\vspace{20pt}
\begin{center}


{\huge\bf Massless Higher Spins and Holography}\\


\vspace{30pt}
{\sf\Large E. Sezgin}\\[5pt]

{\it\small Center for Theoretical Physics, Texas A\&M University,
College Station, TX 77843, USA\\ {\sf Email:
sezgin@physics.tamu.edu}}\vspace{15pt}

{\sf\Large P. Sundell}\\[5pt]
{\it\small Department for Theoretical Physics, Uppsala
Universitet, Sweden\\ {\sf Email: P.Sundell@teorfys.uu.se}}


\vspace{30pt} {\bf\Large Abstract}\end{center}

We treat free large $N$ superconformal field theories as
holographic duals of higher spin (HS) gauge theories expanded
around AdS spacetime with radius $R$. The HS gauge theories
contain massless and light massive AdS fields. The HS current
correlators are written in a crossing symmetric form including
only exchange of other HS currents. This and other arguments point
to the existence of a consistent truncation to massless HS fields.
A survey of massless HS theories with $32$ supersymmetries in
$D=4,5,7$ (where the 7D results are new) is given and the
corresponding composite operators are discussed. In the case of
$AdS_4$, the cubic couplings of a minimal bosonic massless HS
gauge theory are described. We examine high energy/small tension
limits giving rise to massless HS fields in the Type IIB string on
$AdS_5 \times S^5$ and M theory on $AdS_{4/7}\times S^{7/4}$. We
discuss breaking of HS symmetries to the symmetries of ordinary
supergravity, and a particularly natural Higgs mechanism in $AdS_5
\times S^5$ and $AdS_4\times S^7$ where the HS symmetry is broken
by finite $g_{\rm YM}$. In $AdS_5 \times S^5$ it is shown that the
supermultiplets of the leading Regge trajectory cross over into
the massless HS spectrum. We propose that $g_{\rm YM}^2=0$
corresponds to a critical string tension of order $1/R^2$ and a
finite string coupling of order $1/N$. In $AdS_7 \times S^4$ we
give a rotating membrane solution  coupling to the massless HS
currents, and describe these as limits of Wilson surfaces in the
$A_{N-1}(2,0)$ SCFT, expandable in terms of operators with
anomalous dimensions that are asymptotically small for large spin.
The minimal energy configurations have semi-classical energy $E=s$
for all $s$ and the geometry of infinitely stretched strings with
energy and spin density concentrated at the endpoints.

\setcounter{page}{1}

\pagebreak


\section{Introduction}


The strong form of the Maldacena conjecture states that Type IIB
closed string theory on $AdS_5\times S^5$ with $N$ units of
five-form flux and string coupling $g_s$ corresponds to $d=4$,
$\cN=4$ SYM theory with $SU(N)$ gauge group and Yang-Mills
coupling $g_{\rm YM}^2=g_s$ \cite{malda,klebanov,witten}. This
conjecture has been primarily tested for $N>> g_{\rm YM}^2 N>>1$,
where supergravity is a valid approximation \cite{rev,df}. It is
natural to study the correspondence for $N>>1>>g_{\rm YM}^2 N$,
and possibly $g_{\rm YM}^2=0$, where the SYM theory becomes a
theory of a free $SU(N)$ valued $d=4$, $\cN=4$ vector singletons.
At weak 't Hooft coupling $\lambda\equiv g_{\rm YM}^2 N<<1$ the
natural gauge invariant operators are composite single-trace
operators which can be arranged into `trajectories' according to
the value of the twist $E-s$, where $E$ is the conformal dimension
and $s$ is the spin. The twist is the anomalous contribution to
$E$, which becomes small at weak 't Hooft coupling and large $N$.

A basic observation \cite{polyakov} is the non-intersection
principle in a CFT which states that as the coupling varies there
cannot be any mixing between operators that are not mixing already
at the free level. This applies to both the spectrum of composite
operators of 4d SYM in the limit $N>>1>>g_{\rm YM}^2 N$ and the
spectrum of vertex operators of the sigma model for $N>> g_{\rm
YM}^2 N>>1$. Thus an important test of Maldacena conjecture is to
verify that the trajectories of SYM operators with constant twist
cross over into the closed string Regge trajectories.

In this paper we shall show that this is indeed the case for the
leading trajectories, which consist of the states with minimal $E$
for fixed $s$. In fact, on the SYM side the leading trajectory,
i.e. the operators with minimal twist, consists of bilinear higher
spin (HS) tensors. In the free limit, these have twist $2$ and the
$s\geq 1$ sector coincides with the space of conserved HS
currents. General aspects of these currents have been discussed in
\cite{anselmi,vcurrents}. The precise spectrum of twist $2$
operators and the corresponding HS symmetry algebra extension of
the conformal/AdS group was constructed in \cite{us1,us2} using
group theoretic methods which shows that the twist $2$ operators
in fact form an irreducible `gauge' multiplet of the HS algebra.

In \cite{us1,us2} it was also shown how to describe the HS gauge
multiplet on the bulk side at the level of a linearized AdS field
theory containing HS gauge fields as well as other interesting HS
fields generalizing the self-dual two-form of the supergravity
multiplet contained in the spectrum. This immediately raises the
following questions; is it possible to extend this picture to an
interacting theory of massless HS fields in $AdS_5$, and if so, is
this theory the result of a consistent truncation of the full
closed string theory in the limit $N>>1>> g_{\rm YM}^2 N$?

There is increasing evidence for the existence of an interacting
5D massless HS gauge theory \cite{us1,us2,5dv1,5dv2,mikhailov}.
This theory has been constructed at the linearized level
\cite{us1,us2}, and certain cubic interactions of the minimal
bosonic theory have already been constructed \cite{5dv2}. The
structures involved are natural generalizations of those in
$D=4$, and we expect that a similar development will unfold in
$D=5$.

For those readers not too familiar with massless HS gauge
theories, in this paper we review some of their basic properties
in dimensions of interest, namely, $D=4,5,7$. The general
formulation of interacting massless HS gauge theory has been known
in $D=4$ for quite some time \cite{4dv1} (see, \cite{vr2} for a
review). In testing the free CFT/HS gauge theory correspondence
ideas, it is important to exhibit the couplings of the HS gauge
theory. The $D=4$, $\cN=8$ theory has been examined in great
detail in \cite{us3,us4}. In $D=4$ the basic interactions are
contained in a minimal bosonic model which can be embedded as a
consistent truncation into HS gauge theories with $\cN\geq 0$. The
explicit couplings of the minimal bosonic model in $D=4$ are given
in a generally covariant curvature expansion scheme in
\cite{ssc,cubic}. Here we shall summarize the results of
\cite{cubic} at the level of cubic couplings. The analogous
bosonic truncation in $D=5$ was given in \cite{us1} and in $D=7$
in \cite{us7d}, though the full interactions still remains to be
found. In this paper we also give the symmetry algebra and
massless spectrum of the $D=7$, $\cN=2$ HS gauge theory.

The issue of consistent truncation is crucial since the subleading
trajectories in the gauge theory correspond to massive AdS fields
which are light, meaning that their AdS energies are not separated
from the massless ones by a mass-gap. Here it is important to note
that regardless of the detailed structure of the bulk
interactions, it is still possible \cite{edseminar} to arrange the
effective bulk action into a $1/N^2$ expansion such that its
extremum reproduces the $1/N^2$ expansion of the correlators of
the composite operators of the $SU(N)$ invariant singleton theory.
In fact, this expansion remains highly nontrivial even in the
limit $g_{\rm YM}^2 =0$ \cite{su1,su2}. In particular, if one sets
to zero all the massive fields on the boundary, then the extremum
of the full effective action should reproduce the correlators of
the bilinear twist $2$ operators. The massive fields may still
become excited in the bulk, if massless fields act as sources for
massive fields. If this is the case, then the massless HS gauge
theory cannot serve as a good approximation for studying these
processes, not even as an effective theory since it is not
possible to eliminate the light massive fields while preserving
locality (the non-localities which one encounters in massless HS
theory are not that bad). Thus, for the massless HS gauge theory
to be relevant, it must be possible to consistently set the
massive fields to zero in the full theory, at least in the leading
nontrivial order in the $1/N^2$ expansion.

There are several ways to test this consistent truncation.
Firstly, it requires consistent interactions among massless
fields, for which there are many indications as already mentioned.
Given the consistent equations of motion or action for the
massless fields, one must then compute the bulk tree amplitudes,
which by definition will only contain massless excitations in the
internal lines, and check that they correspond to the correlators
of bilinear composite operators computed in the singleton theory
\cite{cubic,wip}. This direct method is technically rather
involved, however, and in this paper we instead provide indirect
evidence for consistent truncation by examining the nature of the
correlators between bilinear operators in singleton theories with
large $N$. We also suggest that the arguments given in
\cite{min2,rastelli,pioline} for the consistent truncation of Type
IIB and eleven-dimensional supergravities on $AdS_{4/7}\times
S^{7/4}$ to gauged supergravity carry over to the HS context.

In this paper, we also emphasize the fact that the relations
between the closed string parameters $g_s$ and $\alpha'$ in
$AdS_5\times S^5$ and the SYM parameters $g_{\rm YM}^2$ and $N$
have so far been tested only in the limit $N>>g_sN>>1$. In this
regime the relations can be derived e.g. by first identifying the
gauge theory parameters with the closed string parameters in flat
10D spacetime, and then use D3-brane soliton description to
interpolate from flat spacetime down to $AdS_5\times S^5$. Since
only $16$ supersymmetries are preserved globally by the D3 brane,
there may be string corrections to this computation.

It is important to note that the strong coupling tests of the
AdS/CFT duality which are based on exact calculations on the SYM
side ( see, for example, \cite{zarembo} and references therein)
are still limited on the bulk side in that they do not go beyond
the leading $\sqrt{\l}$ approximation to the closed string theory
in AdS background. As $g_{\rm YM}^2N$ becomes small (keeping $N$
large), we do not know  the precise relations between closed
string parameters in AdS and the gauge theory parameters. It is
clear that the string coupling $g_s$ decreases and the sigma model
coupling $\alpha'/R^2$ increases as $g_{\rm YM}^2$ decreases. We
shall speculate that the bulk parameters approach critical values
as $g_{\rm YM}^2=0$ where the bulk theory is described by closed
string theory with coupling $1/N^2$ and a singleton worldsheet CFT
based on critical level $k$ affine $PSU(2,2|4)$ algebra, and that
the left- and right-moving singleton spin fields can be used in
the construction of vertex operators describing massless HS fields
in the bulk. The level $k$ is related to the worldsheet sigma
model coupling constant, i.e. $\alpha'/R^2=l_s^2/R^2$. The
corrected relations between the closed string parameters in
$AdS_5\times S^5$ and the gauge theory parameters we propose are
given by

\bea g_{\rm s}&=&f_1(\l)g_{\rm YM}^2\ ,\quad l_{\rm
s}=f_2(\l)R\ ,\la{eq11}\w2 f_1(\l)&\sim& 1\ ,\quad
f_2(\l)\sim\l^{-1/4} \quad \mbox{for $\l>>1$} \ ,\nn\w2
f_1(\l)&\sim& {1\over \l}\ ,\quad  f_2(\l)\sim 1\ ,\quad \mbox{for
$\l<<1$}\ .\la{eq12}\eea

Another aspect of massless higher spins and holography which we
emphasize in this paper is a Higgs mechanism by which the HS
symmetries are spontaneously broken \cite{edseminar} down to the
symmetries of ordinary supergravity. This phenomenon is best
studied in the case of $AdS_5 \times S^5$, primarily due to the
fact that the free boundary SYM theory can be continuously
deformed by switching on the coupling constant $g_{YM}$. As a
result, the HS currents with spin $s>2$ will no longer be
conserved. The resulting anomalies in the conservation laws for
these currents are encoded in operators which can be coupled to
Higgs fields which undergo their landmark shift transformations.
Consequently, the Higgs mechanism mentioned above is expected to
take place.

Given that the full interacting HS theory theory in 5D is still
not known, of course we cannot work out the details of the Higgs
mechanism here. However, we do provide kinematic framework for it
which suggests that the Higgsing phenomenon takes place in an
infinite number of massless ${\cal N}=8$, $D=5$ multiplets
containing HS fields, which together with the supergravity
multiplet make up the spectrum of the massless HS gauge theory. In
particular, we focus on the Higgsing of the Konishi multiplet,
which has $s_{\rm max}=4$ and is expected to play an important
role in study of the first massive Type IIB closed string level,
and outline how the Higgs mechanism can be extended to all the HS
multiplets. This phenomenon of anomalies in the HS current
conservation laws in the boundary having a holographic dual
description in the bulk as spontaneous breaking of the
corresponding HS gauge symmetries is similar to a phenomenon of a
chiral $U(1)$ anomaly having its gravity dual in a particular
$AdS_5$ supergravity, as has been recently shown in \cite{kw}.

Switching our discussion to the case of M theory, we first recall
that the Maldacena conjecture \cite{malda,rev} states the
equivalence between M theory on on $AdS_{4/7}\times S^{7/4}$ with
$N$ units of $7/4$-form flux on $S^{7/4}$, and superconformal
field theories (SCFT) with $16$ supercharges describing the low
energy dynamics of $N$ parallel coinciding M$2/5$ branes in flat
eleven dimensional spacetime. Apparently, these SCFTs are isolated
fixed points of the renormalization groups (RG) that do not admit
any marginal deformations, with or without preservation of
supersymmetry. Consequently they do not admit any coupling
constants and Lagrangian descriptions. The main window for viewing
these strongly coupled theories is therefore through the bulk
supergravity, which is a valid approximation to M theory at fixed
energies provided $N>>1$. This corresponds to a subset of the SCFT
operators with fixed conformal dimensions as $N>>1$
\cite{malda,rev}. Recently other limits of the correspondence
based on considering large internal spin have been proposed
\cite{ppwave}.

In analogy with Type IIB closed string theory on $AdS_{5}\times
S^{5}$, it is natural to ask whether M theory on $AdS_{4/7}\times
S^{7/4}$ has an unbroken phase in which M theory corrections
become relevant at fixed energy and the effective description of
the bulk theory becomes a HS gauge theory with holographic dual
given by a free SCFT in $d=3$ or $d=6$. In other words, we wish to
examine whether it is possible to have a `phase diagram' with two
fixed points, one corresponding to the free singleton SCFT
describing the unbroken HS phase and another one corresponding to
the strongly coupled SCFT describing the broken phase. From the
bulk point of view the broken phase is described by membranes
interacting in the flat eleven dimensional center of
$AdS_{4/7}\times S^{7/4}$, while the unbroken phase, which is
specific to $AdS$, is described by membranes interacting close to
the boundary of $AdS$.

By examining the RG flows on M2/5 branes and D2/4 branes we are
led to propose that the relevant free SCFTs in $d=3,6$ are
described by free $SU(N)$ valued $OSp(8|4)$ singletons and free
$SU(N)$ valued  $d=6$, $\cN=(2,0)$ tensor singletons. These
theories have of course figured in the literature before (see, for
example, \cite{pioline}), and have been used in many circumstances
in order to unravel information about the strongly coupled SCFTs
\footnote{Free singletons, which form $N-1$ plets of the Weyl
group of $SU(N)$, appear in various `trivial' IR limits describing
stacks of separated branes sitting at certain orbifold
singularities \cite{seiberg}. These free singletons should not be
confused with the $SU(N)$ valued singletons, though they are
curious from the HS perspective and they should presumably be
included into the phase diagram as separate HS phases. }.
Our point here is that due to the salient features of the large
$N$ limit the free SCFTs make sense on their own as holographic
images of the interesting unbroken phases of M theory. Technically
speaking, large $N$ implies factorization and $1/N$ expansion of
correlators which can be matched with the expansion of the bulk
amplitudes in terms of the fundamental Planck scale.

As in the case of the Type IIB theory, an important issue is
whether there is a consistent truncation down to a massless
sector. The ideas for examining this are similar to those
described above for the Type IIB theory. The $D=4$ case is
particularly tractable as in this case we already know the full
form of the interactions among massless HS fields, which makes it
possible to test directly the consistent truncation without first
having to construct the interactions.

An intriguing feature of the proposed unbroken phases of M theory
on $AdS_{4/7}\times S^{7/4}$ is that the spectrum is discrete and
that there is a finite coupling, $1/N$. Thus the unbroken phases
of M theory appears to be on the same footing as the unbroken
phase of the Type IIB theory on $AdS_5\times S^5$. This suggests
that the unbroken phases in $AdS_{4/7}\times S^{7/4}$ are theories
of M2 branes with fixed tension.

To gather further evidence for this, we examine a family of
rotating membrane solutions in $AdS_7\times S^4$ that are curved
space generalizations of those given in flat spacetime in
\cite{rot1,rot2} and membrane analogs of the string solutions
found recently in \cite{gkp2} (which in fact describe the leading
Regge trajectory states). The minimal energy configurations have
semi-classical energy $E=s$ for all $s$, and the geometry of
infinitely stretched membranes of zero width, whose energy and
spin densities are concentrated in the asymptotic region. By
examining the supersymmetry enhancement in this region we can
further show that the rotating membranes indeed couple to the
bilinear HS currents in the SCFT.

There is an important difference between the membrane solitons and
the string solitons given in \cite{gkp2}. The string solitons
couple to operators whose anomalous dimensions become
asymptotically small only for large $s$, $(E-s)/s\ra 0$ as
$s\alpha'/R^2\ra\infty$. The membrane solitons, on the other
hand, couple to anomaly free operators for any value of $s$. This
is because they arise by taking the limit of zero width which has
the dual interpretation of shrinking a Wilson surface which means
that the holographic dual flows to the free singleton SCFT in
$d=6$.

We find it rather compelling that relatively simple, free SCFTs
contain information about the unbroken, and perhaps more
fundamental, phases of Type IIB closed string and M theory.
Moreover, this means that the results on free SCFT which are
scattered over the literature can now be given a more direct
physical interpretation.

In AdS/CFT correspondence, it is important that both the bulk and
the boundary theories admit $1/N$ expansions which define the
physically relevant, i.e. asymptotically convergent, expansions.
In the unbroken HS phase, the bulk side may also admit a strongly
coupled closed string/membrane sigma model description, which we
propose has large, but fixed, coupling given by a critical
tension, as mentioned above. In any event, consistent truncation
makes it possible to directly test the AdS/CFT correspondence
using only the action for the massless HS fields which does not
require strongly coupled sigma model computations.

The breaking of the HS symmetries requires the inclusion of Higgs
fields whose interactions require us to go beyond the consistent
truncation to massless fields. Whether this can be done at the
level of some effective field theoretical construction in the
bulk or whether it requires extracting information from the
strongly coupled sigma model is not clear at present. Here we can
only speculate that the large amount of symmetry present in the
unbroken phase should make the critical string and membrane sigma
models amenable to exact methods.

This paper is organized as follows. In Section 2, the properties
of HS gauge theories in $D=4,5,7$ are reviewed, including their
underlying symmetry algebras and field contents. The results for
the HS superalgebra and spectrum in $7D$ are new. In Section 3,
the composite singleton operators corresponding to the massless
states of HS gauge theories, their KK towers and Higgs multiplets
are discussed. In Section 4, important aspects of the CFT/HS gauge
theory correspondence, and in particular the $1/N$ expansion in
the free CFT on the boundary are described. In Section 5, the $5D$
HS gauge theory as the bulk theory arising in the critical limit
of Type IIB string theory and a Higgs mechanism breaking the HS
gauge symmetries down to those of ordinary supergravity are
discussed. In Section 6, first the $CFT_3$/HS gauge theory
correspondence for M theory on $AdS_4\times S^7$ is described.
Then, the minimal bosonic truncation of the theory and its cubic
interactions are described. In Section 7, first the $CFT_6$/HS
gauge theory correspondence for M theory on $AdS_7\times S^4$ is
discussed. Then our rotating membrane solution in $AdS_7\times
S^4$ is given and its properties and relevance to the 7D HS gauge
theory are described. Section 8 is devoted to a summary and
discussion. In Appendix A, we present several tables which show
various sectors of the massless HS gauge theory spectra in
$D=5,7$. In Appendix B, we summarize the UIRs and BPS states of
the maximal AdS superalgebras in $D=4,5,7$. In Appendix C and D,
we collect further group theoretical information that is useful
for Section 2 and 3.


\section{Massless Higher Spin Gauge Theories in $D=4,5,7$}


HS gauge theories are generally covariant theories which admit AdS
as a vacuum and have an infinite number of local HS
supersymmetries based on HS superalgebras which are infinite
dimensional extension of the finite dimensional AdS superalgebras
\cite{kv2} . The fundamental UIRs of the HS super algebras in
$D=d+1=4,5,7$ dimensions are ultra-short $d$-dimensional conformal
supermultiplets, which we will refer to as singletons
\footnote{In $d=4,6$, these are usually referred to as doubletons,
due to the fact that their oscillator construction is based on two
sets of oscillators as opposed to a single set of oscillators used
in $d=3$.}.
Gauging of such a HS superalgebra yields a $D$-dimensional theory
based on a massless HS supermultiplet given by the symmetric
product of two singletons. In this paper we shall focus our
attention on the HS extension of the AdS superalgebras in
$D=4,5,7$ with 32 real supersymmetries because these are the most
natural ones to explore from the string/M theory point of view. In
Section 8, we shall comment on possible extensions to higher D and
higher number of supersymmetries.

The massless HS multiplet is an infinite tower of massless AdS
supermultiplets with supergravity at the lowest level. One key
property is the fact that a HS gauge theory in $D>3$ cannot be
consistently truncated to an AdS supergravity. Basically, this is
due to the fact that derivatives of lower spin fields serve as
sources for HS fields, and it can also be seen from the structure
of the OPE of free field theory stress-energy tensors in $d>2$
\cite{anselmi}. However, in $D=4,5,7$ there exist minimal bosonic
truncations which have remarkably simple physical field content,
namely massless fields of spin $s=0,2,4,6,...$ described by doubly
traceless, symmetric tensors $\phi_{\mu_1\dots\mu_s}$. The
embedding of these theories in their supersymmetric extensions is
explained in Tables \ref{tfos}, \ref{tfis} and \ref{tss}.

As for the full and covariant (i.e. background independent)
interactions among the massless fields, they are known in the 4D
theory \cite{4dv1,ssc,cubic}. A condensed account of how to
extract cubic couplings in $D=4$ will be given in Section 6.2. The
fully interacting theories in $D=5,7$ have not yet been
constructed, though the results obtained so far are promising
\cite{5dv2,us1,us2,us7d}.

We next list the HS superalgebras in $D=4,5,7$, their singleton
and massless representations, and how the latter ones are
assembled into master 1-form and master 0-form fields. The results
in $D=4,5$ were obtained in \cite{us3,us2}. The minimal bosonic HS
algebra in $7D$ was obtained in \cite{us7d}. The results presented
here for its supersymmetric extension are new.


\subsection{The $D=4$, $\cN=8$ Massless HS Gauge Theory}


The 4D HS algebra $hs(8|4)$ is realized in terms of oscillators
obeying the following algebra \cite{kv2,us3,us4}
\footnote{The algebra $hs(8|4)$ is called $shs^E(8|4)$ in
\cite{us1} and $shs^E(8,4|0)$ in \cite{kv2}, where it is also
shown to be isomorphic to $ho(8;8|4)$.}

\be y_\a\star y_\b=y_\a y_\b+i\e_{\a\b}\ ,\quad y_\a\star
\yb_{\ad}= y_\a \yb_{\ad}\ ,\quad (y_\a)^\dagger=\yb_{\ad}\ , \ee
\be \th^i\star\th^j=\th^i\th^j+\d^{ij}\ ,\quad
(\th^i)^\dagger=\th^i\ , \ee

where $y_\a$ $(\a=1,2)$ is a Weyl spinor which is a Grassmann even
generator of a Heisenberg algebra, and $\th^i$ ($i=1,...,8$) is a
Grassmann odd generator of an $SO(8)$ Clifford algebra. The
$\star$ denotes the associative product between oscillators. The
products on the right hand sides are Weyl ordered, so that for
example $y_\a y_\b=y_\b y_\a$ and $\th^i\th^j=-\th^j\th^i$. Using
the above contraction rules it is straightforward to compute the
$\star$ product between two arbitrary Weyl ordered polynomials of
oscillators.

The algebra $hs(8|4)$ consists of arbitrary Grassmann even and
anti-hermitian polynomials $P(y,\yb,\th)$ that are sums of
monomials of degree $4\ell+2$ where $\ell=0,1,2,...$, which will
be referred to as the level index. The Lie bracket between $P,Q\in
hs(8|4)$ is given by $[P,Q]_\star$. Thus, denoting by $P^{(\ell)}$
an $\ell$th level monomial, the commutation relations have the
schematic form

\be [P^{(\ell_1)},P^{(\ell_2)}]_\star =\sum_{|\ell_1-\ell_2|\leq
\ell \leq \ell_1+\ell_2} P^{(\ell)}\ .\la{ell}\ee

In particular, the zeroth level of $hs(8|4)$ is the maximal finite
subalgebra $OSp(8|4)$ whose generators schematically take the form

\bea Q_{\a i}&=&y_\a\th_i\ , \quad\quad  {\bar Q}_{\ad
i}=\yb_{\ad}\th_i\ ,
\quad\quad U_{ij}=\th_i\th_j\ , \nn\\
M_{\a\bd}&=&y_\a \yb_{\bd}\ ,\quad\ M_{\a\b}=y_\a y_\b\ , \quad\
M_{\ad\bd}=\yb_{\ad}\yb_{\bd}\ . \eea

A generator $P^{(\ell)}$ in the $\ell$th level of $hs(8|4)$ can be
expanded as

\be P^{(\ell)}(y,\yb,\th)=
\sum_{\footnotesize\ba{c}m+n+p\\=4\ell+2\ea}{1\over m!\,n!\,p!}~
\yb^{\ad_1}\cdots \yb^{\ad_m} y^{\b_1}\cdots y^{\b_n}
\th^{i_1}\cdots \th^{i_p}
P_{\ad_1\dots\ad_m\,\b_1\dots\b_n\,i_1\dots i_p}\ .\la{p4}\ee

The spins of the components are given by $s=\ft12(m+n)$. The
components with integer spin are Grassmann even and those with
half-integer spin are Grassmann odd. Bosons are in the $1$, $28$
and $35_\pm$ irreps of $SO(8)$ and fermions in the $8$ and $56$
irreps. The reality properties follow from $P^\dagger=-P$.

A UIR of $OSp(8|4)$ is denoted by $D(E_0,s;a_1,a_2,a_3,a_4)$,
where the notation is explained in Appendix B. The fundamental UIR
of $OSp(8|4)$, which is also a UIR of $hs(8|4)$, is the
ultra-short singleton \cite{g4,ns}

\be D(\ft12,0;0,0,0,1)\oplus D(1,\ft12;0,0,1,0)\ . \la{4ds}\ee

By taking products of singletons we obtain further unitary
representations of $hs(8|4)$. Two singletons yield $OSp(8|4)$
weight spaces corresponding to massless $AdS_4$ fields with
$E_0=s+1$ \cite{ff,gw,bsst}.

The massless sector of the $hs(8|4)$ gauge theory is formulated in
terms of an $hs(8|4)$ valued master gauge field $A_\m(y,\yb,\th)$
(with expansion given by \eq{p4}) and a master zero-form
$\Phi(y,\yb,\th)$ in a quasi-adjoint representation of $hs(8|4)$
with expansion

\be \Phi(y,\yb,\th)= \sum_{\footnotesize\ba{c}-m+n+p\\=0\mbox{ mod
$4$} \ea}{1\over m!\,n!\,p!}~ \yb^{\ad_1}\cdots \yb^{\ad_m}
y^{\b_1}\cdots y^{\b_n} \th^{i_1}\cdots \th^{i_p}
\Phi_{\ad_1\dots\ad_m\,\b_1\dots\b_n\,i_1\dots i_p}\ .\la{phi4}\ee

The reality condition on $\Phi$ is discussed in detail in
\cite{us3,us4}. The gauging gives rise to a set of field equations
for physical fields (the action still remains to be found) whose
spectrum is given by the symmetric product of two singletons which
is given in Table \ref{tfos}. The physical spin $s\geq 1$ fields
are the gauge fields in $A_\m(y,\yb,\th)$ that correspond to
$hs(8|4)$ generators in \eq{p4} satisfying $|m-n|\leq 1$. Those
with $m=n$ contain the vierbein and its HS generalizations, while
those with $|m-n|=1$ contain the gravitini and their HS
generalizations. The physical fields with $s\leq \ft12$ arise in
$\Phi(y,\yb,\th)$ as the components in \eq{phi4} with $m+n\leq 1$.
The remaining fields in $A_\m$ and $\Phi$ are auxiliary and given
in terms of derivatives of the independent fields.\\

So far we have discussed the free massless HS gauge theory. The
general formulation of {\it interacting} massless HS gauge theory
has been given in $D=4$ \cite{4dv1} (see, \cite{vr2} for a
review), and examined in detail for $\cN=8$ \cite{us3,us4}. There
exists a minimal bosonic truncation of this theory whose spectrum
consist the physical states with spin $s=0,2,4,...$, each
occurring once. This theory exhibits the basics of any HS gauge
theory rather well and it will be discussed in considerable detail
in Section 6.2, which is based on \cite{cubic}.

\tfos


\subsection{The $D=5$, $\cN=4$ Massless HS Gauge Theory}


The 5D HS superalgebra $hs(2,2|4)$ \cite{us2} is realized in terms
of the following oscillators
\footnote{The algebra $hs(2,2|4)$ is called $ho_0(1,0|8)$ in
\cite{5dv2}.}

\be y_\a\star \yb_\b =y_\a \yb_\b+C_{\a\b}\ ,\quad y_\a\star y_\b=
y_\a y_\b\ ,\quad (y^\dagger i\C^0C)_\a=\yb_\a\ , \ee \be
\th^i\star\bar{\th}_j=\th^i\bar{\th}_j+\d^i_j\ ,\quad \th^i\star
\th^j=\th^i\th^j\ ,\quad  (\th^i)^\dagger=\bar{\th}_i\ ,  \nn \ee

where $y_\a$ $(\a=1,\dots,4)$ is a Grassmann even Dirac spinor and
$\th^i$ ($i=1,...,4$) is a Grassmann odd $SO(6)\simeq SU(4)$
spinor. The charge conjugation matrix $C_{\a\b}$ is
anti-symmetric. The algebra $hs(2,2|4)$ consists of Grassmann even
and anti-hermitian polynomials $P(y,\yb,\th,\bar{\th})$ that are
sums of monomials of degree $4\ell+2$ ($\ell=0,1,2,...$) that are
invariant under the $U(1)_Z$ generated by

\be Z=\ft12 (\yb y+\bar{\th}\th);\ee

and traceless in their spinor indices:

\be P^{(\ell)}(y,\yb,\th,\bar{\th})=
\sum_{\footnotesize\ba{c}m+n+p+q\\=4\ell+2\\
m+p=n+q \ea}{1\over m!\,n!\,p!\,q!}~ \yb^{\a_1}\cdots \yb^{\a_m}
y^{\b_1}\cdots y^{\b_n} \th^{i_1}\cdots
\th^{i_p}\bar{\th}_{j_1}\cdots \bar{\th}_{j_q}\, P_{\a_1\dots
i_p}{}^{j_1\dots j_q}\ ,\la{p5}\ee

where

\be C^{\a_1\b_1}P_{\a_1\dots\a_m\,\b_1\dots\b_n\,i_1\dots
i_p}{}^{j_1\dots j_q}=0\ ,\qquad  P_i{}^i=0\ .\ee

The tracelessness of $P_i{}^j$ means the removal of the outer
$U(1)_Y$ automorphism generator

\be Y=\bar{\th}\th\ .\ee

The Lie bracket between $P,Q\in hs(2,2|4)$ is given by
$[P,Q]_\star/\cI$ where $\cI$ is the ideal generated by elements
of the form

\be \sum_{n=1}^{\infty} P_n(y,\yb,\th,\bar{\th})\star
\underbrace{Z\star\cdots\star Z}_{\mbox{$n$ factors}}\
,\la{ideal}\ee

where $P_n$ are polynomials which are traceless in their spinor
indices. The structure of the Lie bracket is similar to \eq{ell}.

The zeroth level of $hs(2,2|4)$ is the maximal finite subalgebra

\be PSU(2,2|4)=PU(2,2|4)/U(1)_Z\ ,\ee

where $PU(2,2|4)$ is the centrally extended superalgebra (with
$31$ bosonic generators). The $PSU(2,2|4)$ generators are realized
schematically as

\be Q_{\a i}= y_\a\tb_i\ ,\quad {\bar Q}_\a^i = \yb_\a\th^i\
,\quad M_{\a\b} = \yb_\a y_\b -\ft14 C_{\a\b} (\yb y)\ , \quad
U^i{}_j= \tb^i\th_j-\ft14 \d^i_j (\tb\th) \ . \ee

The Lorentz spin of a generator in \eq{p5} is given by
$(j_L,j_R)=(\ft12 m,\ft12 n)$ and the $U(1)_Y$ charge by $Y=p-q$.
The components with integer $j_L+j_R$ are Grassmann even and those
with half-integer $j_L+j_R$ are Grassmann odd. Bosons are in the
$1_0$, $15_0$, $20'_0$, $6_2$, $10_2$ and $1_4$ irreps of
$SU(4)\times U(1)_Y$ and fermions in the $4_1$, $4_3$ and $10_3$
irreps. The reality properties follow from the condition
$P^\dagger=-P$ which, in particular, implies that the irreps with
$Y=0$ are real. The generators of the algebra are summarized in
Table \ref{tfa} and Table \ref{tfb} in Appendix A.

A UIR of $SU(2,2|4)$ is denoted by $D(E_0,j_L,j_R;a_1,a_2,a_3)_Y$
where the notation is explained in Appendix B. The fundamental
UIRs of $SU(2,2|4)$ are the ultra-short singletons given in Table
\ref{tfid} \cite{g5} in Appendix A. Due to the modding out of the
ideal $\cI$ generated by elements of the form \eq{ideal} the
fundamental UIR of $hs(2,2|4)$ is the singleton with vanishing $Z$
charge, i.e. the Maxwell supermultiplet \cite{dp,gm,nst,g5}

\bea && D(1,0,0;0,1,0)_{_0} \oplus D(\ft32,\ft12,0;1,0,0)_{_{-1}}
\oplus D(\ft32,0,\ft12;0,0,1)_{_{1}} \nn\w2 && \oplus
D(2,1,0;0,0,0)_{_{-2}} \oplus D(2,0,1;0,0,0)_{_2}\ .
\la{maxwell}\eea

By taking products of this multiplet we obtain further unitary
representations of $hs(2,2|4)$. In particular, the product of two
singletons yields massless $AdS_5$ fields whose energies, which
are given by $E_0=2+j_L+j_R$ saturate the unitarity bound of a
continuous series (denoted as series A in Appendix B)\cite{dp,g5}.

\tfis

The massless sector of the $hs(2,2|4)$ gauge theory is formulated
in terms of an $hs(2,2|4)$ valued master gauge field
$A_\m(y,\yb,\th,\bar{\th})$ and a master zero-form
$\Phi(y,\yb,\th,\bar{\th})$ in a certain quasi-adjoint
representation of $hs(8|4)$ \cite{us1,us2}, which contains the
Weyl tensors and the extra `matter' fields given in Table
\ref{tfc} in Appendix A. The gauging gives rise to physical fields
whose spectrum is given by the symmetric product of two singletons
given in Table \ref{tfis}. The fields with $|Y|\leq 1$ and
$|Y|\geq 2$, $j_L+j_R\geq \ft12$ carry $SO(4,1)$ weights such that
the analysis of the curvature constraints in this sector is
analogous to that in $D=4$. In the $|Y|\geq 2$, $j_L+j_R\geq 1$
sector the fields carry $SO(4,1)$ weights that require a separate
analysis. One finds that \cite{us2} the physical fields arise as
two-form potentials in $\Phi$ obeying the odd-dimensional
self-duality equation $B_2=\star dB_2$ or higher-spin analogs of
this equation (such equations have been more recently studied in
the lightcone gauge in \cite{met1}). The gauge fields in $A_\mu$
with $|Y|\geq 2$ are auxiliary fields which are related to the
independent two-forms in $\Phi$ by generalized Hodge dualization
rules \cite{us2}.

So far we have discussed the free massless HS gauge theory. The
full interacting theory based on $hs(2,2|4)$ has not been
constructed yet. However, the kinematics established in \cite{us2}
and summarized above, together with the already established
principles \cite{4dv1} that govern the structure of the
interacting HS gauge theory in $D=4$, suggest that the full $5D$
interacting theory is perfectly within reach. Indeed certain cubic
interactions of the minimal bosonic HS theory in $5D$ have already
been constructed by Vasiliev \cite{5dv2}.


\subsection{The $D=7$, $\cN=2$ Massless HS Gauge Theory}


The linearized gauge theory of the minimal bosonic HS subalgebra,
$hs(8^*)$ in $D=7$ was introduced in \cite{us7d}. Here we shall
construct its supersymmetric extension $hs(8^*|4)$. This algebra
is realized in terms of the following oscillators :

\be y_\a\star \yb_\b=y_\a \yb_\b+C_{\a\b}\ ,\quad y_\a\star y_\b=
y_\a y_\b\ ,\quad (y^\dagger i\C^0C)_\a=\yb_\a\ , \ee \be
\th^i\star\bar{\th}^j=\th^i\bar{\th}^j+\O^{ij}\ ,\quad \th^i\star
\th^j=\th^i\th^j\ ,\quad  (\th^i)^\dagger=\bar{\th}^j\O_{ji}\ ,
\ee

where $y_\a$ $(\a=1,\dots,8)$ is a Grassmann even Dirac spinor and
$\th^i$ ($i=1,...,4$) is a Grassmann odd Dirac spinor of
$SO(5)\simeq USp(4)$. The charge conjugation matrix $C_{\a\b}$ is
symmetric and $\O^{ij}$ is the antisymmetric $USp(4)$ invariant
tensor. The algebra $hs(8^*|4)$ consists of Grassmann even and
anti-hermitian polynomials $P(y,\yb,\th,\bar{\th})$ that are sums
of monomials of degree $4\ell+2$ ($\ell=0,1,2,...$) which are
invariant under the $SU(2)_Z$ generated by

\be Z_3=\ft14( \yb^\a y_\a+\bar{\th}^i\th_i)\ ,\quad Z_+=\ft14(
y^\a y_\a+\th^i\th_i)\ ,\quad
Z_-=\ft14(\yb^\a\yb_\a+\bar{\th}^i\bar{\th}_i)\ . \ee

and traceless in their spinor indices. The Lie bracket between
$P,Q\in hs(8^*|4)$ is given by $[P,Q]_\star/\cI$ where $\cI$ is
the ideal generated by elements of the form

\be \sum_{n=1}^{\infty}P_n^{I_1\dots
I_n}(y,\yb,\th,\bar{\th})\star Z_{I_1}\star \cdots \star Z_{I_n}\
,\la{e}\ee

where $P_n^{I_1\dots I_n}(y,\yb,\th,\bar{\th})$ has an expansion
in terms of traceless, Weyl ordered multispinors and the $SU(2)_Z$
indices $I_1\dots I_n$ are symmetric. The structure of the Lie
bracket is again similar to \eq{ell}. The zeroth level of
$hs(8^*|4)$ is the maximal finite subalgebra $OSp(8^*|4)$ realized
schematically as

\be Q_{\a i}=y_\a\bar{\th}_i-\yb_\a\th_i\ ,\quad
M_{\a\b}=\yb_{[\a}y_{\b]}\ ,\quad U_{ij}=\th_{(i}\bar{\th}_{j)}\ .
\ee

An $\ell$th level generator $P^{(\ell)}$ in $hs(8^*|4)$ can be
expanded as

\be P^{(\ell)}(y,\yb,\th,\bar{\th})=\sum_{\footnotesize
\ba{c}m+n+p+q\\=4\ell+2\\
m+p=n+q \ea}{1\over m!\,n!\,p!\,q!}~ \yb^{\a_1}\cdots \yb^{\a_m}
y^{\b_1}\cdots y^{\b_n} \th^{i_1}\cdots
\th^{i_p}\bar{\th}^{j_1}\cdots \bar{\th}^{j_q} P_{\a_1\dots j_q}\
,\la{7de}\ee

where the components are traceless in their Lorentz spinor indices
and belong to super Young tableaux with two rows of length
$2\ell+1$. A single box in the super Young tableaux represents the
superoscillator $\x^A=(y^\a,\th^i)$ or
$\bar{\x}^A=(\yb^\a,\bar{\th}^i)$. An arbitrary Weyl ordered
monomial in these superoscillators corresponds to a super Young
tableaux with two rows. The restriction $m+n=p+q$ in \eq{7de}
(i.e. equal number of $\xi^A$ and $\bar{\xi}^A$) follows from the
condition $[Z_3,P]_\star=0$, while the condition
$[Z_{\pm},P]_\star=0$ rules out super Young tableaux with rows of
unequal length. The resulting super Young tableaux of width
$2\ell+1$ splits into a set of Young tableaux of  spinors. Each
$SO(6,2)$ Young tableaux branches into a set of Young tableaux of
$SO(6,1)$ spinors. The spinorial $SO(6,1)\times SO(5)$ Young
tableaux can be converted into tensorial ones by multiplying with
appropriate Dirac matrices of both groups. The resulting $SO(5)$
irreps are $1_0, 5_0, 10_0, 14_0, 1_2, 5_2, 10_2, 1_4 $ in the
bosonic sector and $4_1, 16_1, 4_3$ in the fermionic sector, where
the subscripts denote the $U(1)_Y$ charge defined as

\be Y= n_{\bar\th}-n_\th\ ,\ee

with $n_{\bar\th}=q$ and $n_\th=p$, as specified in the expansion
\eq{7de}. The $SO(6,1)$ highest weights $(m_1,m_2,m_3)$ are given
by

\be m_1=2\ell+1-\ft12(n_\th+n_{\bar\th})\geq m_2\geq m_3=\ft12
|Y|\ .\ee

Note that since $P$ is assumed to be Grassmann even the components
in \eq{7de} with integer weights are Grassmann even and those with
half-integer weights are Grassmann odd. The reality properties
follow from $P^\dagger=-P$. As a result, all $SO(6,1)\times SO(5)$
representations obey symplectic reality conditions. For example,
the supercharge $Q_{\a i}$ obey a symplectic Majorana condition so
that it has 32 real components:

\be \bar{Q}_{\a i}\equiv (Q_{\b j})^\dagger(i\C^0C)_{\b\a}\O_{ji}
= Q_{\a i}\ .\ee
These results are summarized in Tables \ref{tsa} and \ref{tsb} in
Appendix A.

A UIR of $OSp(8^*|4)$ is denoted by $D(E_0,J_1,J_2,J_3;
a_1,a_2)_Y$ where the notation is explained in Appendix B. The
fundamental UIRs of $OSp(8^*|4)$ are the singletons given in Table
\ref{tsd} \cite{g7} in Appendix A. The singleton which is singlet
of $SU(2)_Z$ also forms an UIR of $hs(8^*|4)$. This singleton is
the $(2,0)$ tensor multiplet \cite{g7,nst,gvw,min}

\bea&&  D(2,0,0,0;0,1)_{_0} \oplus D(\ft52,1,0,0;1,0)_{_1} \oplus
D(\ft52,0,0,1;1,0)_{_{-1}}  \nn\w2 && \oplus D(3,0,0,2;0,0)_{_2}
\oplus D(3,2,0,0;0,0)_{_{-2}} \ .\eea

By taking products of this singleton one obtains further unitary
representations of $hs(8^*|4)$. In particular, the square yields
massless $AdS_7$ fields with energy $E_0=4+s$ where $s\equiv J_1$.
These energies belong to an isolated series ( denoted as series B
in Appendix B)\cite{min}, unlike in $D=4,5$ where the massless
fields have energies that saturate a continuous series (the
continuous series is saturated by lowest weight spaces arising in
the product of three singletons).

The superalgebra $hs(8^*|4)$ has a minimal bosonic HS subalgebra
$hs(8^*)$ whose representation theory and gauging was described in
\cite{us7d}. We shall assume that the massless sector of the
$hs(8^*|4)$ gauge theory is formulated in terms of an $hs(8^*|4)$
valued master gauge field $A_\m(y,\yb,\th,\bar{\th})$ and a master
zero-form $\Phi(y,\yb,\th,\bar{\th})$ in a quasi-adjoint
representation
\footnote{This representation was defined for $hs(8^*)$ in
\cite{us7d}. Its generalization to $hs(8^*|4)$ will not be given
here.}
of $hs(8^*|4)$ and that the gauging gives rise to physical fields
whose spectrum is given by the symmetric product of two tensor
singletons listed in Table \ref{tss}. The gauge fields with $Y=0$
carry $SO(6,1)$ weights which are similar to those in the minimal
bosonic theory \cite{us7d}. The gauge fields with $|Y|\geq 1$
carry $SO(6,1)$ weights which are analogous to those carried by
the $|Y|\geq1$ fields in the $hs(2,2|4)$ theory in $D=5$. Thus we
expect that the physical fields with $|Y|\leq 1$, $s\geq 1$ arise
in $A_\m$. The remaining physical fields, which have $s\leq
\ft12$, or $s\geq 1$ and $|Y|\geq 2$, must arise in $\Phi$ and be
those given by Table \ref{tsc} in Appendix A. In particular we
expect that the physical fields with $|Y|\geq 2$, $s\geq 1$ arise
as three-form potentials in $\Phi$ obeying the odd-dimensional
self-duality equation $B_3=\star dB_3$ and their HS analogs, and
that the corresponding gauge fields in $A_\mu$ are related to
these three-forms by generalized Hodge dualization rules analogous
to those found in the $hs(2,2|4)$ theory in $D=5$ \cite{us2}.

So far we have discussed the free massless HS gauge theory. The
interacting theory has not been constructed yet. However, the
kinematics of theory established here, together with what we know
in lower dimensions about interacting HS gauge theories should
help a great deal in such a construction. In particular, there are
many parallels with the kinematics of the $5D$ HS gauge theory
which is noteworthy.

\tss


\section{Composite Operators in Singleton Theories}


In this section we describe the singleton theories in $d=3,4,6$
with $16$ supersymmetries that are of relevance to the HS gauge
theories in $D=4,5,7$ described in the previous section. We shall
also identify the superfield realization of the HS currents in
terms of these singletons, whenever possible.

Explicit expressions for the supersymmetric currents have been
constructed so far in $d=3$, and for a minimal bosonic truncation,
in arbitrary dimensions. The bosonic currents are formed out of a
set of real scalar singletons and that are primary fields carrying
$SO(d,2)$ lowest weights $(E_0;m_1,\dots m_{[d/2]})=(d-2+s;
s,0,...,0)$, where $s=0,2,4,...$. These tensors are conserved
currents for $s\geq 2$. The minimal bosonic HS theories are still
`maximal' in the sense that the twist $d-2$ currents with even
spin are the only composites which are both conserved and primary.
There are conserved currents with $E_0-s> d-2$ as well as
$E_0-s=d-2$ and odd spin, though these can be shown to be
descendants of those with twist $d-2$ and even spin.


\subsection{The $d=3$, $\cN=8$ Singleton and Its Composites}


The fundamental UIR of $OSp(8|4)$, which is also a UIR of
$hs(8|4)$, is the ultra-short singleton specified in  \eq{4ds}.
This is just the $d=3, {\cN}=8$ scalar multiplet, and its
superfield realization has been known for sometime. In particular,
it has arisen in the superembedding formulation of $M2$-branes
\cite{sb}. Following \cite{fs1}, let us work with a realization
related to the one in \cite{sb} by triality. The singleton
superfield is then carries a spinor representation of $SO(8)$ and
obeys the constraint

\be D_{\a i} \Phi_A = (\C_i)_A{}^{\dB}\,\chi_{\a \dB}\ ,
\la{4dsc}\ee

where $\chi_{\a\dB}$ is a spinor superfield, $i,A,\dB =1,...,8$
label the $8_v,8_s,8_c$ representations of $SO(8)$, respectively,
and $\C$-matrices are the chirally projected $SO(8)$ Dirac
matrices. The singleton superfield $\Phi_A$ carries the irrep
$D(1/2,0;0,0,0,1)$, which belongs to series B and it is BPS 1/2
multiplet.  See Appendix B for notation and further details.

Several composite operators built out of two singletons
superfields $\Phi_A$ and their derivatives are known
\cite{fs1,fs2,fs3,af1,minwalla}. Let us identify those which
correspond to the spectrum of massless field in the $D=4$ HS gauge
theory based on $hs(8|4)$ as shown in Table \ref{tfos}. The level
$\ell=0$ supercurrent is realized as \cite{fs1}

\be J_{AB}= \Phi_A\Phi_B -\ft18\, \d_{AB}\,\Phi^2\ ,\quad\quad
\Phi^2 := \Phi^A \Phi_A\ .\ee

The superfield $J_{AB}$ carries the irrep $D(1,0;0,0,2,0)$ which
belongs to series B and it is BPS 1/2 multiplet. Its lowest
component carries the $35_s$ irrep of $SO(8)$. Together with the
scalars in $35_c$ that arise in the $\theta$-expansion, they form
the $70$-plet corresponding to the $70$ scalar fields of level
$\ell=0$ supergravity multiplet in $AdS_4$ which has $2^8$ degrees
of freedom.

At level $\ell=1$, we have the supercurrent \cite{fs1}

\be J = \Phi^2, \ee

which, as a consequence of the basic singleton constraint
\eq{4dsc}, obeys \cite{fs1}

\be D^{ij} J -trace =0 \ ,\quad\quad\quad  D^{ij} := D^{\a i}
D_\a^j\ . \ee

The superfield $J$ carries the irrep $D(1,0;0,0,0,0)$, which is
semi-short IUR that saturates the unitarity bound of series A. Its
lowest component is a scalar, and another scalar arises in the
$\theta$-expansion. Altogether, $2\times 2^8$ degrees of freedom
arise \cite{af1} and they correspond to the massless fields of
level $\ell=1$ shown in Table \ref{tfos}.

Finally, the level $\ell\ge 2$ supercurrents can be realized in
terms of the singleton superfield as follows \cite{fs1}

\bea J_{\a_1...\a_{4\ell-4}} &=& \sum_{k=0}^{2\ell-2}\,(-1)^k
\large[32i
\partial_{(\a_1\a_2...}\partial_{\a_{2k-1}\a_{2k}}\Phi^A \
\partial_{\a_{2k+1}\a_{2k+2}}
...\partial_{\a_{4\ell-5}\a_{4\ell-4})}\Phi_A \w3 &&
+(\C^i\C^j)_{AB} \,\partial_{(\a_1\a_2...}
\partial_{\a_{2k-1}\a_{2k}}D^i_{\a_{2k+1}}\Phi^A\
D^j_{\a_{2k+2}} \partial_{\a_{2k+3}\a_{2k+4}}
...\partial_{\a_{4\ell-5}\a_{4\ell-4)}}\Phi^B \large]\ \ .\nn \eea

These currents obey the constraint \cite{fs1}

\be D^{i\a} J_{\a\a_2...\a_{4\ell-4}}=0\ . \quad\quad \ell \ge 2\
,\ee

The superfield  $J_{\a\a_2...\a_{4\ell-4}}$ carries the irreps
$D(2\ell-1,2\ell-2;0,0,0,0)$, which is semi-short IUR that
saturates the unitarity bound of series A. Its lowest component is
the current with spin $s_{min}=2\ell-2$ and higher components go
up to $s_{\max}=2\ell+2$. They correspond to the level $\ell\ge 2$
massless multiplets listed in Table \ref{tfos}.

As discussed in the introduction, and to be elaborated further in
Section 5, if interactions can be switched on in the 3d CFT such
that the HS gauge symmetry breaks down to $OSp(8|4)$, then the HS
currents for level $\ell \ge 1$ will no longer be conserved.
Assuming such breaking, we can characterize the anomalies in
conservation law for these currents as

\bea && D^{ij} J -trace = g \S^{ij}-trace \ ,\la{3da}\w2 &&
D^{i\a}J_{\a\a_2...\a_{4\ell-4}}= g \S^i{}_{\a_2...\a_{4\ell-4}}\
, \eea

where $g$ is some coupling constant, and the right hand sides
denote superfields which are to be determined. These superfields
carry the following irreps

\bea \S^{ij}\ : & D(2,0;2,0,0,0)\ , \la{r1}\w2
\S^i{}_{\a_2...\a_{4\ell-4}}\ :& \quad\quad\quad\quad
D(2\ell-\ft12,2\ell-\ft52; 1,0,0,0)\ . \la{r2}\eea

This means that $\S^{ij}$ describes a BPS 1/8 multiplet and it
satisfies the unitarity condition of series B. In \cite{fs1}, a
BPS short multiplet of this type is built out of four singletons
using harmonic superspace technique. In terms of ordinary
superfields we write it as

\be \S^{ij} =
\left(\C_{imnp}\right)_{AB}\,\left(\C_j{}^{mnp}\right)_{CD}\,
\Phi^A\Phi^B\Phi^C\Phi^D-trace\ . \ee

This is just the $35$-plet contained in the symmetric product
$(8_s\times 8_s\times 8_s\times8_s)_S$.  Since the superfield
$\S^{ij}$ represents a BPS 1/8 multiplet, its components go up to
$s_{max}= 7/2$. Therefore, it is natural to consider this
superfield as a candidate for coupling to Higgs superfield in the
bulk which can be eaten by the massless Konishi multiplet to
become massive.

Turning to  the candidate anomaly superfield
$\S^i{}_{\a_2...\a_{4\ell-4}}$ given in \eq{r2}, we observe that
it carries a semi-short IUR that saturates the unitarity bound of
series A. In general, such multiplets have been constructed as
\cite{fs1}

\bea {\cal S}^{[a_i]} &=& \Phi^2\, BPS^{[a_i]}\ , \la{4dc1}\w2
{\cal S}^{\{\m_1...\m_s\} [a_i]}&=&J^{\{\m_1...\m_s\}}\,
BPS^{[a_i]}\ , \la{4dc2} \eea

where $BPS^{[a_i]}$ is any one of the BPS short multiplets listed
in \eq{3dbps1}-\eq{3dbps3}, and $J^{\{\m_1...\m_s\}}$ is a spin
$s$ current. Assuming that the candidate anomaly superfield
$\S^i{}_{\a_2...\a_{4\ell-4}}$ belongs to an irreducible
representation of $OSp(8|4)$, since it is an $8$-plet of $SO(8)$,
it requires the BPS 1/8 multiplet $D(1,0;1,0,0,0)$ and a spin
$s=2\ell-\ft12 $ current in \eq{4dc2}. However, the BPS 1/8
multiplet cannot be built out of one type of singleton field.
Thus, the construction of $\S^i{}_{\a_2...\a_{4\ell-4}}$, which is
important for a  Higgs mechanism that can work at all levels $\ell
\ge 1$, remains an open problem.

The BPS multiplets that can be constructed from the product of one
type of singletons are all the BPS 1/2 and BPS 1/4 multiplets
listed in \eq{3dbps1} and \eq{3dbps2}, and all those BPS 1/8
multiplets listed in \eq{3dbps3} with integer $s$ \cite{fs2} .
These multiplets, as well as the semi-short multiplets discussed
above which make use of them, are likely to play a significant
role in the description of the full HS gauge theory based on
$hs(8|4)$. In particular, the KK supermultiplets associated with
level $\ell$ supermultiplets of the massless HS theory are
expected to be BPS 1/2 multiplets. For example, the level $\ell=0$
multiplet and its KK towers are realized as \cite{af1}

\be D(k/2,0;0,0,k,0)\ :\quad\quad \Phi_{(A_1}\Phi_{A_2}\cdots
\Phi_{A_k)} - {\rm traces} \ ,\quad k=2,3,... \ee

Taking $k=2$ gives the massless supergravity multiplet and
$k=3,4,...$ give their massive KK descendants. Similarly, the
semi-short multiplets \eq{4dc1} and \eq{4dc2} with BPS 1/2
composites carrying the irrep $D(k/2,0;0,0,k,0)$ are candidates
for KK descendants of the level $\ell > 0$ massless multiplets of
the HS gauge theory based on $hs(8|4)$.


\subsection{The $d=4$, $\cN=4$ Singleton and Its Composites}


The fundamental  UIR of $SU(2,2|4)$, which is also a UIR of
$hs(2,2|4)$, is the ultra-short singleton specified in
\eq{maxwell}. This is the $d=4, {\cN}=4$ Maxwell multiplet
realized in terms of superfield $W_{ij}$, where $i=1,...,4$ labels
the $4$-plet of $SU(4)$ and $W^{ij}=-W{ji}$.
\footnote{ This is the unique singleton multiplet of $PSU(2,2|4)$
and it has vanishing $U(1)_Z$ central charge. The centrally
extended $PU(2,2|4)$ superalgebra admits an infinite number of
singleton multiplets. These have $j_R=0$ (their complex conjugates
have $j_L=0$), and $E_0=j_L+1$. Each singleton multiplet forms a
massless UIR of the $d=4, {\cN}=4$ Poincar\'e superalgebra and is
characterized by central charge $\ell = 2|Z|=0,1,2,...$. Viewed as
massless states of $d=4$ Poincar\'e group, they carry Lorentz spin
(i.e. maximum $SO(2)$ helicity) $s=j_L$. The first three levels of
the singleton spectrum shown in Table \ref{tfid} are special
because they are the only singleton multiplets which contain
scalar fields. They are $D(1,0,0;0,1,0)$, $D(1,0,0;1,0,0)$ and
$D(1,0,0;0,0,0)$ with $Z$ charges $(0,1/2,1)$ and they can be
described by superfields $(W_{ij},W^i, W)$, respectively. The
level $\ell\ge $ singletons $D(\ft{\ell}2,\ft{\ell}2-1,0;0,0,0)$
have central charge $Z=\ft{\ell}2$ and can be described by
superfield $\omega_{\a_1...\a_{\ell-2}}$. The constraints
satisfied by all singleton superfields can be found in
\cite{fs1}.}.
It satisfies the following constraints and reality condition

\bea &&D_{\a}^{(i} W^{j)k} = 0 \ ,\quad  {\bar D}_{\a i}
W^{jk}-{\rm trace}=0\ , \nn\w2
&&W_{ij}\equiv(W^{ij})^\dagger=\ft12\e_{ijkl}W^{kl}\ .\la{sc4}
\eea

The singleton superfield $W_{ij}$ carries the irrep
$D(1,0,0;0,1,0)$. It belongs to series C and it describes a BPS
1/2 multiplet. There are several papers which deal with the
construction of the composite operators built out of the Maxwell
(or SYM) singleton. See, for example,
\cite{berg,sc,af2,ffz,af3,flz,fz0,fz,fs6,afsz,hh1}. Here we shall
follow closely the treatment of \cite{fs1}.

For each state in the spectrum of the HS gauge theory listed in
Table \ref{tfis}, one can construct the corresponding conserved
current out of two Maxwell singletons and their derivatives. To
begin with, the level $\ell=0$ supercurrent is contained in the
superfield $J_{ij,kl}$, which is in $20'$ of $SU(4)$ and is given
by \cite{berg,sc}

\be J_{ij,kl}=W_{ij}W_{kl}-\ft1{12}\e_{ijkl}W^{mn}W_{mn}\ .
\la{sugra} \ee

Defining $W^a \equiv (\C^a)_{ij}\,W^{ij}\,(a=1,2,...,6)$, where
$\C^a$ are the chirally projected $SO(6)$ Dirac matrices, the
current superfield \eq{sugra} can equivalently be written as
$J_{ab}=W_aW_b -\ft16 \d_{ab}\,W_cW_c$.

Defining $J_{ij}^{kl}= \epsilon^{klmn} J_{ij,mn}$, on the other
hand, it obeys the constraint \cite{sc}

\be D_{\a}^m J_{ij}^{kl}=\chi_{\a ij}^{mkl}+\d^m_{[i}\l_{\a
j]}^{kl}+\d_{[i}^{[k} \l_{\a j]}^{l]m}\ ,\la{setc} \ee

where $\l$ and $\chi$ are both totally anti-symmetric in lower and
upper indices and totally traceless. The superfield $J_{ij,kl}$
carries the irrep $D(2,0,0;0,2,0)$. It belongs to belongs to
series C and it describes a BPS 1/2 multiplet. Its components can
be shown to contain the composite operators that correspond to the
level $\ell=0$ supergravity multiplet shown in Table \ref{tfis},
and that the components with spin $s\ge 1$ are conserved currents.

The level $\ell=1$ supercurrent is also a special one and is known
as the massless Konishi multiplet. It has the simple form
\cite{sc}

\be J=W_{ij}W^{ij}\ . \la{kon1} \ee

As a result of the basic singleton constraint \eq{sc4}, this
current obeys the constraint

\be D^{ij} J=0\ , \quad\quad\quad  D^{ij} := D^{\a (i} D_\a^{j)}\
. \la{kc1} \ee

This multiplet has $5\times 2^8$ and they precisely correspond to
the level $\ell=1$ massless states shown in Table \ref{tfis}. It
is characterized by the irrep $D(2,0,0;0,0,0)$ carried by its
lowest component. It is a semi-short multiplet which saturates the
unitarity bound of series A.

In the Poincar\'e limit, the states are labelled by the little
group $SO(3)\times SU(4)$. Denoting the irreps by $R_{s}$, where
$R$ is denotes an $USp(8)$ irrep (which should be decomposed into
$SU(4)$ irreps)and $s$ is the $SO(3)$ spin, the level $\ell=1$
massless Konishi multiplet can be obtained by tensoring the level
$\ell=0$ supergravity multiplet with an $SU(4)$ singlet spin $s=2$
state as follows:

\bea {\rm Massless\ Konishi:}\quad &&
(42_0+48_{1/2}+27_{1}+8_{3/2}+1_{2})\times 1_{2}= \nn\w2 &&
1_{0}+8_{1/2}+(27+1)_{1}+(48+8)_{3/2}+(42+27+1)_{2}\nn\w2 &&
+(48+8)_{5/2} + (27+1)_{3}+8_{7/2}+1_{4}  \la{k5} \eea

The massless multiplets arising at level $\ell \ge 2$ in the
spectrum shown in Table \ref{tfis} are generic in their structure.
The corresponding conserved  currents are contained in a
superfield

\be J_{\m_1\m_2...\m_{2\ell-2}}\ ,\quad\quad \ell\ge 2\ ,\ee

which obey the constraints \cite{fs1}

\be ({\bar\s}^{\m_1})_{\ad\b}\,D^{i\b}\,
J_{\m_1\m_2...\m_{2\ell-2}}=0\ , \quad\quad
(\s^{\m_1})_\a{}^{\bd}\,{\bar D}_{i\bd}\,
J_{\m_1\m_2...\m_{2\ell-2}}=0\ . \ee

The superfield $J_{\m_1\m_2...\m_{2\ell-2}}$ carries the irrep
$D(4\ell-2,2\ell-2,2\ell-2;0,0,0)$ and its components have spins
that range from $(2\ell-2)$ to $(2\ell+2)$. This superfield
saturates the unitarity bound of series A and it describes a
semi-short multiplet.

The explicit construction of all the supercurrents in terms of
Maxwell singleton is straightforward but tedious exercise which
apparently has not been carried so far. They are known, however,
for the minimal bosonic truncation of the massless HS gauge theory
in $D=5$ discussed above. They take the form
\cite{vcurrents,mikhailov}

\be j_{\m_1\cdots\m_{2\ell-2}} = \sum_{k=0}^{2\ell-2} {(-1)^k\over
(k!)^2!((s-k)!)^2}\,\del_{\m_1}\cdots
\del_{\m_k}\phi^*\,\del_{\m_{k+1}}\cdots \del_{\m_{2\ell-2}}\phi -
{\rm traces}\ . \ee

So far we have considered free SYM singletons. Switching on the
SYM interactions, the currents listed above for $\ell \ge 1$ will
no longer be conserved. The resulting anomalies can be
characterized as follows

\bea && D^{ij} J = \sqrt \l \S^{ij}\ ,\nn\w2 &&
({\bar\s}^{\m_1})_{\ad\b}\,D^{i\b}\,
J_{\m_1\m_2...\m_{\ell-2}}=\sqrt \l
\S^i_{\m_2...\m_{2\ell-2},{\dot\a}}\ , \nn\w2
&&(\s^{\m_1})_\a{}^{\bd}\,{\bar D}_{i\bd}\,
J_{\m_1\m_2...\m_{2\ell-2}}=\sqrt \l  \S_{i
\m_2...\m_{\ell-2},\a}\ ,\eea

where the constant normalization factor is introduced for later
convenience (see Section 5).

The superfields on the right hand side carry the following UIRs of
$SU(2,2|4)$

\bea \S^{ij}\  & : & \ \ \ D(3,0,0;2,0,0)\ , \la{5r1} \w2
\S_{\m_2...\m_{2\ell-2},{\dot\a}}\ & :& \ \ \
D(2\ell-\ft32,\ell-\ft32,\ell-1;0,0,1)\ ,\la{5r2}\w2
\S_{\m_2...\m_{2\ell-2},\a}\ &:& \ \ \
D(2\ell-\ft32,\ell-1,\ell-\ft32;1,0,0)\ . \la{5r3} \eea

In the interacting SYM singleton theory the anomaly superfield
$\S^{ij}$ takes the well known form (see, for example,
\cite{fz0,hh1}):

\be \S^{ij}= {4\over N^{3/2}}\,{\rm
Tr}\,W^{k(i}W^{j)\ell}W_{k\ell}\ , \ee

where the constant normalization factor is introduced for later
convenience (see Section 5). This superfield belongs to series B
and it describes a BPS 1/8 multiplet. Consequently its components
go up to $s_{max}=7/2$ and therefore it is a candidate for
coupling to Higgs superfield in the bulk which can be eaten by the
massless Konishi multiplet to become massive. All the components
of the massive Konishi multiplet of $PSU(2,2|4)$ have been
tabulated in \cite{af2}.

The candidate  anomaly superfields
$\S^i_{\a_2...\a_{2\ell-2},\a}$, on the other hand carries a
semi-short IUR that satisfy the unitarity bounds of series A or B.
In general, such multiplets have been constructed as \cite{fs1}

\bea {\cal S}^{[a_i]} &=& \Phi^2\, BPS^{[a_i]}\ ,\nn\w2 {\cal
S}^{\{\m_1...\m_s\} [a_i]}&=&J^{\{\m_1...\m_s\}}\, BPS^{[a_i]}\
,\la{4dss} \eea

where $BPS^{[a_i]}$ is any one of the BPS operators listed in
\eq{4dbps1}-\eq{4dbps3}, and $J^{\{\m_1...\m_s\}}$ is a spin $s$
current, to be constructed out of the free SYM singleton in our
case. For the BPS 1/2 and BPS 1/4 cases, both of the above
operators saturate the series A unitarity bound \eq{4ub1}, while
in the case of BPS 1/8, they belong to series B. Assuming that the
candidate anomaly superfield $\S^i_{\a_2...\a_{2\ell-2},\a}$
carries an irreducible representation, and given that it is in
$(100)$ of $SU(4)$, attempting to construct it as in \eq{4dss}
requires the use of BPS 1/8 multiplet $D(3/2,0,0;1,0,0)$, as
follows from \eq{4dbps3}. However, these BPS multiples cannot be
built out of SYM singletons alone \cite{fs1}.

The BPS multiplets that can be constructed out of products of SYM
singleton alone are all the BPS 1/2 and BPS 1/4 multiplets listed
in \eq{4dbps1} and \eq{4dbps2}, and all those BPS 1/8 multiplets
listed in \eq{4dbps3} with integer r \cite{fs2}. These multiplets,
and the semi-short multiplets discussed above which make use of
them, are likely to play a role in finding the massive states of
the full HS gauge theory based on $hs(2,2|4)$. In particular, the
KK supermultiplets associated with level $\ell$ supermultiplets of
the massless HS theory are expected to make use of the BPS 1/2
states. For example, the level $\ell=0$ multiplet and its KK
towers are realized as \cite{witten,ffz,af3}

\be D(k,0;0,k,0)\ :\quad\quad W_{(a_1}W_{a_2}\cdots W_{a_k)} -
{\rm traces} \ , \quad k=2,3,... \ee

Setting $k=2$ gives the massless supergravity multiplet and
$k=3,4,...$ their massive KK descendants. Similarly, the
semi-short multiplets \eq{4dc1} and \eq{4dc2} involving the BPS
1/2 composites carrying the irrep $D(k,0;0,k,0)$ are candidates
for KK descendants of the level $\ell > 0$ massless multiplets of
the HS gauge theory based on $hs(2,2|4)$.


\subsection{The $d=6$, ${\cN}=(2,0)$ Tensor Singleton and Its Composites}


The fundamental UIRs of $OSp(8^*|4)$ are the singletons given in
Table \ref{tsd} in Appendix A \cite{min,g7}. Each row in the Table
denotes an irreducible singleton multiplet. The superfield
realization of the $6d$ singletons have been studied by several
authors. Here we shall follow \cite{fs1,fs2} where several
references to earlier literature can also be found. There exist
several papers on the construction of the composite operators out
of the 6d singletons as well; see \cite{hst, fs1,fs2,fs7}, for
example.

There exist an infinite set of singletons of $OSp(8^*|4)$. They
are shown in Table \ref{tsd} and listed in Appendix B. The $(2,0)$
tensor singleton is the only one which is singlet under an
$SU(2)_Z$ defined in Section 2.3. Here we shall focus our
attention to the level $\ell=0$ singleton described by the
superfield $W^{ij}$ which forms the tensor multiplet of
$d=6,{\cN}=(2,0)$ Poincar\'e supersymmetry, since all the HS gauge
theory states will be formed out of them. To begin with, we shall
take a single copy of the tensor multiplet. Abelian nature of the
singletons is essential for the construction of conserved
currents. The superfield $W_{ij}$ satisfies the following
constraints and reality condition \cite{hst}

\be D_{\a}^{(i} W^{j)k} = 0 \ ,\quad \quad {\bar
W}_{ij}=\O_{ik}\O_{jl} W^{kl}\ . \ee

The singleton superfield $W_{ij}$ carries the irrep
$D(2;0,0,0;0,1)$ which belongs to series D and it is BPS 1/2
supermultiplet. For each state in the spectrum of the HS gauge
theory listed in Table \ref{tss}, one can construct the
corresponding conserved current out of two tensor singletons and
their derivatives. To begin with, the level $\ell=0$ supercurrent
is contained in the superfield $J_{ij,kl}$, which is in $14$-plet
of $USp(4)$ and is given by

\be J_{ij,kl}=W_{ij}W_{kl}-\ft16\O_{k[i} \O_{j]\ell}W^{mn}W_{mn}\
. \la{7dsugra} \ee

Defining $W^a \equiv (\C^a)_{ij}\,W^{ij}\, (a=1,2,...,5)$, where
$\C^a$ are the $SO(5)$ Dirac matrices, the current superfield
\eq{7dsugra} can equivalently be written as $J_{ab}=W_aW_b -\ft16
\d_{ab}\,W_cW_c$.

The superfield $J_{ij,kl}$ carries the irrep $D(4;0,0,0;0,2)$,
which belongs to series D and it describes a BPS 1/2 multiplet.
This is the level $\ell=0$ supergravity multiplet shown in Table
\ref{tss}.

The level $\ell=1$ supercurrent is similar to the ones in $d=3,4$
and it takes the form

\be  J=W_{ij}W^{ij}\ . \la{kon6} \ee

This current obeys the constraint \cite{fs1}

\be \e^{\a\b\c\d}\,D_\a^{(i}D_\b^jD_\c^{k)}\,J=0\ . \ee

The superfield $J$ carries the irrep $D(4;0,0,0;0,0)$. It has
$14\times 2^8$ components and it can be obtained group
theoretically by tensoring the level $\ell=0$ supergravity
multiplet with the graviton state which has $14$ degrees of
freedom. It is a semi-short multiplet which belongs to series B.

The massless multiplets arising at level $\ell\ge 2$ in the
spectrum shown in Table \ref{tfis} are generic and the
corresponding conserved currents are contained in the superfield

\be J_{\a_1...\a_{2\ell-2},\b_1...\b_{2\ell-2}}\ ,\quad\quad\quad
\ell \ge 2\ , \la{kon7}\ee

where the $\a$ and $\b$ indices are symmetrized separately. These
current superfields obey the constraint \cite{fs1}

\be \e^{\d\c\a_1\b_1}\,D_\c^i
J_{\a_1...\a_{2\ell-2},\b_1...\b_{2\ell-2}}=0\ . \ee

The superfield $J_{\a_1...\a_{2\ell-2},\b_1...\b_{2\ell-2}}$
carries the irrep $D(2\ell+2;0,2\ell-2,0;0,0)$. It is a semi-short
multiplet which belongs to series B.

An explicit construction of these supercurrents in terms of the
$(2,0)$ tensor singleton apparently has not been carried out so
far. They are known, however, for the minimal bosonic truncation
of the massless HS gauge theory in $D=7$ discussed earlier. They
take the form \cite{vcurrents,mikhailov}

\be j_{\m_1\cdots\m_{2\ell-2}} = \sum_{k=0}^{2\ell-2} {(-1)^k\over
k!(k+1)!(s-k)!(s-k+1)!}\,\del_{\m_1}\cdots
\del_{\m_k}\phi^*\,\del_{\m_{k+1}}\cdots \del_{\m_{2\ell-2}}\phi -
{\rm traces}\ . \ee

So far we have considered free $(2,0)$ tensor singletons.
Interactions for multi-copies of these singletons are not known
and they are expected to be radically different than those
familiar from  ordinary field theory. These interactions are also
expected to break the HS gauge symmetries down to those of level
$\ell=0$ supergravity. Let us characterize the break-down in the
conservation laws of the supercurrents of level $\ell\ge 1$ as
follows

\bea && \e^{\a\b\c\d}\,D_\a^{(i}D_\b^jD_\c^{k)}\,J=g \Sigma^{\d
ijk}\ , \la{6danomaly1}\w2 && \e^{\d\c\a_1\b_1}\,D_\c^i
J_{\a_1...\a_{2\ell-2},\b_1...\b_{2\ell-2}}=g \S^{\d
i}_{\a_2...\a_{2\ell-2},\b_1...\b_{2\ell-2}}\ ,
\la{6danomaly2}\eea

where $g$ is some coupling constant. Unlike in the cases of
$d=3,4$, here we see that the representation content of the
candidate anomaly superfields do not correspond to any BPS short
or semi-short multiplets listed in Appendix B. Of course, here we
are assuming that these anomaly superfields are irreducible. Their
computation from first principles may in principle reveal that
they are reducible, and possibly derivatives of some irreducible
superfields. The nature of the anomaly superfields should also
reflect the fact that there there are no local non-abelian
interactions for tensor fields that can be described by continuous
deformations of the free theory \cite{as}. This is a qualitative
difference between $d=6$ and $d=3,4$, where the free fields admit
SYM deformations (after dualization of a scalar in $d=3$).

The semi-short multiplets, as in 3d and 4d cases, have also been
constructed in terms of building blocks discussed above, and they
take the form \cite{fs1}

\bea {\cal S}^{[a_i]} &=& \Phi^2\, BPS^{[a_i]}\ , \la{6dss1}\w2
{\cal S}^{\{\m_1...\m_s\} [a_i]}&=&J^{\{\m_1...\m_s\}}\,
BPS^{[a_i]}\ , \la{6dss2}\eea

where $BPS^{[a_i]}$ is any one of the BPS operators listed in
\eq{6dbps1} and \eq{6dbps2}, and $J^{\{\m_1...\m_s\}}$ is a spin
$s$ current, which is to be constructed out of the free $(2,0)$
tensor singleton in our case. Both of these saturate the unitary
bound of  series B.

The BPS multiplets that can be constructed out of products of the
tensor singleton alone are all the BPS 1/2 multiplets listed in
\eq{6dbps1} and all those BPS 1/4 multiplets listed in \eq{6dbps2}
with integer q\cite{fs2}. In particular, the level $\ell=0$
multiplet and its KK towers are realized as \cite{fs1}

\be D(2k,0,0,0;0,k)\ :\quad\quad W_{(a_1}W_{a_2}\cdots W_{a_k)} -
{\rm traces} \ ,\quad k=2,3,... \ee

As in the cases of 3d and 4d, here too, setting $k=2$ gives the
massless supergravity multiplet and $k=3,4,...$ give their massive
KK descendants. Similarly, the semi-short multiplets \eq{6dss1}
and \eq{6dss2} with BPS 1/2 composites carrying the irrep
$D(2k,0,0,0;0,k)$ are candidates for KK descendants of the level
$\ell > 0$ massless multiplets of the HS gauge theory based on
$hs(8^*|4)$.


\section{Higher Spin Gauge Theory and Holography}


We shall first discuss some general features of HS gauge
theory/singleton correspondence before we turn to the cases of
interest in Type IIB string theory and M theory. In particular,
the properties of the free boundary CFT's which indicate that the
massless HS gauge theories in the bulk provide effective
descriptions of the full HS gauge theories truncated to their
massless sector will be emphasized.

Consider a CFT$_d$ consisting of $N'$ supersingletons $W^{i}$,
where $i=1,...,N'$ is an internal index and each $W^i$ belongs to
some singleton representation of the superconformal group. Let
each singleton belongs to an irreducible representation of some
internal symmetry group $G$ and consider $G$ invariant composite
operators $\cO$. Our first basic assumption is that the
correlation functions of invariant composite operators factorize
as $N'\ra\infty$. For example, by using the operator product
expansion, a four-point function $<\cO_1\cO_2\cO_3\cO_4>$ can be
decomposed as

\bea &&<\cO_1\cO_2\cO_3\cO_4> =
<\cO_1\cO_2><\cO_3\cO_4>+<\cO_1\cO_2\cO_3\cO_4>_{\rm conn}\ ,\w2
&& <\cO_1\cO_2\cO_3\cO_4>_{\rm conn}= \sum_r
{<\cO_1\cO_2\cO_r><\cO_r\cO_3\cO_4>\over <\cO_r\cO_r>}\ ,\eea

where the disconnected terms are the contributions from the unit
operator and the connected terms are the contributions from the
remaining operators. The factorization means that the the
connected terms are suppressed by powers of $1/N'$:

\be  {<\cO_1\cO_2\cO_3\cO_4>_{\rm conn}\over
<\cO_1\cO_2><\cO_3\cO_4>}\ra 0\quad \mbox{as $N'\ra\infty$\ .}\ee

In general, there can be several parameters in addition to $N'$ in
CFT$_d$. Fortunately, supersymmetry puts considerable amount of
constraint on these possibilities. With application to Type IIB
string and M theory in mind, we shall assume that $G=SU(N)$ and
consider $SU(N)$ valued singleton scalar superfields denoted by
$W^I$, $I=1,...,n$. In this case we have $N'=N^2-1$ and the
singletons transform in the fundamental representation of the
$R$-symmetry group $SO(n)$. For the cases of interest, namely in
$d=3,4,6$, we have in mind the $R$-symmetry groups $SO(8), SO(6)$
and $SO(5)$, respectively, which correspond to $16$ ordinary plus
$16$ special supersymmetries in the CFT$_d$. The $SU(N)$ valued
singletons in $d=4$ are adequate for discussing the tensionless
limit of the Type IIB theory on $AdS_5\times S^5$. The extent to
which $SU(N)$ valued singletons in $d=3,6$ may encode the
properties of (an unbroken phase of) M theory on $AdS_{4/7}\times
S^{7/4}$ is discussed in Sections 6 and 7.

The basic composite operators in CFT$_d$ are primary {\it
bilinear} single-trace operators $\cO_{(2)r}$, where the index $r$
labels collectively the set of $SO(d,2)\times R$-representations
involved \cite{anselmi,vcurrents}. These operators do not mix with
any other operators and provide conserved HS currents with spin
$s\ge 1$, and certain composite operators of lower spin $s<1$.
Together they form an HS multiplet that corresponds in a
one-to-one fashion to an HS multiplet of physical massless bulk
fields
\footnote{In the minimal bosonic truncation this dictionary has
been extended to also include local currents corresponding to the
auxiliary HS gauge fields of the bulk theory \cite{vcurrents}.
This offers an opportunity to compute bulk amplitudes in a first
order formalism.},
$\phi_{(2)r}$. In the supersymmetric singleton models of special
interest to Type IIB/M theory the bilinear primaries are discussed
in Section 3 and the corresponding massless spectra are listed in
Tables \ref{tfos}, \ref{tfis} and \ref{tss}.

The free CFT$_d$ also contains composite operators which are $p$th
order monomials in the basic singleton and its derivatives. Those
composites which are not normal ordered products of other
composites as $N'\ra \infty$ are interpreted as massive
single-particle states in AdS. We shall denote these operators and
the corresponding massive bulk fields by ${\cal O}_{(p)r}$ and
$\phi_{(p)r}$, respectively, where $p\geq 3$ and $r$ is an
additional set of indices labeling the $SO(d,2)\times R$ weights.
The massiveness means that there is no shortening of the
associated $SO(d,2)$ weight spaces. This implies that the massive
operators are not conserved and hence there are no gauge
symmetries associated with the corresponding massive AdS fields.
However, as discussed in the previous section, some of the massive
operators belong to shortened supermultiplets, provided that the
superconformal weights saturate certain unitarity bounds or belong
to discrete series. This is the case, for example, for 1/2 BPS KK
modes and the Higgs multiplets listed in the previous section.

For fixed $p$ the space of massive operators $\cO_{(p)r}$ clearly
decomposes into irreducible HS multiplets, though the
representation theory of HS algebras, such as their root
structure, has not yet been developed far enough to characterize
the precise `lowest' weights carried by these multiplets ( see
\cite{us7d} for a discussion of this point).

Composite operators which are normal ordered products of other
composite operators as $N'\ra\infty$ are interpreted as
many-particle states. In the case of $SU(N)$ valued singletons,
the single-particle states, $\cO_{(p)r}$ ($p=2,3,...$) are given
in the large $N$ limit by single-trace operators. The $n$-particle
states, which we shall denote by $\cO_{(p_1,\dots,p_n)r}$ are
given in this limit by multi-trace operators in the form of normal
ordered products of single trace operators $\cO_{(p_i)r_i}$ and
their derivatives, $p_i=2,3,...$, $i=1,...,n$.

For finite $N$ there is mixing between the single-trace and
multi-trace operators \cite{anselmi,su2}. This is because
$n$-particle states in the bulk couple to operators that
diagonalize the two-point function:

\be <\cO_{R}\cO_{S}>=\eta_{RS}\ ,\la{etars}\ee

where $R=(p_1,\dots,p_n)r$ and $\eta_{RS}$ is an $N$-independent
diagonal matrix. For example, consider the minimal bosonic
truncation based on a single $SU(N)$ valued singleton field $W$.
The bilinear and tri-linear composites, which have to be
single-traces, do not mix. However, the quartic composites do mix,
and they do so as follows. The diagonal scalar states of energy
$\D=2d-4$ are given schematically by

\be \cO_{(4)}=J_{(4)}+ f J_{(2,2)}\ ,\quad
\cO_{(2,2)}=J_{(2,2)}-{2 f\over 1+f^2}J_{(4)}\ ,\la{422}\ee

where  $f(N)={a\over N}+{b\over N^3}$, with $a$ and $b$ being some
constants, and

\be J_{(4)}\sim {1\over N^2}:\tr(W^4):\ ,\quad\quad J_{(2,2)}\sim
:\tr(W^2)\tr(W^2):\ , \ee

are assumed to be normalized such that

\be <J_{(4)}J_{(4)}>=\D^4\ ,\quad <J_{(4)}J_{(2,2)}>=f(N) \D^4\
,\quad <J_{(2,2)}J_{(2,2)}>=\D^4\ ,\ee

where $\D=|x|^{-d+2}$ is the singleton propagator.

Having introduced the main notation and kinematics, we now
continue with the discussion of the factorization of correlators
as $N\ra \infty$. From \eq{etars} it follows that as $N\ra \infty$
a general $n$-point correlator either vanishes if $n$ is odd or
can be written as the sum of products of $n/2$ two-point
functions. Thus, in the limit $N\ra \infty$ the singleton CFT$_d$
describes an `anti-holographic' bulk theory of free $n$-particle
states corresponding to $\cO_{(p_1,\dots,p_n)r}$. For finite $N$,
the $1/N$ corrections to the singleton CFT give rise to nontrivial
connected parts of the correlation functions which we wish to
represent as anti-holographic interactions. To be more precise we
wish to examine whether the $SU(N)$ valued singleton field theory
is the holographic dual of an interacting $(d+1)$-dimensional
theory based on an effective action, consisting of a bulk term
plus a boundary term

\be \C_{\rm eff}[\phi_{(p)r}]=\C_{\rm
eff,bulk}[\phi_{(p)r}]+\C_{\rm eff,boundary}[\phi_{(p)r}]\
,\la{geff}\ee

which admit perturbative expansions in powers of $1/N$ around an
$AdS_{d+1}$ vacuum\cite{edseminar}. The boundary term plays a role
in representing certain correlators, such as the extremal
correlators discussed below, that cannot be reproduced from a bulk
action. This boundary term is needed because the variational
principle requires $\C_{\rm eff} [\f_{(p)r}]$ to be stationary
when the fields are varied subject to Dirichlet conditions. The
variation of $\C_{\rm eff, boundary}[\f_{(p)r}]$ should therefore
cancel the total derivatives from the variation of $\C_{\rm eff,
bulk}[\f_{(p)r}]$ that give rise to boundary terms that involve
normal derivatives of the variations. For example, $\C_{\rm eff,
bulk}[\f_{(p)r}]$ is expected to contain an ordinary $R$-term for
the spin $2$ fields and consequently that $\C_{\rm eff,
boundary}[\f_{(p)r}]$ contains the corresponding Brown-York term.

In the case of $16$ supersymmetries \eq{geff} can be expressed
formally as

\be e^{i\C_{\rm eff}[\phi_{(p)r}(V)]}=<e^{i\sum_{p,r}\int d^dx
d^{16}\th \cO_{(p)r}V_{(p)r}}>\ ,\la{genfunc}\ee

where the effective action on the left hand side is evaluated
subject to boundary conditions dictated by superconformal tensors
$V_{(p)r}$. These superfields are prepotentials for super Weyl
multiplets containing the boundary conditions on the AdS
curvatures.

The correlators of composite operators on the right hand side of
\eq{genfunc} are well-behaved functions of the insertion points as
long as they are separated. However, as these points coincide, the
correlators are in general rather badly behaved distributions.
Thus a more careful definition of the generating functional of
correlators requires the choice of a regularization scheme. This
leaves room for anomalous effects, even though the singleton
theory is free, which may serve the purpose of selecting critical
field content and number of supersymmetries. In other words,
consistency of the right hand side of \eq{genfunc} in the case of
a free SCFT in $d$ dimensions with finite sources for composite
operators should be about as restrictive as consistency of an
interacting SCFT in $d$ dimensions. Moreover, the fact that a
successful definition of \eq{genfunc} in principle would give rise
to a consistent bulk theory including quantum gravity
\footnote{As the basic mechanism behind holography is general
covariance, this raises the question whether holography exhibits
any new features as general covariance is extended by HS
symmetries. To analyze this, we presumably need to refine our
present, mainly algebraic, understanding of HS symmetries by
formulating these in a more geometric language, perhaps by
extending the set of spacetime coordinates as to realize HS gauge
transformations as extended reparametrizations \cite{5dv1}.}
suggests that only the special supersymmetric singletons
corresponding to limits of string/M theory will be viable in the
above sense. Thus we shall assume that ultimately \eq{genfunc}
makes sense only for free SCFTs in $d\leq 6$ with less than or
equal to $16$ supersymmetries
\footnote{ Massless HS fields admit background independent
self-interactions in $D=4$, and it is most likely that this is the
case for all $D$ (though interactions in $D>7$ bring in symplectic
spacetime symmetries). However, the theories of massless HS fields
in higher dimensions are presumably not consistent truncations of
quantum consistent theories.}.
We address these issues further below when we discuss the
subleading $1/N$ corrections to the definition of the vacuum used
in the correlator on the right hand side of \eq{genfunc}.

The generating functional makes sense only as an asymptotic
expansion in $1/N$ in which a given order is a formal power series
expansion in $\phi_{(p)r}$, which has a finite radius of
convergence by the combinatorial counting rules for double line
diagrams of fixed topology. From the normalization \eq{etars} and
assuming that $<\cO>=0$ it follows that as far as the $1/N$
counting goes the effective action has the form

\bea  && \C_{\rm eff}[\phi]=\f^2+{1\over
N}f_3\left(\ft1{N^2}\right)\f^3 +{1\over N^2}
f_4\left(\ft1{N^2}\right)\f^4+\cdots\ ,\nn\w2 &&
f_n\left(\ft1{N^2}\right)\sim 1+\cO\left(\ft1{N^2}\right)\ . \eea

The singleton field theory determines $\C_{\rm eff}[\Phi_{(p)r}]$
up to non-linear field redefinitions of the type $\phi\ra
\phi+\ft1{N}\phi^2+\cdots $. After rescaling the fields as

\be \f=N\Phi\ ,\ee

we define the classical action as follows

\bea \C_{\rm eff}[\Phi] &=& \C_{\rm cl}[\Phi]+\cO(1/N^2)\
,\la{scl2}\w2 \C_{\rm cl}[\Phi]&=& {N^2\over R^{d-1}}\int d^{d+1}x
\cL(\F,R\partial\F,(R\partial)^2\F,\dots)+\mbox{boundary term}\
,\eea

where $R$ is the AdS radius. We can now state the properties of
the HS gauge theories as follows. They possess:

\begin{itemize}
\item[a)] a set of one-particle states forming HS multiplets
\item[b)] a corresponding set of `vertex operators' of a free CFT$_d$;
\item[c)] a fundamental mass scale, $1/R$ where $R$ is the AdS radius,
and a fundamental expansion parameter, $l_{\rm Pl}/R$ where the
Planck length $l_{\rm Pl}$ determines the normalization of the
effective AdS action to be
\footnote{In the case of $SU(N)$ valued singletons $N'=N^2-1$,
which means that the Planck constant in the bulk is given by
$\hbar=1/N^2$. The $1/N$ corrections to the bulk theory are
therefore weighted by positive integer powers of the Planck's
constant.}

\be {1\over l_{\rm Pl}^{d-1}}={N'\over R^{d-1}}\ .\la{lpl}\ee
\end{itemize}

Given these facts we would like to determine the effective action
$\C_{\rm eff}[\phi_{(p)r}]$ from a set of bulk interactions,
without any direct reference to the boundary singleton. The basic
issue is whether the interactions can be derived from a string or
membrane sigma model, that can be coupled to the HS background
fields. The mass-scale of the HS spectrum is set by the AdS radius
$R$, which is suggestive of a sigma-model with a fixed critical
tension of order $1$ in units where the AdS radius $R=1$, as we
shall discuss further in Section 5,6 and 7.

Due to the absence of mass-gap it is not possible to separate the
massless fields, $\f_{(2)r}$, from the massive AdS fields,
$\f_{(p)r}$, $p>2$, by taking a low energy limit. In a local
process in AdS with energies of the order $E\sim n/R$, $n>>1$, the
massive modes with $E_0< n/R$ behave essentially as the KK modes
which arise in an AdS compactification of string/M theory. Thus
the only reasonable possibility in which the massless modes can be
separated from the massive modes in a HS theory is by consistent
truncation to the massless sector
\footnote{We thank L. Rastelli for helpful discussions on this
point.},
which is similar to what happens in the (maximally supersymmetric)
sphere compactifications of Type IIB and eleven-dimensional
supergravities. There are examples, however, of compact manifolds,
such as $T^{1,1}$, where the higher dimensional supergravity
theory does not admit a consistent truncation despite the fact
that there does exist a lower-dimensional gauged supergravity.
\footnote{We thank C. Pope for pointing this to us.}

Thus we propose that the HS gauge theories in $D=4,5,7$ with gauge
groups $hs(8|4)$, $hs(2,2|4)$ and $hs(8^*|4)$ admit consistent
truncation down to the corresponding massless theories, which we
described in Section 2. This consistent truncation can be directly
tested by verifying that the massless bulk theory reproduces
exactly the correlators of the corresponding bilinear operators in
the singleton theory. This is a nontrivial test since nothing is
known about higher-dimensional covariant description of the HS
theory so far.

Consistent truncation of the full HS gauge theory to its massless
sector requires that there are no terms in the effective bulk
action of the form $\int \f_{(p)}\f_{(2)}\cdots \f_{(2)}$ for
$p\geq 3$. Let us show this in the case of scalar bulk fields.
Then the corresponding singleton correlators are non-zero provided
that $\D_{(p)}\leq n\D_{(2)}$ where $n\geq 2$ is the number of
massless fields. The case $\D_{(p)}= n\D_{(2)}$ is called an
extremal correlator. The extremality condition implies $p=2n$ and
in that case it is straightforward to use free field contraction
rules to show that

\be < \cO_{(p)}(x)\cO_{(2)}(x_1)\cdots \cO_{(2)}(x_n)>=
\prod_{i=1}^n (\D(x-x_i))^2\ ,\ee

where $\D(x)=|x|^{-d+2}$ is the singleton propagator. Consider, on
the other hand, the bulk integral

\be I=\int {d^{d+1}z\over z_0^{d+1}}
K_{\D_{(p)}}(z,x)K_{\D_{(2)}}(z,x_1)\cdots K_{\D_{(2)}}(z,x_n)\
,\ee

where $K_{\D}(z,x_i)$ is the standard bulk-to-boundary propagator.
$K_{\D}(z,x_i)\sim z_0^{d-\D}\d^d(z-x_i)$ for small $z_0$, and as
$z\ra x_i$,

\be  I\sim \int {dz_0\over z_0} z_{0}^{-\D+n\D_{(2)}}
\prod_{i=1}^n (\D(x-x_i))^2\ .\ee

Thus, in the extremal case this integral diverges logarithmically,
and the residue of the pole, treating $\D$ as a variable, has the
same structure as the extremal correlation function. By
assumption, the anti-holographic dual should, however, give rise
to finite amplitudes. The resolution is that a term which diverges
logarithmically is scale-invariant, which means that it can be
represented equivalently by a boundary term which is finite. Thus
extremal correlators give rise to couplings that are boundary
terms and therefore they do not upset consistent truncation.

A similar argument applies to the near-extremal case, when
$d-2<\D<n\D_{(2)}$. Here the integral $I$ is finite, but the
dependence on the $x$'s is not of the same form as the singleton
CFT correlator. There are exchange diagrams, though, with the
correct structure of the $x$- dependence \cite{df}. Thus the
near-extremal correlators must be represented anti-holographically
in terms of exchange diagrams, and there cannot be any contact
term in the bulk action that can upset consistent truncation we
are examining.

The above evidence for consistent truncation is similar to the one
given for ordinary Type IIB supergravity $AdS\times S^5$
\cite{min2,rastelli} and eleven dimensional supergravity on
$AdS_{4/7}\times S^{7/4}$ \cite{pioline}. The main difference is
that whereas the arguments in SUGRA only holds for $1/2$ BPS
states, the arguments given here for HS theory hold for more
general operators since the holographic dual is by assumption a
singleton.

To provide further evidence for consistent truncation, we examine
the correlator of four massless scalar operators
$\cO_i=\cO_{(2)}(x_i)$, $i=1,...,4$. Using free field theory
contraction rules it can be written on manifestly crossing
symmetric form as

\bea < \cO_1 \cO_2 \cO_3 \cO_4> &=& \eta_{12} \eta_{34} +
\eta_{14} \eta_{23} + \eta_{13} \eta_{24} + < \cO_1 \cO_2 \cO_3
\cO_4>_{\rm conn}\ ,\w2 < \cO_1 \cO_2 \cO_3 \cO_4>_{\rm conn} &=&
A^{(s,t)}_{1234} + A^{(t,u)}_{1324} + A^{(u,s)}_{1243}\
,\la{o1o2o3o4}\eea

where

\be A^{(x,y)}_{ijkl}=< :\cO_i \cO_k:~: \cO_j \cO_l:
>_{\rm conn}\ee

and $x$ and $y$ denote in which of the $s$-, $t$- and $u$-channels
the quantity $A^{(x,y)}_{ijkl}$ has singularities. In the limit
$x_{12}$, $x_{34}\ra0$, the correlator can be expanded in the
$s$-channel by using the OPE

\be \cO_{1}\cO_{2}=\eta_{12} + C_{12}{}^{(2)r} \cO_{(2)r}(x_2) +
C_{12}{}^{(4)r}\cO_{(4)r}(x_2) + C_{12}{}^{(2,2)r}
\cO_{(2,2)r}(x_2)\ ,\la{schope}\ee

where we recall that $\cO_{(2)r}$ denotes the set of all primary
bilinear single-trace operators labeled by an index $r$, and
$\cO_{(4)r}$ and $\cO_{(2,2)r}$ are as given in \eq{422}. The
resulting $s$-channel expansion is given by

\be < \cO_1 \cO_2 \cO_3 \cO_4
>_{s-{\rm ch}}= \eta_{12}\eta_{34}+
C_{12}{}^{(2)r}C_{34,(2)r} + < \cO_1 \cO_2 \cO_3 \cO_4
>_{s-{\rm ch, finite}}\ ,\la{sch}\ee

where $C_{RST}= C_{RS}{}^U\eta_{UT} = <\cO_R\cO_S\cO_T>$, and $<
\cO_1 \cO_2 \cO_3 \cO_4>_{s-{\rm ch, finite}}$, which is finite in
the $s$-channel, is given by

\bea < \cO_1 \cO_2 \cO_3 \cO_4>_{s-{\rm ch, finite}}&=& <:\cO_1
\cO_2:~: \cO_3
\cO_4:>=\eta_{13}\eta_{24}+\eta_{14}\eta_{23}+A^{(t,u)}_{1324}\
,\w2 A^{(t,u)}_{1324}&=& C_{12}{}^{(4)r} C_{34,(4)r} +
C_{12}{}^{(2,2)r} C_{34,(2,2)r}\ .\la{schfin}\eea

The structure of \eq{schfin} appears to be problematic for
consistent truncation, and cannot be ignored in the large $N$
limit as follows from

\be C_{(2)(2)}{}^{(2)r}\sim \ft1{N}\ ,\quad
C_{(2)(2)}{}^{(4)r}\sim \ft1{N}\ ,\quad C_{(2)(2)}{}^{(2,2)r}\sim
1\ .\ee

It is possible, however, to write \eq{schfin} in a more tractable
form as a manifestly crossing symmetric sum of terms involving
only exchange of bilinear operators. To this end we first note
that the crossing symmetry of the singleton theory implies that
the complete $s$-channel expansion \eq{sch} is equal in the sense
of analytical continuation to the complete $t$- and $u$-channel
expansions in the limits $x_{14}$, $x_{23}\ra0$ or $x_{13}$,
$x_{24}\ra0$, respectively. Thus \eq{sch} must contain
contributions that are singular in the $t$- and $u$-channels. From
the form of \eq{o1o2o3o4} we therefore deduce that the singular
part of \eq{sch} actually must consist of two separate
contributions, one which becomes singular in the $t$-channel and
another one which becomes singular in the $u$-channel. We also see
that the problematic term in \eq{sch} must have singularities in
both the $t$- and $u$-channels, which by crossing symmetry should
describe massless exchanges. In fact, from \eq{schope} and free
field theory contraction rules it follows that

\bea A^{(s,t)}_{1234}&\equiv&<:\cO_1\cO_3:~:\cO_2\cO_4:>_{\rm
conn}=\ft12 C_{12}{}^{(2)r}C_{34,(2)r}+\ft12
C_{32}{}^{(2)r}C_{14,(2)r}\ ,\la{cross1}\w2
A^{(t,u)}_{1324}&\equiv&<:\cO_1\cO_2:~:\cO_3\cO_4:>_{\rm
conn}=\ft12 C_{13}{}^{(2)r}C_{24,(2)r}+\ft12
C_{14}{}^{(2)r}C_{23,(2)r}\ ,\la{cross2}\w2
A^{(u,s)}_{1243}&\equiv&<:\cO_1\cO_4:~:\cO_2\cO_3:>_{\rm
conn}=\ft12 C_{12}{}^{(2)r}C_{43,(2)r}+\ft12
C_{13}{}^{(2)r}C_{42,(2)r}\ .\la{cross3}\eea

To compute the first term in \eq{cross1} we use
\eq{schope} to expand the single contraction connecting $\cO_1$ to
$\cO_2$ in terms of $C_{12}{}^{(2)r}\cO_{(2)r}$ and similarly for
$3$ and $4$. The remaining two contractions that contribute to the
connected part give rise to $\ft12 \eta_{rs}$, where the factor of
$\ft12$ arises due to the normal ordering prescription which
forbids contractions connecting $1$ with $2$ and $3$ with $4$,
respectively. The second term in \eq{cross1} contains
the single contractions connecting $1$
to $4$ and $3$ to $2$. The relations \eq{cross2}and \eq{cross3}
are obtained analogously. Eqs. \eq{cross1}and \eq{cross3} imply
that the finite contribution \eq{schfin} can be rewritten in terms
of partial wave expansions involving only exchange of bilinear
operators in the crossed channels. Thus the complete
four-point correlator can be written in a manifestly crossing
symmetric form involving only massless partial waves:

\bea < \cO_1 \cO_2 \cO_3 \cO_4> &=&
\eta_{12}\eta_{34}+\eta_{23}\eta_{14}+\eta_{24}\eta_{13}+ < \cO_1
\cO_2 \cO_3 \cO_4>_{\rm conn}\ ,\w2 < \cO_1 \cO_2 \cO_3
\cO_4>_{\rm conn}&=& C_{12}{}^{(2)s}C_{34,(2)s} +
C_{23}{}^{(2)t}C_{14,(2)t}+ C_{13}{}^{(2)u}C_{42,(2)u}\
.\la{stuinv}\eea
This generalizes so that any correlator of bilinear operators can
be written as a manifestly channel duality invariant sum of
conformal blocks involving only exchange of bilinear operators.

The test of holography requires that the result \eq{stuinv} is
consistent with that obtained from the corresponding Witten
diagram that uses the classical action of the HS gauge theory
truncated to its massless sector. Now it has been shown in
\cite{liu} that a Witten diagram with four external scalars and
exchange of an internal scalar $\phi$ equals the sum of the
conformal block with exchange of the scalar operator $\cO$
coupling to $\phi$ {\it plus} terms which have the same structure
as, but do not exactly agree with, the conformal blocks with
exchange of operators corresponding to the two-particle states
formed out of the external scalar states. In \cite{liu} it has
also been shown that a Witten contact diagram with four external
scalars has the form of two-particle exchange.

Thus \eq{stuinv} has the form required by consistent truncation,
provided that the quartic bulk interactions in the cases of
interest lead to cancellation of the parts in bulk four-point
amplitudes that have the structure of conformal blocks with
massless two-particle state exchange. The remaining terms, which
come from the Witten diagrams with exchange of massless bulk
fields, can then be written in manifestly $s$-$t$-$u$ channel
duality invariant form as conformal blocks with exchange of the
corresponding bilinear operators, as in \eq{stuinv}. Thus, the
bulk side of the story remains to be established. It would be
interesting to examine to what extent the requirement that
two-particle partial waves must cancel determines the structure of
higher order interactions in the action for massless fields. We
shall return to this point below in discussing the interaction
ambiguity in the massless sector.

Having gathered evidence for the consistent truncation, let us now
proceed to explore some of its consequences. In the above
discussion, we have implicitly made the assumption that the
correlators in the free singleton theory are given by ordinary
vacuum expectation values on a conformal plane. Let us assume that
this is indeed correct in the large $N$ limit. The generating
functional for correlators of bilinear operators is then  given
for large $N$ by a one-loop functional determinant, i.e. the
connected $n$-point correlators are planar diagrams that scale
like $N^{-(n-2)}$. Thus, the effective bulk action for the
massless fields is `classical' and takes the form

\be  \C_{\rm cl}[\Phi_{(2)r}]= {N^2\over R^{d-1}} \int d^{d+1}x
\cL(\Phi_{(2)r},R\partial\Phi_{(2)r},(R\partial)^2\Phi_{(2)r},\dots)
+\mbox{boundary terms}\ ,\la{scl}\ee

where $R$ is the AdS radius. The Lagrangian  contains higher
derivative interactions and the quadratic part is ghost and
tachyon free. It is important to note that the quantity $R\nabla$
is not small in an expansion around AdS.

By construction, both $\C_{\rm cl}[\F_{(2)r}]$, and the full
classical action $\C_{\rm cl}[\F_{(p)r}]$ defined in \eq{scl2},
reproduce the correlators of bilinear operators holographically to
the leading orders in the $1/N$ expansion, i.e. the extrema of the
two actions are equal provided the massive modes $\F_{(p)r}$,
$p\geq 3$ are set to zero at the boundary of $AdS$. The consistent
truncation can now be phrased as the stronger condition

\be \C_{\rm cl}[\Phi_{(2)r}]= \C_{\rm
eff}[\F_{(p)r},\F_{(3)}=0,\dots]\ .\la{constr}\ee

This offers the following possibility to test consistent
truncation directly. Based on the results in $D=4$, we expect that
HS gauge symmetry together with the requirement of manifest local
Lorentz symmetry determines a family of actions

\be S[\phi_{(2)r};\cV]\ee

for massless fields where $\cV$ represents a set of arbitrary
parameters. As explained in Section 6.2, and in more detail in
\cite{cubic}, there exist an interaction ambiguity in the $4D$ HS
gauge theory which involves the introduction of an odd function
$\cV(x)=\sum_{n=1}^\infty b_{2n+1} x^{2n+1}$. Already the simplest
choice $ \cV(x) = b_1 x $ gives rise to a highly nontrivial model
with a structure of the type indicated in \eq{scl}. The $n$'th
order term in $\cV(x)$ results in higher order derivative
corrections starting at order $2n+2$ in the Lagrangian. Thus, in
$D=4$ the consistent truncation \eq{constr} implies a specific
choice $\cV(x)=\cV_{\C}(x)$ such that

\be \C_{\rm cl}[\phi_{(2)r}]=S[\phi_{(2)r};\cV_{\C}]\
.\la{shsc}\ee

Thus, consistent truncation means that there exists a set of
parameters $\cV_{\C}$ for which the extremum of
$S[\phi_{(2)r};\cV_{\C}]$ corresponds to the generating functional
of correlators of bilinear operators in the singleton theory. A
perturbative scheme for obtaining the interactions in $D=4$ to any
desired order is given in \cite{ssc,cubic}, and described in
Section 6 for the case of quadratic terms in the field equations.
We are still lacking the description of the full interactions for
massless fields in $D>4$, though we expect that the basic building
blocks are of the kind described in \cite{us1,5dv2,us2,us7d}.

The supersymmetric HS gauge theories in $D=4,5,7$ can be truncated
consistently to a minimal bosonic HS theory with massless and
massive fields. Moreover the massless minimal bosonic theory is a
consistent truncation of the massless supersymmetric HS theory.
Hence, if the truncation of the massive modes is consistent in the
supersymmetric theory then this must also be the case in the
minimal bosonic theory. In particular, in $D=4$ the interaction
ambiguities in the supersymmetric theory and the minimal bosonic
theory are parametrized by the same function $\cV$.

Let us now examine more closely the qualitative behavior of the
$1/N$ dependence of some singleton correlators involving higher
than second order traces of singletons. For example, the connected
part of the correlator of four cubic scalar single-trace
operators, $\cO_{(3)}\sim \ft1{N^{3/2}}:\tr (W^3):$ (these
operators do not mix with any other operators) contains both
planar and non-planar double-line graphs which scale like $1/N^2$
and $1/N^4$, respectively, in the large $N$ limit. Another
interesting example is the correlator of three $\cO_{(4)}$
operators, which contain $1/N$ and $1/N^3$ contributions. In the
supersymmetric case, one can arrange the cyclic orders of
R-symmetry indices carried by the singletons to cancel the leading
$1/N$ contribution, and thus the corresponding cubic coupling in
the effective action \cite{su1}. Hence, the full singleton theory
encodes information about a nontrivial $1/N$ expansion of the
anti-holographic dual.

As we have already mentioned, we think of these corrections as
being generated by a quantum theory in the bulk which is generated
by a string theory or some other sigma model which can be coupled
to the massless HS fields. From this point of view, it would be
natural to have subleading $1/N$ corrections also to the
interactions in the massless sector, so that \eq{scl} would only
be valid for large $N$. We would also expect corrections to
$\C_{\rm eff}[\F_{(p)r}]$ which violate the consistent truncation
\eq{constr}. These effects do not arise, however, if we treat the
correlators in the singleton theory as ordinary vacuum expectation
values of operators inserted on the conformal plane.

We conclude this section by speculating on possible subleading in
$1/N$ corrections to the free singleton correlators on the right
hand side of \eq{genfunc}. To this end, let us assume that the
free singleton theory in question is an actual limit of a CFT
describing the low energy dynamics of open string modes in string
theory or `open membrane' modes in M theory. For concreteness, let
us consider the case of the $SU(N)$ invariant singleton theory
that arises as a limit of the $d=4$, $\cN=4$ SYM theory. For
finite open string length the prescription for computing open
string theory amplitudes is to attach open string vertex operators
to open string boundaries and sum over all open string
fluctuations. This includes virtual processes including formation
of closed string loops. A closed string loop can be created by
inserting a `sewing operator'

\be R_s=\sum_s V_s(z) \bar V^s(0) \ee

on the string worldsheet where the sum runs over a complete set of
physical closed string states. In taking the low energy limit
leading to the conformal SYM theory, the physical effect of the
sewing operation is included into the $1/N$ expansion of the SYM
theory with finite $g^2_{\rm YM}$. Thus the limit $g^2_{\rm YM}\ra
0$ is not smooth in the sense that the closed string sewing
operations, which are present for any finite $g^2_{\rm YM}$ are
absent for $g^2_{\rm YM}=0$, simply because there are no virtual
processes in the singleton theory that leads to the addition of
internal `handles' in the $1/N$ expansion. This is reminiscent of
the fact that the deformation of the free singleton theory
corresponding to switching on finite $g_{\rm YM}^2$ cannot be
described directly at the level of the composite operators built
from the singleton superfield, which contains the abelian field
strength but not any explicit gauge potential. In fact, this
requires that we introduce gauge couplings by hand, after which
$g_{\rm YM}^2$ can be shifted to any finite value by marginal
deformations.

The above arguments suggest that we modify the definition of the
generating functional in the singleton theory by working with full
singleton correlators given schematically by

\be <\cO_1 \cdots \cO_n>_{\rm full}~ = ~\sum_k {1\over
k!}<R^k\cO_1 \cdots \cO_n>\ ,\ee

where the (super)conformally invariant singleton sewing operator
$R$ is defined as the sum over a complete set single trace
operators describing a virtual closed string process:

\be R~=~\sum_{p}\int {d^d x d^d y \over |x-y|^{2d}}
\eta^{rs}(x-y)\cO_{(p)r}(x)\cO_{(p)s}(y)\ .\ee

Since each power of $R$ adds an extra power of $1/N^2$, the above
definition does not affect the classical limit though it yields
the desired nontrivial subleading $1/N$ corrections to the
correlators. The insertion of $R$ formally corresponds to taking a
trace, which in turn implies that the correlation function becomes
periodic along a cycle on the conformal plane. In string theory,
$R_s$ acts similarly, and has the geometric effect of adding a
handle to the two-dimensional worldsheet. This suggests that $R$
insertions describe large fluctuations of the D3 brane worldvolume
in the singleton limit. As in the closed string theory, the
consistency of the sewing operation in the free singleton theory
may lead to restrictions on the spacetime superdimension.

In summary, we propose to use HS symmetries in diverse dimensions
to determine actions (or field equations) for massless HS
multiplets up to certain well-defined interaction ambiguities and
then to compare the resulting Witten amplitudes with correlators
of bilinear operators in corresponding large $N$ singleton
theories. The next step in this program is to explain the
consistent singleton/HS correspondences as limits of string and M
theories, which in particular require the identifications of
possible schemes for breaking HS symmetries.

We emphasize that the tests of CFT/AdS in the HS regime involve a
free CFT on the boundary, unlike the tests in the supergravity
regime where the boundary CFT is strongly coupled. This is
possible due to the proposed consistent truncation and the fact
that there still remains the expansion parameter $1/N$.

It is not clear exactly how the state of affairs will change once
the HS symmetries are broken. In Section 3 we have identified
candidate Higgs multiplets in $d=3,4$. Presumably this can be done
also in $d=6$ provided that we develop the proper mathematical
language for describing the interactions on the M5 brane. In
general, we expect that the Higgsing upsets the consistent
truncation to the massless sector alone. Moreover, it is not
obvious if there exists a generalized consistent truncation scheme
that retains the massless, Higgs and other relevant massive
fields. In any event, it will be interesting to see whether HS
field theoretic methods can be used to describe the Higgsing or
one has to resort to some more basic definition of the bulk
interactions, based on some sigma model. We believe it is too
early to make any conclusive remarks on this, though it seems
possible to describe couplings between massless HS fields and
Higgs fields, which should form HS multiplets fitting into master
fields of the type discussed in Section 2.

In Sections 5-7 we shall discuss these issues in more detail, and
case by case for the theories described in Section 2.


\section{Type IIB on $AdS_5\times S^5$ and 5D Higher Spin Gauge Theory}


According to the strong version of the Maldacena conjecture
\cite{malda,rev,df} $d=4$, $\cN=4$ SYM theory with $SU(N)$ gauge
group, gauge coupling $g_{\rm YM}^2$ and 't Hooft coupling
$\l=Ng_{\rm YM}^2$ is equivalent to Type IIB string theory on
$AdS_5\times S^5$ of radius $R$ with string coupling $g_s$ and
string length $l_s$ given by

\bea g_s&=&f_1(\l)g_{\rm YM}^2\ ,\quad\quad  f_1(\l)\sim 1
\quad\quad\quad  \mbox{for $\l>>1$} \ ,\nn\w3 l_s&=&f_2(\l)R\
,\quad\quad\quad f_2(\l)\sim \l^{-1/4}\quad \mbox{for $\l>>1$}\
.\la{lstr}\eea

For large $\l$, these relations are deduced by interpolating
between the $AdS_5\times S^5$ vacuum with radius $R$ and dilaton
$e^\phi=g_s$, and the ten-dimensional Minkowski vacuum, using the
classical D3-brane solution with harmonic function $H(r)=1+4\pi
Ng_sl_s^4 r^{-4}$. The functions $f_{1,2}(\l)$ account for
possible string corrections to the interpolating region, where
only $16$ supersymmetries are preserved. The Type IIB string/4d
SYM correspondence is an AdS/CFT correspondence whereby the 4d SYM
theory is identified as the holographic dual of the Type IIB
closed string theory. The closed string theory is based on a
non-linear sigma-model with coupling constant $l_s/R$. A
(dimensionless) closed string amplitude $A^{(\rm str)}$ has the
doubly asymptotic expansion

\be A^{(\rm str)} = \sum_{g=0}^\infty g_s^{2g-2}A^{(\rm
str)}_g(l_s/R)\ ,\la{as}\ee

where the amplitude $A^{(\rm str)}_g(l_s/R)$, which is obtained
from worldsheet perturbation theory on a Riemann surfaces of fixed
genus $g$, is given by an asymptotic expansion in $l_s/R$. The 5D
Planck length is given by

\be {1\over l_{\rm Pl}^3}={N^2\over R^3}\ .\ee

Thus the perturbative string expansion in $AdS_5\times S^5$ makes
sense provided that

\be N>>1\ ,\quad g_s<<1\ ,\quad l_s<<R\ .\ee

The 't Hooft expansion of the corresponding correlation function
$A^{(\rm SYM)}$ in the SYM theory reads

\be A^{(\rm SYM)}=\sum_{g=0}^\infty N^{2-2g}A^{(\rm SYM)}_g(\l)\
,\ee

where the amplitude $A^{(\rm SYM)}_g(\l)$ is obtained from
double-line Feynman graphs with fixed topology and is given by an
analytical expansion in $\l$. Hence the conjectured correspondence
$A^{(\rm str)}=A^{(\rm SYM)}$ can be examined order-by-order in
string loop expansion and SYM $1/N$ expansion, leading to a set of
strong/weak coupling dualities between $A^{(\rm str)}_g(l_s/R)$
and $A^{(\rm SYM)}_g(\l)$.

As discussed earlier, it has been proposed that the HS gauge
theory emerges in the description of the Type IIB string theory on
$AdS_5\times S^5$ in the limit \cite{su1,su2,us1,edseminar}

\be g_s\ra0\ ,\quad  l_s\ra\infty\ ;\quad N>>1,\ \ R\
\mbox{fixed}\ .\la{lim}\ee

In this limit the dual free SYM theory is described by an $SU(N)$
valued $d=4$, $\cN=4$ SYM singleton. As discussed in the previous
section, the bulk physics is conjectured to be an HS gauge theory
in $5D$ which admits a consistent truncation to an effective
action $\C_{\rm cl}[\F_{(2)r}]$ for massless fields. The HS gauge
group $hs(2,2|4)$ and its massless gauge theory has been described
in \cite{us2}. We emphasize that there should be direct agreement
between the individual terms in the $1/N$ expansions of massless
gauge theory amplitudes and the correlators of bilinear currents
in the free CFT as described in \eq{shsc} (without having to first
obtain strong coupling results).

There still remains the task of constructing the full interacting
HS gauge theory in $5D$, though cubic interactions for massless
spin $s=2,4,6,...$ fields have already been constructed in
\cite{5dv2}. These form a subset of the cubic interactions of the
minimal bosonic truncation $S_{\rm bos}$ of $S[\f_{(2)};\cV]$
provided that it is consistent to set the scalar field $\phi$ in
$S_{\rm bos}$ equal to zero at the cubic level. This requirement
means that $S_{\rm bos}$ must not have any cubic interactions that
are linear in $\phi$ and quadratic in spin $s\geq 2$ fields. On
the other hand, from the known stress-energy tensor OPEs (see, for
example,  eq. (4.58) in \cite{anselmi}), it follows that the
effective action $\C_{\rm eff}[\f_{(2)}]$ should give rise to a
non-zero cubic graviton-graviton-scalar amplitude. Thus the scalar
can only be consistently truncated at the cubic level if this
amplitude is represented by a boundary term in $\C_{\rm
eff}[\f_{(2)}]$, i.e. if the correlator in question is extremal or
near-extremal. Whether or not this is the case remains to be seen.

We next discuss breaking of the HS symmetry. The level $\ell=0$
supergravity multiplet of the massless spectrum of the $hs(2,2|4)$
theory contains a dilaton, $\varphi$ which is an $SU(4)$ singlet
with energy $\D=4$ and AdS mass $m^2=0$. Since $m^2=0$ it is
consistent to give $\varphi$ a VEV in the linearized theory, and
we shall assume that this is possible also in the full HS gauge
theory. This corresponds to switching on a finite $g_{\rm YM}^2$
in the 4d SYM theory. As result the 4d supercovariant derivative
$D_\a^i$ becomes also gauge covariant. This does not upset the
stress-energy conservation law \eq{setc}, as it is first order in
the superderivative, while it breaks the Konishi multiplet
conservation law \eq{kc1}, which is second order in derivatives.
Using  the relation

\be D^{ij} W^{kl}=-2g_{\rm YM}[W^{k(i},W^{j)l}]\ ,\ee

which follows from the superspace formulation of the ${\cN}=4$ SYM
system in 4d, one finds that the anomalous conservation law for
the Konishi current is given by (see, for example,
\cite{fz0,hh1}):

\be D^{ ij}J= {4 g_{\rm YM}\over N}\tr\,W^{k(i}W^{j)l}W_{kl}\equiv
\sqrt{\l} \S^{ij} \ .\la{ano}\ee

The operator $\S^{ij}$ belongs to the massive Higgs multiplet with
$s_{\rm max}=\ft72$ discussed in Section 3. Thus, for finite
$g_{\rm YM}^2$ the anomalous conservation law \eq{ano} describes
how $\S^{ij}$ is `eaten' by the massless Konishi operator $J$ to
form a massive operator which belongs to the long massive Konishi
multiplet with $s_{\rm max}=4$ containing $2^{16}$ states. The
coupling between the corresponding bulk fields, which are
described on the boundary by prepotentials $V$ and $V_{ij}$, and
the massless Konishi operator $J$ and its Higgs descendant
$\S^{ij}$ is described by

\be S_{\rm boundary}=\int d^4x d^{16}\th \left( JV + \S^{ij}V_{ij}
\right) \ . \ee

For finite $g_{\rm YM}^2$, the action $S_{\rm boundary}$ is
invariant under modified gauge transformations involving a
St\"uckelberg shift transformation of the massive Higgs field,

\be \d V=D^{ ij}\L_{ij}\ ,\quad \d V_{ij}=-\sqrt{\l}\L_{ij}\
.\la{stu}\ee

We thus expect that for finite $<\varphi>=g_s$ the effective
action $\C_{\rm eff}[\f_{(p)r}]$ contains kinetic terms of the
schematic form $|d\f_{(2)}|^2+|d\f_{(3)}+\sqrt{\l} \f_{(2)}|^2$,
describing a single massive gauge field with non-critical mass
\cite{edseminar}

\be m^2-m^2_{\rm crit}\sim {\l\over R^2}\ ,\la{ncm}\ee

where $(D^2-m^2_{\rm crit})\phi=0$ for an AdS massless field
$\phi$.

As discussed in Section 3, the massive spectrum also contains
$1/2$ BPS massive states that have the interpretation of KK modes
built on the massless HS multiplets. We shall assume that the
Higgs mechanism can be described at the level of KK towers as
well, and that the remaining massive HS multiplets can be
organized into massive HS multiplets and their KK towers. This
picture is suggestive of a covariant theory in $D=10$ with
`critical' length scale $l_{10}$ and coupling constant $g=1/N$
which admits $AdS_5\times S^5$ with radius $R=l_{10}$ as a vacuum.
Since HS interactions in AdS spaces blow up in the flat limit for
finite $g$, we do not expect the 10D HS theory to admit 10D
Minkowski space as a vacuum for finite $g>0$. For $g=0$ we get a
quadratic Lagrangian, however, which is second order in
derivatives, and as it contains no positive powers of $R$, it does
admit a flat space limit. Thus, the tensionless limit of the Type
IIB string theory in 10D flat spacetime is trivial.

Higgsing of the critical theory leads to a non-critical theory
with $l_{10}<R$ which for $l_{10}<<R$ should be identified with
Type IIB string theory in $AdS_5\times S^5$ with $l_s\sim l_{10}$.
For small $l_s/R$ the spectrum of string states with AdS energy
(measured in units of $1/R$) satisfying the condition
$E<<R^2/l^2_s$, and spin $s<<R^2/l^2_s$, can be obtained by KK
reducing the 10D Minkowski space spectrum on $S^5$ by means of
group theoretical methods (at the classical level these states are
described by `short' strings with energy $E=Rl/l_s^2$ and length
$l<<R$). In particular, for fixed $SO(4)\times SO(6)$ highest
weight, the worldsheet Hamiltonian has a ground state which is the
`lightest' state carrying that highest weight. The lightest states
states correspond to the leading Regge trajectory in 10D Minkowski
space and form supermultiplets in both 10D Minkowski space and in
$AdS_5\times S^5$ with $s_{\rm max}=2,4,6,..$. In 10D Minkowski
space these arise at closed string level $\ell=\ft12 s_{\rm
max}-1$ (see, for example, \cite{ramond}), where all multiplets
are massive except for level $\ell=0$ where the supergravity
multiplet resides. For example, the lightest $s_{\rm max}=4$
multiplet is the massive Konishi multiplet which resides at level
$\ell=1$.

As $l_s/R$ varies from $l_s/R<<1$ to $l_s/R>>1$ the different
Regge trajectories do not mix \cite{polyakov} even though the
five-form flux and other terms of order $1$ in units of $R$ will
become comparable to the mass-term. This follows from the fact
that in an exact CFT that admits a perturbative formulation, such
as the worldsheet theory and the boundary SYM theory, there cannot
be mixing between two operators that do not mix in the free
theory. Note that such an admixture would require the introduction
of a mass-parameter in the perturbative formulation, which is not
compatible with conformal invariance.

Indeed, there is an exact agreement between the supermultiplet
structures of the leading Regge trajectory for large string
tension and the set of massless states of the critical $hs(2,2|4)$
theory, such that the level $\ell$ multiplet on the leading Regge
trajectory flows, after reversed Higgsing, to the level $\ell$
multiplet of the massless spectrum given in Table \ref{tfis}.

We have already argued in Section 4 and 5 that there should exist
a consistent truncation of the full $hs(2,2|4)$ theory down to its
massless sector. There is no analogous truncation of the
non-critical string theory down to the leading Regge trajectory
because the lightest states of level $\ell\geq 1$ consist of
massless states plus Higgs states. The Higgs states belong to the
massive sector of the $hs(2,2|4)$ theory and therefore break the
consistent truncation.

Since the HS symmetries are broken spontaneously it would be
interesting to construct a HS field theoretic description in AdS
of the couplings between the massless fields and the Higgs fields.
Clearly the master field formalism described in Section 2 should
be useful in doing this, though one presumably needs to invoke
some additional information, perhaps from the structure of the
factorization of the SYM correlation functions for $\l<<1$. Thus
we should try to find  a HS action $S (\F_{(2)r},H_{r};\cV,\cM)$
for massless fields $\F_{(2)r}$ and Higgs fields $H_r$, where
$\cV$ are the parameters describing the gauge interactions, as
will be discussed in Section 6.2, and $\cM$ are the parameters
describing the coupling of the gauge multiplet to the massive
Higgs fields. We can then study the issue of whether the
`weak/weak' version of the AdS/CFT correspondence, which is valid
for the massless sector at $\l=0$, can be generalized to include the
leading Regge trajectory for $\l>0$.

The connection between the leading Regge trajectory at small $l_s$
and the bilinear HS currents in the SYM theory at small $\l$ has
also been made by the authors of \cite{gkp2} who give `long
string' solutions to the worldsheet sigma model in the limit
$l_s/R<<1$. These solutions describe states on the the leading
Regge trajectory with spins $s>>R^2/l_s^2>>1$ and AdS energies
$E\sim s+R^2/l_s^2 \log(sl_s^2/R^2)\sim s$, which couple to
bilinear HS currents in the SYM theory. These operators arise in
the OPE of Wilson lines making up the boundaries of worldsheets of
infinitely long strings on the leading Regge trajectory. The
leading Regge trajectory also contains `short strings' which have
spins $s<<R^2/l_s^2$ and energies $E\sim \sqrt{s}R/l_s$. Since
$E/s\sim 1$ for long strings, and $E/s>>1$ for short strings, the
long ones cannot decay into large number of short ones.
Furthermore, in any Regge trajectory there should be long string
with large spins and asymptotically small anomalous dimensions,
suggesting that string interactions in the limit $s>>R^2/l_s^2>>1$
have a consistent description in terms of long strings with $E\sim
s$. Upon increasing $l_s/R$, we expect a short string state with
fixed $s$ to become long for large enough $l_s/R$. In the limit in
which $\l\ra 0$ on the SYM side, there should exist a $hs(2,2|4)$
invariant worldsheet sigma model describing closed string
interactions in the bulk corresponding to the free SYM theory.

From the above discussions we are led to propose that there is a
cross-over from large to small $\l$ in the expressions for the AdS
string length and string coupling in terms of the gauge theory
quantities given in \eq{lstr}, such that

\be f_1(\l)\sim 1/\l\ ,\quad f_2(\l)\sim 1+\cO(\l)\quad \mbox{for
$\l<<1$}\ .\ee

Then $l_s/R\sim 1$ and $g_s\sim 1/N$ as $\l\ra 0$. This suggests
that the $hs(2,2|4)$ higher spin gauge theory is described by a
string theory which has a left-moving and right-moving
$PSU(2,2|4)$ KM algebra with critical level $k=k_{\rm crit}\sim 1$
which admits a singleton representation and an affine $hs(2,2|4)$
extension. To be more precise, the critical value for the level
should be such that there exists a maximally reducible Verma
module based on the singleton which contains a maximal number of
null-states. In fact, it has been shown \cite{ds1} that the affine
$SO(3,2)\simeq Sp(4)$ algebra admits singleton-like
representations for $k=5/2$. It would be interesting to generalize
this result to $SO(D-1,2)$ and supersymmetric cases. For critical
level the closed string spectrum then contains physical massless
HS states states formed by multiplying a left-moving and a
right-moving singleton. The algebra $hs(2,2|4)$ can be identified
with the following coset

\be hs(2,2|4)=Env(PSU(2,2|4))/\cR\ ,\la{env}\ee

where $\cR$ is a certain ideal generated by elements in
$Env(PSU(2,2|4))$ which vanish identically when the $PSU(2,2|4)$
generators are realized in terms of a single super-oscillator as
described in Section 2.1. For $k=k_{\rm crit}$ this construction
should lift to the affine case. The symmetry enhancement from AdS
group to HS algebra for critical level, i.e. critical radius in
units of fixed string length, would be similar in spirit to the
$SU(2)$ enhancement occurring at the self-dual radius for string
theory on a circle.

The possibility to realize massless higher spins directly in the
bulk as products of left-moving and right-moving singleton
representations at critical KM level is rather appealing. Perhaps
the close resemblance between the HS gauge theories in $D=4,5,7$
is an indication of that singletons play a similar role on
critical membranes in $D=4,7$.


\section{M Theory on $AdS_4\times S^7$ and 4D Higher Spin Gauge Theory}


\subsection{Holography}\la{sec:m2}


Already in \cite{duff} it was observed that the $OSp(8|4)$
singleton may play a role in the description of the supermembrane
on $AdS_4\times S^7$. In \cite{bsst} the quantization of the
$d=3$, $\cN=8$ singleton theory corresponding to a single membrane
was shown to yield the infinite set of massless HS fields
contained in the symmetric tensor product of two singleton weight
spaces \cite{ff}. Moreover, it was conjectured in \cite{bsst} that
these massless states, as well as the massive states contained in
the higher order tensor products, arise in the supermembrane
theory
\footnote{To describe the $S^7$ compactified M theory all higher
tensor products are needed. The resulting theory lives on the
double cover of $AdS_4$ times $S^7$. It is consistent to truncate
the theory to only even powers of the singleton. This corresponds
to M theory on the single cover of $AdS_4$ times $S^7/{\sf
Z}_2\simeq {\sf R}{\bf P}^7$.  }.
Subsequently, the group theoretical HS/singleton connection was
utilized in \cite{kv2} and the fully interacting massless HS field
equations in $D=4$ were constructed in \cite{4dv1}. In the light
of \cite{malda}, the 4D HS/singleton connection found in
\cite{bsst} was revived as an actual AdS/CFT correspondence in
\cite{us3,us4}.

Importantly, the role of large $N$ discussed in Section 4 was not
emphasized in these early formulations of the correspondence. Thus
we need to refine the formulation of the correspondence by
identifying the appropriate dependence on $N$ of the free
$OSp(8|4)$ singleton.

Let us first recall the Maldacena conjecture \cite{malda} on the
correspondence between M theory on $AdS_4\times S^7$ with $N$
units of $7$-form flux on $S^7$ and the low energy dynamics of $N$
parallel $M2$ branes in flat eleven-dimensional spacetime, which
is described by a strongly coupled $d=3$, $\cN=8$ CFT with
$SO(8)_R$ symmetry \cite{malda,rev}. This theory defines a
nontrivial IR fixed point of $d=3$, $\cN=8$ SYM theory with
$SU(N)$ gauge group. The resulting $SO(7)_R$-invariant flow has an
anti-holographic description as a D2 brane near-horizon geometry,
which is reliable in the UV where the dilaton is small. In the IR
the dilaton blows up and the IIA solution lifts to the $SO(8)_R$
invariant $AdS_4\times S^7$ near horizon region of a stack of $N$
coinciding M2 branes. The resulting anti-holographic description
of the strongly coupled SCFT is conjectured to be M theory on
$AdS_4\times S^7$. For large $N$ the membrane tension scales like

\be T_{M2}={1\over l^3_{M2}}\sim {\sqrt{N}\over R^3}\
,\la{4dm2}\ee

where $R$ is the AdS radius, and the  4D Planck length is given by

\be {1\over l_{\rm Pl}^2}={N^{3/2}\over R^2}\ .\la{4dpl}\ee

Hence, for large $N$,

\be R>>l_{M2}>>l_{\rm Pl}\ .\la{hier}\ee

For AdS energies $E$ obeying $1<< E<< R/l_{\rm M2}$ the low-energy
dynamics of the anti-holographic dual is conjectured \cite{malda}
to be described by $D=4$, $\cN=8$ gauged supergravity. In
particular, it follows from the normalization \eq{4dpl} that the
strongly coupled SCFT has $\sim N^{\ft32}$ massless degrees of
freedom for large $N$ \cite{kt,hs}.

In the UV limit of the D2 brane geometry the dilaton $e^{\f}$
vanishes and the 10D gravitational curvature diverges (which one
might interpret as the appearance of the new massless HS states
that we shall define below). The D2 brane field theory becomes a
$SU(N)$ invariant theory of free 3d super Maxwell multiplets. Here
we note that the Yang-Mills coupling in the dual SYM theory on the
stack of N coinciding D2-branes, $g^2_{\rm YM}=g_s/l_s$, is held
fixed in taking the near-horizon limit. This coupling also
coincides with the `local' Yang-Mills coupling on a stack of probe
D2-branes placed at energy scale $u$ in the near-horizon region,
$g_{\rm YM}^2(u)\equiv e^{\f(u)}\sqrt{-g_{00}(u)/ l_s^2}=g^2_{\rm
YM}$, as required for interpreting the stack of probe branes as
describing a Higgs branch of the dual SYM theory. Thus both the
dilaton and running string length vanishes in the UV limit, which
is why we can trust the free $SU(N)$ field theory even though the
gravitational curvature diverges.

Dualizing the vector fields and using $g_{\rm YM}^2$ to rescale
the fields, we obtain a free $SU(N)$ valued $OSp(8|4)$ singleton
$\Phi^i\in {\bf 8}_{v}$ described by the $SO(8)_R$ invariant
Lagrangian
\footnote{The singleton consists of $8$ scalars in $8_v$ and $8$
spinors in $8_s$ of $SO(8)_R$. By triality one can also obtain a
singleton multiplet in which the scalars are in $8_s$ and the
spinors in $8_c$.}

\be  \int d^3 x\, \tr \left((\partial \F^i)^2+{\rm
fermions}\right)\ . \ee

Conversely, assuming that this Lagrangian describes a fixed point
on the membrane we can break $SO(8)_R\ra SO(7)_R$ by taking the M
theory to have a finite radius $R_{11}$ and take $\Phi^8$ to be
periodic:

\be  \Phi^8\sim \Phi^8 + g\ ,\ee

where the radius $g$ is a constant with dimension $1/2$ which we
identify as $g=R_{11}/(l_{11})^{3/2}$ and $l_{11}$ is the
eleven-dimensional Planck length. We recover the free $OSp(8|4)$
invariant singleton in the decompactification limit $R_{11}\ra
\infty$. We may instead use $g$ to dualize $\F^8$ and introduce
Yang-Mills interactions with $g_{\rm YM}=g$. The effective
coupling is $g^2_{\rm eff}=g^2/u$, where $u$ is the 3d energy
scale, and as a result the theory now decompactifies in the IR
\cite{malda,rev,seiberg,int,min2}. Thus we have two
decompactification limits, the free $SU(N)$ valued $OSp(8|4)$
singleton field theory which resides in the UV and the strongly
coupled $SO(8)_R$ invariant $d=3$, $\cN=8$ SCFT  in the IR.

Thus it is natural to describe the low energy dynamics of M2
branes in terms of an UV fixed point of free $SU(N)$ valued
$OSp(8|4)$ singletons and an IR fixed point of strongly coupled
$OSp(8|4)$ singletons. We note that the number of massless degrees
of freedom indeed decreases along the RG flow, from $N^2$ to
$N^{3/2}$.

We conjecture that the free singleton theory at the UV fixed point
mentioned above is the holographic dual of the $hs(4|8)$ gauge
theory which admits the massless $hs(4|8)$ gauge theory described
in Section 2.2 as a consistent truncation. This theory describes
an unbroken phase of M theory with $N$ units of M2 brane charge.
The strongly coupled fixed point is the holographic image of a
broken phase, which admits an effective supergravity description
at low energies.

There are also IR fixed points containing free $OSp(8|4)$
singletons forming $N-1$ dimensional representation of the Weyl
group of $SU(N)$ \cite{seiberg}. These are curious points from the
point of view of HS dynamics, and it may be that one should also
include them as nontrivial points in the phase diagram.

As discussed in the previous section, the unbroken phase of the
Type IIB theory on $AdS_5\times S^5$ arises either as the critical
limit $\l\ra0$ at fixed $E$ and $s$, or as the high energy limit
$s>> \sqrt{\l}$ at fixed $\l$ and $N>>1$. Moreover, as we shall
see in the next section, the unbroken phase of M theory on
$AdS_7\times S^4$ arises at high energies whereby certain membrane
solitons propagate close to the boundary of $AdS_7$. This suggests
that also the unbroken phase of M theory on $AdS_4\times S^7$
arises in a high energy limit in which bulk membranes couple to HS
operators in the strongly coupled SCFT$_3$ with asymptotically
small anomalous dimensions, $(E-s)/s\ra 0$, as $s\ra\infty$. The
four-form flux in the $AdS_4$ directions results ensures the
M2-brane equations admit spherical membrane solutions in
$AdS_4\times S^7$ \cite{bdps1,bdps2,dps}. These solutions carry
internal $SO(8)$ spin, and are hence closely related to the
matrix-model found in the $pp$-wave limit \cite{ppwave}. It is
natural to expect that these solutions can be deformed into
time-dependent membrane solutions carrying also $AdS$ spin, in
analogy with the string solutions in $AdS_3$ with NS-fluxes
\cite{ooguri}. We also expect the anomalous part of the energy to
be minimized and certain fractional supersymmetry to be restored
by taking large AdS radius, i.e. large bulk energies, such that
the solution couples to the conserved HS currents of the $hs(8|4)$
theory. The fact that the holographic dual resides at a UV fixed
point should be encoded into the local geometry of the solution
and to how it minimizes the AdS energy, as in the case of the
rotating membrane in $AdS_7\times S^4$.

It will be interesting to examine the above picture in more detail
and in particular to examine the fluctuation spectrum about this
solution, where we expect to find some critical membrane theory
with fixed tension, and perhaps singletons in the worldvolume,
giving rise to the massless HS states.

We expect that the Higgsing of the massless HS fields and the
resulting spontaneous breaking of the $hs(4|8)$ is described by a
radially dependent solution to the HS theory which is the
anti-holographic dual of the 3d SYM flow obtained by switching on
a finite $g_{\rm YM}^2$ as discussed above. It will be interesting
to see whether HS field theoretic methods are still relevant for
describing this solution, which would then yield `weak/weak'
correspondence between the HS theory coupled to Higgs sector and
the SYM theory with expansions in both $1/N$ and $g_{\rm YM}^2$.
It may also be necessary to exhibit in more detail the nature of
the above-mentioned critical membrane.


\subsection{Cubic Couplings in the 4D Higher Spin Gauge Theory}


In this section we shall outline the structure of the minimal
bosonic HS gauge theory in $D=4$ which is a consistent truncation
of the supersymmetric HS theory discussed in Section 2.1. The
spectrum consists of massless fields with spin $s=0,2,4,...$, each
occurring once. The underlying algebra, called $hs(4)$, is an
infinite dimensional extension of the bosonic $AdS_4$ group.
Similar truncation exists also in $D=5,7$ at the spectrum level
but only in $D=4$ a full interacting theory is known, both
supersymmetric and minimal bosonic.

The $4D$ minimal bosonic model is of great interest because it is
the simplest interacting HS gauge theory (with propagating HS
degrees of freedom), and yet it exhibits all the essential
principles that underlie such theories. It is a very good starting
point for finding ways to construct the $D=5,7$ HS gauge theories
as well. Moreover, it is amenable to calculations and it is
possible to test directly in this model the consistent truncation
of the kind discussed in Section 4 which is required for the
holography picture to make sense. Here, we will not go as far as
carrying out these tests \cite{wip} but we will nonetheless
exhibit the structure the couplings to give the reader an idea
about how they actually look like, as well as providing enough
ingredients to facilitate the required holography computations.

Here we shall focus our attention on the quadratic terms in  all
the field equations, which, of course, mean all the cubic
couplings at the action level. In an accompanying paper
\cite{cubic}, we shall give a more detailed treatment involving an
expansion scheme where the gravitational gauge fields are treated
exactly and the gravitational curvatures and the HS gauge fields
as weak perturbations to all orders. The 4D HS/3d singleton
correspondence in the $hs(4)$ theory at the level of quadratic
field equation/cubic action will be provided elsewhere \cite{wip}.

The massless field equations (including general interaction
ambiguities) have been given in \cite{4dv1} and studied in more
detail in \cite{us3,us4,ssc} and more recently in \cite{cubic}.
These studies are based on a curvature expansion scheme. The most
important step in the expansion scheme is the linearized analysis
which shows that all auxiliary fields are non-propagating. As a
result it is possible to solve iteratively for the auxiliary
fields and obtain the physical field equations to any order. In
fact, this scheme yields field equations in terms of only the
physical fields.

The HS spin algebra $hs(4)$ is obtained from $hs(4|8)$ defined in
Section 2.1 by setting the fermionic generators $\th^i$ equal to
zero. To describe the field equations in 4D spacetime, which has
coordinates $x^\mu$, one introduces an auxiliary set of
coordinates $(z^\a,\bar{z}^{\ad})$ which are Grassmann even
spinors that are non-commutative in nature, and consider
extensions $\varphi(x;z,\zb)$ of the basic spacetime fields
$\varphi(x)$. One then imposes an integrable curvature constraint
in the extended space, whose $(x;z,\zb)$-components determine the
$(z,\zb)$ dependence of the extended fields $\varphi(x;z,\zb)$ in
terms of ``initial'' conditions $\phi(x)$. Setting $z=\zb=0$ in
the remaining $x$-components of the curvature constraint leads to
reduced curvature constraints in spacetime, which are integrable
by construction and one can show that they contain the physical
field equations of the HS gauge theory. Since $(z,\zb)$ are
non-commutative, the reduced constraints contain interactions even
though the original constraint in $(x;z,\zb)$ space has a simple
form.

The basic building blocks of the theory are a master 0-form $\Phi$
and a master 1-form

\be \widehat{A}=dx^\mu \widehat{A}_\mu + dz^\a \widehat{A}_\a+
d\zb^{\ad}\widehat{A}_{\ad}\ ,\ee

where the hats are used to indicate quantities that depend on
$(z,\zb)$. The hatted fields are given as expansions order by
order in $z$ and $\zb$, with expansion coefficients which are
functions of $(x,y,\yb)$ with the $(y,\yb)$ expansions determined
by suitable group theoretical conditions. These conditions are
engineered such that at $z=\zb=0$, the pulled-back components

\be A_{\mu}=\widehat{A}_\mu|_{Z=0}\ ,\quad
\Phi=\widehat{\Phi}|_{Z=0}\la{ic}\ee

define an $hs(4)$ valued spacetime one-form and a spacetime
zero-form in a certain quasi-adjoint representation of $hs(4)$
\cite{ssc}:

\bea A_\mu(x;y,\yb)&=& {1\over 2i} \sum_{m+n=2\ {\rm mod}\ 4}
{1\over m!n!}\, \yb^{\ad_1}\cdots\yb^{\ad_m} y^{\a_1}\cdots
y^{\a_n} A_{\m \a_1\dots\a_n\ad_1\dots\ad_m}(x)\ ,\la{aexp}\w2
\Phi(x;y,\yb)&=& \sum_{|m-n|=0\ {\rm mod}\ 4} {1\over m!n!}
\yb^{\ad_1}\cdots\yb^{\ad_m} y^{\a_1}\cdots y^{\a_n} \Phi_{
\a_1\dots\a_n\ad_1\dots\ad_m}(x)\ .\la{fexp}\eea

The curvature constraints giving rise to the spacetime field
equations read

\bea
\widehat{F}&=&d\widehat{A}+\widehat{A}\star\widehat{A}=\ft{i}4
dz^\a\wedge dz_\a {\widehat \Phi} \star \k + \ft{i}4
d\zb^{\ad}\wedge d\zb_{\ad} {\widehat \Phi} \star \bar{\k}\
,\la{c1}\w3
\widehat{D}\widehat{\Phi}&=&d\widehat{\Phi}+\widehat{A}\star
\widehat{\Phi}-\widehat{\Phi}\star \bar{\pi}(\widehat{A})=0\
,\la{c2}\eea

where the operators $\k,{\bar\k}$ are defined as

\be \k=\exp(iy^\a z_\a)\ ,\quad\quad
\bar{\k}=\k^\dagger=\exp(-i\yb^{\ad}\zb_{\ad})\ ,\ee

the $\pi$-map, and its complex conjugate ${\bar\pi}$, acting on an
arbitrary polynomial $f(y,\yb;z,\zb)$ are defined as

\be \pi(f(y,\yb;z,\zb))= f(-y,\yb;-z,\zb)\ , \quad\quad
{\bar\pi}(f(y,\yb;z,\zb))= f(y,-\yb;z,-\zb)\ , \ee

and the $\star$-product between two arbitrary polynomials
$f(y,\yb,x;\zb)$ and $g(y,\yb;z,\zb)$ is defined as

\be f*g=f \exp\left[ i\left({\overleftarrow{\partial}\over
\partial z_\a} + {\overleftarrow{\partial}\over \partial
y_\a}\right) \left({\overrightarrow{\partial}\over \partial z^\a}
- {{\overrightarrow\partial}\over \partial y^\a}\right)
+i\left({\overleftarrow{\partial}\over \partial {\zb}_{\ad}}
-{\overleftarrow{\partial}\over \partial {\yb}_{\ad}}\right)
\left({\overrightarrow{\partial\over \partial {\zb}^{\ad}}} +
{\overrightarrow{\partial}\over \partial {\yb}^{\ad}}\right)
\right]g\ . \ee

The constraints \eq{c1} and \eq{c2} have the gauge symmetry

\be \d \widehat{A}=
d\widehat{\e}+[\widehat{A},\widehat{\e}]_{\star}\ ,\quad\quad\quad
\d\widehat{\Phi}=
\widehat{\e}\star\widehat{\phi}-\widehat{\Phi}\star\widehat{\e}\
.\la{cgs}\ee

Given the initial conditions \eq{ic}, the components of the
constraints \eqs{c1}{c2} which have at least one $\a$ or $\ad$
index can be solved by expanding $\widehat{A}$ and
$\widehat{\Phi}$ in powers of $\Phi$, which contains curvatures
and the scalar field, as follows:

\be \widehat{\Phi}=\sum_{n=1}^\infty \widehat{\Phi}^{(n)}\ ,\quad
\widehat{A}_\a=\sum_{n=1}^\infty \widehat{A}_\a^{(n)}\ ,\quad
\widehat{A}_\mu= \sum_{n=0}^\infty \widehat{A}_\mu^{(n)}\
,\la{ces}\ee

where

\bea \widehat{\Phi}^{(n)}|_{Z=0}&=&\mx{\{}{ll}{\Phi\ ,&n=1\\0\
,&n=2,3,...}{.}\la{phiic}
\\
\widehat{A}_\m^{(n)}|_{Z=0} &=& \mx{\{}{ll}{A_\m\ ,&n=0\\0\
,&n=1,2,3,...}{.}\la{amic}\\ \widehat{A}_\a|_{Z=0}&=&0\
,\la{aaic}\eea

and $\widehat{\Phi}^{(n)}$ ($n=2,3,...$), $\widehat{A}^{(n)}_\a$
($n=1,2,3,...$) and $\widehat{A}_\m^{(n)}$ ($n=2,3,...$) are $n$th
order in $\Phi$. Note that $\widehat{A}_\m^{(n)}$ are linear in
$A_\m$. The condition \eq{aaic} is a physical gauge condition,
which can be imposed by using the gauge symmetry \eq{cgs}.

As shown in detail in \cite{cubic}, one first solves iteratively
the constraints ${\widehat F}_{\m\a}={\widehat F}_{\a\b}=
{\widehat F}_{\ad\bd}=0$ and $D_\a {\widehat \Phi}=0$ to determine
the $\Phi$ expansions of ${\widehat A}_\mu, {\widehat A}_\a $ and
${\widehat \Phi}$, which  schematically take the form

\be {\widehat A}_\m = {\widehat A}_\m [A_\m,\Phi]\ ,\quad\quad
{\widehat A}_\a= {\widehat A}_\a[\Phi]\ ,\quad\quad {\widehat
\Phi}={\widehat \Phi}[\Phi]\ .\ee

Having solved the $Z$-space part of \eq{c1} and \eq{c2}, the
remaining constraints $\widehat{F}_{\m\n}=0$ and $\widehat{D}_\m
\widehat{\Phi}=0$ yield spacetime field equations of the form

\bea F_{\m\n} &=&-\sum_{n=1}^\infty\sum_{j=0}^{n} \left(
\widehat{A}_{[\m}^{(j)}\star
\widehat{A}_{\n]}^{(n-j)}\right)|_{_{Z=0}}\ , \la{cn}\w2 D_\m \Phi
&=& \sum_{n=2}^\infty  \sum_{j=1}^{n}
\left(\widehat{\Phi}^{(j)}\star
\pi(\widehat{A}^{(n-j)}_\m)-\widehat{A}^{(n-j)}_\m\star
\widehat{\Phi}^{(j)} \right)|_{Z=0}\ , \la{bn} \eea
where $F=dA+A\star A$ and $D\Phi=d\Phi+A\star
\Phi-\Phi\star\bar{\pi}(A)$.

Next, we define the physical scalar $\phi$ and expand the master
gauge field $A_\m(x,y,\yb)$ as

\be \Phi|_{Y=0}=\phi .\ee

As for the vierbein and HS gauge fields, requiring that they
transform homogeneously under Lorentz transformation, one is led
to the following expansion scheme for the master gauge field

\be A_\m= e_\m +\o_\m + W_\mu +\left(i
\o_\m^{\a\b}\,\widehat{A}_\a\star\widehat{A}_\b -
\mbox{h.c}\right)_{Z=0}\ , \ee

where the vielbein and the Lorentz connections are defined as

\be e_\mu = {1\over 2i} e_\mu^{\a\ad}y_\a\yb_{\ad}\ , \qquad
\o_\m={1\over 4i}\left( \o_\m^{\a\b}y_\a y_\b+\mbox{h.c} \right)\
, \ee

and $W_\m$ contains the fields with spin $s=4,6,8,...$ and their
corresponding auxiliary fields.

We are now ready to state the result for the cubic couplings
\cite{cubic}. They give rise to quadratic terms in the field
equations given by \cite{cubic}

\bea (\nabla^2+2)\,\phi &=& \left( \nabla^\mu P^{(2)}_\mu -\ft{i}2
(\s^\mu)^{\a\ad} {\partial\over \partial y^\a}
{\partial\over\partial {\yb}^{\ad}}\, P^{(2)}_\mu\right)_{Y=0}\ ,
\la{sfe}\w4 \left(\s^{\m\n\r}\right)_\a^{\bd}\,{\cal
R}_{\n\r\,\bd\cd}&=& \left(\s^{\m\n\r}\right)_\a^{\bd}\, \left(
{\partial \over\partial {\yb}^{\bd}} {\partial \over \partial
{\yb}^{\cd}}J^{(2)}_{\n\r}\right)_{Y=0}\ , \la{efe}\w4
\left(\s^{\m\n\r}\right)_{\a_1}^{\bd}\,
F^{(1)}_{\n\r\,\a_2...\a_{s-1}\bd \cd \ad_2 ...\ad_{s-1}}&=&
\left(\s^{\m\n\r}\right)_{\a_1}^{\bd}\,\left( {\partial \over
\partial y^{\a_2}}\cdots {\partial\over \partial y^{\a_{s-1}}}
{\partial\over
\partial {\yb}^{\bd}}\cdots {\partial\over {\yb}^{\ad_{s-1}}}\,
J^{(2)}_{\n\r}\right)_{Y=0} \la{hsfe} \eea

where  ${\cal R}_{\m\n\ad\bd} \equiv  F_{\m\n\ad\bd}$ is the
(self-dual part of) the $AdS_4$ valued Riemann curvature, while
the curvature associated with spin $s=4,6,8,..$ fields is defined
as

\bea F^{(1)}_{\n\r\,\a_2...\a_{s-1}\bd \cd \ad_2 ...\ad_{s-1}} &=&
2\nabla_{[\n }W_{\r ]\a_2...\a_{s-1}\bd\cd\ad_2...\ad_{s-1}}
\nn\w2 &&-(s-2)(\s_{\n\r}\s_\m)_{\a_2}{}^\d\,
W_{\m\,\a_3...\a_{s-1}\bd\cd{\dot\d}\ad_2...\ad_{s-1}} \nn\w2 &&
-s (\s_\m\s_{\n\r})_{\bd}{}^\c\,
W_{\m\,\c\a_2\a_3...\a_{s-1}\cd\ad_2...\ad_{s-1}}\ . \la{f1} \eea

The covariant derivatives in \eq{sfe} and \eq{f1} are with respect
to lo the Lorentz connection $\o$. Furthermore, in \eq{hsfe} and
\eq{f1}, separate symmetrization in the dotted and undotted
indices is understood. Further definitions are

\bea P_\m^{(2)} &=& \Phi\star\bar{\pi}(W_\m)- W_\m\star \Phi
\nn\w2 && +\left[\Phi\star\bar{\pi}({\widehat e}_\m{}^{(1)})
-\widehat{e}_\m^{(1)}\star \Phi+\widehat{\Phi}^{(2)}\star
\bar{\pi}(e_\m)-e_\m\star\widehat{\Phi}^{(2)}\right]_{Z=0}
\la{p2}\w4 J_{\m\n}^{(2)}&=& -\Bigg(
\left[\widehat{e}_\m^{(1)}\star \widehat{e}_\n^{(1)}+
\{e_\m,\widehat{e}_\n^{(2)}\}_{\star} +\{e_\m,
\widehat{W_\n}^{(1)}\}_{\star}+ \{W_\m,
\widehat{e}_\n^{(1)}\}_{\star} \right]_{Z=0} \nn\w2 && +\left[i
R_{\m\n}{}^{\a\b}\widehat{A}^{(1)}_\a\star\widehat{A}^{(1)}_\b
+\mbox{h.c.} \right]_ {Z=0} +W_\m\star W_\n \Bigg) + \Bigg(\m
\leftrightarrow \n\Bigg)\ , \la{j2} \eea

and the hatted quantities occurring in the above equations are
given by \cite{cubic}

\bea
\hA^{(1)}_\a  &=&   -{i \over 2} z_\a \int_0^1 tdt~
\F(-tz,\yb)\kappa(tz,y)\ , \la{a1}\w4
\widehat{A}^{(2)}_\a &=& z_\a \int_0^1 tdt\left(
\widehat{A}^{(1)\b}\star \widehat{A}^{(1)}_\b\right)_{z\ra
tz,\zb\ra t\zb} \la{a2}\w2 &+&\zb^{\bd}\int_0^1 tdt \left[
\widehat{A}^{(1)}_{\a},\widehat{A}^{(1)}_{\bd}\right]_{z\ra
tz,\zb\ra t\zb} \nn\w4
\widehat{W}^{(1)}_\mu &=& -i\int_0^1{dt\over t}
\Bigg(\left[{\partial W_\m\over \partial
y^\a},\hA^{\a(1)}\right]_* +\left[\hA^{\ad (1)} ,{\partial
W_\m\over \partial \yb^{\ad}}\right]_* \Bigg)_{z\ra tz,\zb\ra
t\zb} \nn\w4
{\widehat\Phi}^{(2)}&=& z^\a\int_0^1dt\left[\Phi\star
{\bar\pi}(\hA_\a^{(1)}) - \hA_\a^{(1)}\star \Phi
\right]_{t\rightarrow tz, \zb \rightarrow t\zb} \la{phi2}\w3
&+&{\zb}^{\ad}\int_0^1 dt \left[\Phi\star {\pi}({\widehat
A}_{\ad}^{(1)}) - \hA_{\ad}^{(1)}\star \Phi \right]_{t\rightarrow
tz, \zb \rightarrow t\zb} \nn\w4
\widehat{e}^{(1)}_\mu &=& -i e_\m^{\a\ad} \int_0^1 {dt\over t}
\Bigg( \left[{\yb}_{\ad},\hA_\a^{(1)}\right]_*
+\left[\hA_{\ad}^{(1)},y_\a\right]_* \Bigg)_{z\ra tz,\zb \ra
t\zb}\ , \nn\w4
\widehat{e}^{(2)}_\mu&=&-i e_\m^{\a\ad} \int_0^1 {dt\over t}
\Bigg( \left[{\yb}_{\ad},\hA_\a^{(2)}\right]_* +\left[{\widehat
A}_{\ad}^{(2)},y_\a\right]_* \Bigg)_{z\ra tz,\zb \ra t\zb}\ ,
\nn\w3
&& -e_\m^{\a\ad} \int_0^1 {dt\over t} \int_0^1 {dt'\over t'}
\Bigg[ \hA^{\b(1)}\star \left({\partial\over\partial
z^\b}-{\partial\over\partial y^\b}\right) \Big(
\left[{\yb}_{\ad},\hA_\a^{(1)}\right]_*
                     +\left[\hA_{\ad}^{(1)},y_\a\right]_*
                     \Big)_{z\ra t'z,\zb \ra t'\zb}
\nn\w3
&& \qquad\qquad\qquad\qquad + \hA^{\bd(1)}\star
\left({\partial\over\partial \zb^{\bd}}+{\partial\over\partial
\yb^{\bd}}\right) \left( \left[{\yb}_{\ad},\hA_\a^{(1)}\right]_*
+\left[\hA_{\ad}^{(1)},y_\a\right]_*
                     \right)_{z\ra t'z,\zb \ra t'\zb}
\nn\w3
&& \qquad\qquad\quad + \left({\partial\over\partial
z^\b}+{\partial\over\partial y^\b}\right)\left(
\left[{\yb}_{\ad},\hA_\a^{(1)}\right]_*
                     +\left[\hA_{\ad}^{(1)},y_\a\right]_*
                     \right)_{z\ra t'z,\zb \ra t'\zb} \star \hA^{\b (1)}
\nn\w3
&& \qquad\qquad\quad + \left({\partial\over\partial
\zb^{\bd}}-{\partial\over\partial \yb^{\bd}}\right)\left(
\left[{\yb}_{\ad},\hA_\a^{(1)}\right]_*
                     +\left[\hA_{\ad}^{(1)},y_\a\right]_*
                     \right)_{z\ra t'z,\zb \ra t'\zb} \star \hA^{\bd(1)}
                     \Bigg]_{z\ra tz,\zb \ra t\zb}\ . \nn
\eea

In the above formulae, the replacement of $(z,\zb)$ by $(tz,t\zb)$
is to be made inside the integrals and {\it after} performing the
$\star$ products. Note also the quantity $ \hcA^{(1)}_\a$ is a
basic building block which occurs in many of the formulae above
and that it is first order in $\Phi$.

It is important to note that not all the fields occurring in
\eq{aexp} and \eq{fexp} are independent. An analysis of the
constraints \eq{c1} and \eq{c2} shows a)  $\Phi_{\a_1...\a_{2s}}
(s=2,4,...)$ are the Weyl tensors which can be in terms of the
curvatures, b) $\Phi_{\a(m)\ad(n)}$ for $m+n >2$ can be solved in
terms of $\phi$, the Weyl tensors and their derivatives, c)
$\o_\m^{\a\b}$ is, of course, the Lorentz spin connection which
can be solved in terms of the vierbein $e_\m^{\a\ad}$, and d)
$W_{\m\a(m)\ad(n)}$ for $|m-n|\ge 2$ are auxiliary gauge fields
which can be solved in terms of the physical fields
$W_{\a(s-1)\ad(s-1)}$ \cite{cubic}.

The general solution for the auxiliary fields is given by
\cite{cubic}

\bea W_{\a\ad,\b_1\dots \b_m\bd_1\dots \bd_n}&=&
\ft2{m+1}\widehat{\nabla} W_{\bd_1\bd_2,\a\b_1\dots\b_m
\ad\bd_3\dots\bd_n} +\e_{\ad\bd_1}\ft{2n}{n+1}
\Bigg[\ft{n-1}{m+n+2}\widehat{\nabla} W_{\bd_2}
{}^{\cd}{}_{,\a\b_1\dots\b_m\cd\bd_3\dots\bd_n} \nn\w2
&&+\ft{n+1}{m+n+2}\widehat{\nabla}W_{(\a}{}^{\c}{}_{,\b_1\dots\b_m)\c\bd_2
\dots\bd_n} -\ft{m}{(m+1)(m+2)}\e_{\a\b_1}\widehat{\nabla} W
^{\c\d}{}_{,\c\d\b_2\dots\b_m \bd_2\dots\bd_n}\Bigg]\nn\\ &&+
m\e_{\a\b_1}\x_{\b2\dots\b_m\ad\bd_1\dots\bd_n}\ ,\qquad n>m\geq
0\ , \la{auxsolv}\w4 \Phi_{\a_1\dots\a_m\ad_1\dots\ad_n}&=& -i
\widehat\nabla_{\a_1\ad_1}\Phi_{\a_2\dots\a_m\ad_2\dots\ad_n} \ ,
\la{phisolv} \eea

where the the modified covariant derivatives are defined by

\bea && \widehat{\nabla} W_{\a\b,\c_1\dots\c_m\cd_1\dots\cd_n}
=\ft12(\s^{\m\n})_{\a\b} \left(\nabla_\m
W_{\n,\c_1\dots\c_m\cd_1\dots\cd_n}-\ft12
J^{(2)}_{\m\n,\c_1\dots\c_m\cd_1\dots\cd_n}\right)\
,\la{nprime}\w3 &&\widehat{\nabla}_{\a_1\ad_1}
\Phi_{\a_2\dots\a_m\ad_2\dots\ad_n} =
\Big(\nabla_{\a_1\ad_1}\Phi_{\a_2\dots\a_m\ad_2\dots\ad_n}
-P^{(2)}_{\a_1\ad_1,\a_2\dots\a_m\ad_2\dots\ad_n}\Big)\ ,
\la{phisolv2} \eea

and separate total symmetrization of dotted and undotted indices
is understood.  Since $J$ and $P$ depend on the auxiliary fields,
eqs. \eq{auxsolv} and \eq{phisolv} must be iterated within the
curvature expansion scheme. This leads to explicit expressions of
all auxiliary components of $W_\mu$ and $\Phi$ in terms of the
remaining physical fields.

Further comments about the above results are in order:

1)\,The z-dependence of all the fields involved are exhibited. The
above results are explicit and the remaining task is reduced to
performing certain star products and doing some elementary
parameter integrals. These steps, as well as the derivation of the
above results and their generalization to all orders will be
provided elsewhere \cite{cubic}.

2)\,It is easy to rewrite the field equation \eq{efe} for the
graviton as
\footnote{We have set the AdS radius $R=1$ but it is
straightforward to re-introduce $R$ by dimensional analysis in
which the master 0-form and the master 1-form fields are
dimensionless.}

\be R_{\m\n}(\o)-g_{\m\n} =  \left[
(\s_\m{}^\l)^{\a\b}\,\left({\partial \over \partial
y^\a}{\partial\over \partial y^\b}\,J^{(2)}_{\l\n} \right)_{Y=0} +
( \m \leftrightarrow \n ) + \mbox{h.c.}\right] \ ,\ee

where $R_{\m\n}(\o)$ is the Ricci tensor obtained from the Riemann
tensor associated with the Lorentz connection $\o_\m$. It is
important to note that this connection contains torsion as can be
seen from \eq{auxsolv} and \eq{nprime} which for $m=0, n=2$ give

\be \o_\m{}^{ab} = \o_\m{}^{ab}(e)+\k_\m{}^{ab}\ , \ee

where $\kappa_\m{}^{ab}$ is the con-torsion tensor related to the
torsion tensor $T_{\m\n}{}^a$ as

\be \k_\m{}^{ab}=T_\m{}^{ab}-T_\m^{ba}+T^{ab}{}_\m\ , \ee

where

\be T_{\m\n}{}^a= \left(\s^a\right)_{\a\bd}\,\left( {\partial
\over\partial y_\a} {\partial \over \partial
{\yb}_{\bd}}J^{(2)}_{\m\n}\right)_{Y=0}\ . \ee

3)\, The elimination of the  auxiliary fields by means of the
equations \eq{auxsolv} and \eq{phisolv} gives rise to higher
derivative interactions. In particular, in a given spin sector,
the auxiliary fields are $W_{\m \a_1...\a_k\ad_{k+1}...\ad{2s-2}}$
with $k=0,1,...,s/2$ and they are related to the physical fields
$W_{\m\a(s-1)\ad(s-1)}$ schematically as

\be W_{\m, \a(m)\ad(n)}\ \sim \
\partial^{|m-n|/2}\, W_{\m,\a(s-1)\ad(s-1)}\ ,\qquad m+n=2s-2\ .\ee

Similarly, the components $\Phi_{\a(m)\ad(n)}$of the master scalar
field are related to the Weyl tensors which are purely chiral,
their derivatives as well as the derivatives of the scalar as
(taking $m>n$ without loss of generality)

\bea \Phi_{\a(m)\ad(m)} \ &\sim&\ \partial^m\,\phi\ ,\nn\w2
\Phi_{\a(m)\ad(n)}\ &\sim& \ \partial^{(m-n)/2}\,\Phi_{\a(m-n)}\
,\qquad m-n=0\, {\rm mod} \,4\ .\eea

4)\, Whether the master constraints \eq{c1} and \eq{c2} are unique
is an important question. In fact, there exist a generalization of
\eq{c1} in which \cite{cubic} we let

\be {\widehat \Phi}\star \kappa \ \ra \ {\cV}({\widehat \Phi}\star
\kappa )\ ,\quad\quad {\widehat \Phi}\star {\bar\kappa} \ \ra \
{\bar \cV}({\widehat \Phi}\star {\bar\kappa}) \ ,\ee

where $\cV(X)$ is a $\star$-function, with its complex conjugate
$\bar{\cV}(X^\dagger)=(\cV(X))^\dagger$. In \cite{cubic} we argue
that this function must be of the form

\be \cV(X)=  \sum_{n=0}^{\infty} b_{2n+1}\,X^{2n+1}\ ,\qquad
|b_1|=1\ . \ee

A similar interaction ambiguity is expected to arise in HS
theories yet to be constructed in $5D$ and $7D$ as well, and
implications of this are discussed in Section 4, in the context of
5D HS gauge theory and holography. In particular, we argued that
the freedom in choosing $b_{2n+1}$ is important in order to find
the precise agreement between the bulk amplitudes and the boundary
correlators required for the massless theory to be a consistent
truncation.


\section{M Theory on $AdS_7\times S^4$ and 7D Higher Spin Gauge Theory}


The low-energy dynamics of a stack of $N$ parallel coinciding M5
branes in flat eleven dimensional spacetime is described by a
strongly coupled $d=6$, $\cN=(2,0)$ SCFT with $SO(5)_R$ symmetry
group \cite{malda,rev}, known as the the $A_{N-1}(2,0)$ theory
\cite{20a,20b,20c,seiberg}. The theory is conjectured to have no
marginal operators, which means that it describes an isolated UV
fixed point of the renormalization group. It is not known whether
the theory has any relevant operators which preserve the
R-symmetry (the supergravity dual description provides relevant
operators which break the R-symmetry). Conversely, starting in the
IR with a number, $N'$ say, of free $d=6$, $\cN=(2,0)$ tensor
singletons, it is not known how to describe non-abelian
interactions among tensor fields; in fact, there are no local
perturbations with this effect \cite{sevrin}
\footnote{A single tensor multiplet admits self-interactions, such
as for example those describing the motion of a single M5 brane
\cite{hs1,hs2,cederwall}.}.
This is believed to reflect the fact that open membranes ending on
coinciding M5 branes give rise to tensionless closed strings and
that the proper language for formulating the dynamics on the
fivebrane is therefore not ordinary field theory but rather some
nonlocal extension of it.

However, if we are willing to give up 6d covariance, then we can
use lower-dimensional RG flows based on ordinary interacting field
theories to define the $A_{N-1}(2,0)$ theory
\cite{malda,rev,seiberg,min2,int}. In particular, circle
reductions of the 6d theory describes RG flows of 4d and 5d SYM
theories with $SU(N)$ gauge group. The $SO(4)_R$ invariant RG$_5$
flow has a Type IIA supergravity dual description in terms of the
near horizon region of a D4 brane solution. In the UV limit the
dilaton diverges and the solution uplifts to the $AdS_7\times S^4$
near horizon region of the stack of M5 branes. The resulting
anti-holographic description of the $A_{N-1}(2,0)$ theory is
conjectured to be M theory on $AdS_7\times S^4$ \cite{malda}. For
large $N$ the membrane tension scales like

\be T_{M2}\sim {N\over R^3}\ ,\la{7dm2}\ee

where $R$ is the AdS radius, and the 7D Planck length is given by

\be {1\over l_{\rm Pl}^5}={N^3\over R^5}\ .\la{7dpl}\ee

For large $N$ the Planck length is much smaller than the M2 length
scale, $l_{\rm M2}$ which in in turn is much smaller than the
radius. Thus, for energies $E$ obeying $1<< E << R/l_{\rm M2}$ the
low-energy dynamics of the anti-holographic dual is described by
$D=7$, $\cN=2$ gauged supergravity. The $A_{N-1}(2,0)$ theory has
been conjectured to admit an expansion in terms of integer powers
of $1/N$ which factorize for large $N$ \cite{rev}
\footnote{From \eq{7dm2} it follows that M theory on $AdS_7\times
S^4$ has an expansion in terms of integer powers of $1/T_{\rm M2}$
rather than integer powers of the 7D Plank's constant. The same
remark applies to M theory on $AdS_4\times S^7$, which has M2
tension given by \eq{4dm2} and has been conjectured to have an
expansion in terms of integer powers of $1/\sqrt{N}$\cite{rev}.}
From \eq{7dpl} it follows that the $A_{N-1}(2,0)$ theory has $\sim
N^3 $ massless degrees of freedom for large $N$ which contain the
$N-1$ massless $(2,0)$ tensor multiplets of the `Higgs branch' of
the theory.

In the IR limit of the D4 brane geometry the dilaton $e^{\f}$
vanishes and the gravitational curvature diverges. As for the
D2-brane discussed in Section 6.1, the dual SYM coupling $g^2_{\rm
YM}=g_s l_s$ is held fixed in taking the near horizon limit and
equals the local Yang-Mills coupling $g_{\rm YM}^2(u)\equiv
e^{\f(u)}/\sqrt{-g_{00}(u)/l_s^2}=g^2_{\rm YM}$. Hence the local
string length diverges in the IR (unlike the case of the D2 brane
where the local string length disappears together with the dilaton
in the UV). Hence, naively the D4 brane field theory becomes a
free $SU(N)$ valued $d=5$, $\cN=2$ Maxwell theory with $SO(5)_R$
symmetry and finite Yang-Mills coupling $g_{\rm YM}^2$. This
theory can be made scale invariant by absorbing $g^2_{\rm YM}$
into the fields, but this symmetry is superficial since it cannot
be lifted to superconformal invariance.

Instead a more natural interpretation is that superconformal
invariance is restored by uplifting to a free $SU(N)$ valued
$d=6$, $\cN=(2,0)$ tensor singleton described by the
superconformal action\footnote{Tensor self-duality and
supersymmetry can be restored at the level of the field equations
\cite{cederwall}.}

\be S_6=\int d^6x\, \tr\left( |d\F^a|^2 +
|dB|^2+\mbox{fermions}\right)\ ,\la{s6}\ee

where $\F^a$ ($a=1,...,5$) and $B_{\m\n}$ have dimension $2$. The
superspace formulation of this theory is described in more detail
in Section 3.3. The 6d superconformal invariance can be
spontaneously broken by compactifying \eq{s6} on a circle of
radius $R_{11}$ (after which $R_{11}$ can of course be used to
rescale the fields):

\be S_{5+1}={1\over R_{11}}\int d^5x\,\tr\left( |d\f^a|+|dA|^2
+\mbox{KK modes and fermions}\right)\ ,\ee

where all bosonic fields have dimension $1$. Taking $S_{5+1}$ as
the generic starting point for describing the fivebrane dynamics
there are thus two ways in which the theory can decompactify and
become superconformally invariant: either by directly taking
$R_{11}\ra \infty$ which yields back $S_6$; or by throwing away
the KK modes and switching on the Yang-Mills interaction with
$g_{\rm YM}^2=R_{11}$, which then reaches the $A_{N-1}(2,0)$ limit
in the UV limit $R_{11}\ra \infty$. Note that the Yang-Mills
deformation does not lead to loss of degrees of freedom since the
KK modes are exchanged with monopoles with mass proportional to
$g_{\rm YM}^2=R_{11}^{-1}$.

We conclude that it is natural to describe the low energy dynamics
of $N$ coinciding M5 branes in terms of an IR fixed point of free
$SU(N)$ valued $d=6$, $\cN=(2,0)$ tensor singletons and a UV fixed
point given by the $A_{N-1}(2,0)$ theory in the unbroken HS phase.
We note that the number of massless degrees of freedom indeed
decreases along the RG flow, from $N^3$ to $N^{2}$.

We conjecture that the free singleton theory at the IR fixed point
mentioned above is the holographic dual of an $hs(8^*|4)$ gauge
theory which admits a consistent truncation to the massless
$hs(8^*|4)$ gauge theory in $D=7$ described in Section 2.2. This
theory describes an unbroken phase of M theory with $N$ units of
M5 brane charge. The strongly coupled fixed point is the
holographic image of a broken phase which admits an effective
supergravity description at low energies.

As in the case of M theory on $AdS_4\times S^7$, there are also
curious IR fixed points consisting of $N-1$ free tensor multiplets
acted upon by the Weyl group of $SU(N)$ \cite{seiberg} which
should also be included as nontrivial points in the phase diagram
of M theory on $AdS_7\times S^4$.

We can  motivate further our proposal  by constructing and
examining the properties of `long membrane' solutions to the
M2-brane action \cite{bst}

\be S_{M2}=N\int d^3\s \sqrt{-\det \gamma} +N\int C_3\ ,\ee

where we have set the fermions equal to zero,
$\c_{\a\b}=\partial_\a X^M\partial_\b X^N g_{MN}$ and $C_3$ is the
pull-back of the M-theory three-form potential which has non-zero
components only in $S^4$. The worldvolume field equations are

\be \partial_\a(\sqrt{-\c}\c^{\a\b}\partial_\b X^M g_{MQ})-{1\over
2} \sqrt{-\c}\c^{\a\b}\partial_\a X^M\partial_\b X^N \partial_Q
g_{MN}+\e^{\a\b\c}\partial_\a X^M\partial_\b X^N\partial_\b X^P
H_{QMNP}=0\ ,\la{m2fe} \ee

where $H_4=dC_3$. In order to describe the solution, which is
similar to the string solution of \cite{gkp2}, we use the global
coordinates in $AdS_7$:

\be ds^2=-\cosh^2\r dt^2+d\r^2+\sinh^2\r
(d\th^2+\sin^2\th(d\phi^2+\sin^2\phi d\O_3^2))\ .\ee

The $M2$ brane worldvolume coordinates are $(\tau,\s,\vf)$ and our
rotating membrane solution is given by

\be t=\tau\ ,\quad \r=\r(\s)\ ,\quad \th=\th(\varphi)\ ,\quad
\phi=\omega \t\ ,\quad \mbox{fixed point in $S^3$\ ,}\la{m2c}\ee

where the membrane has the topology of a cylinder $-1<\s<1$,
$0\leq\varphi<2\pi$, which has been flattened such that the
portion with $0<\varphi<\pi$ is folded on top of the portion with
$\pi<\varphi<2\pi$. The induced metric becomes

\be
ds^2=-(\cosh^2\r-\o^2\sinh^2\r\sin^2\th)d\t^2+(\r')^2d\s^2+\sinh^2\r
(\th')^2d\varphi^2\ ,\ee

where $\rho'\equiv d\rho/d\s$ and $\theta'\equiv d\theta/d\vf$. It
is straightforward to verify the nontrivial components of the
field equations \eq{m2fe} which are the $t,\r,\th,\phi$ component.
The energy and spin of the configuration \eq{m2c} are given by

\bea E&=&4 N \int_0^{\r_0} d\r \int_{\th_1}^{\th_2} d\th
{\cosh^2\r
\sinh\r\over \sqrt{\cosh^2\r-\o^2\sinh^2\r\sin^2\th}}\ ,\la{m2energy}\\
s&=&4 N \o\int_0^{\r_0} d\r \int_{\th_1}^{\th_2} d\th {\sinh^3\r
\sin^2\th \over \sqrt{\cosh^2\r-\o^2\sinh^2\r\sin^2\th}}\
,\la{m2spin}\eea

where

\be \th_1=\th(0)=\th(2\pi)\ ,\quad \th_2=\th(\pi)\ ,\quad
\r_0=\r(\pm 1)\ .\ee

The solution which minimizes the energy for a fixed spin and fixed
width $\ell=\th_2-\th_1$ (we shall minimize the energy with
respect to the width below) is obtained by centering
$\th(\varphi)$ around $\th=0$ and maximizing the extension in the
$\r$-direction by taking

\be \coth\r_0=\o\ .\ee

If we assume that $\ell$ is small then

\bea E&=& {\ell N\over \o^2} {}_2F_1[2,1;3/2;1/\o^2]\ ,\\
s&=&{2\ell N\over 3\o^3} {}_2F_1[2,2;5/2;1/\o^2]\ .\eea

For $\o>>1$ this describes short membranes with length $\r_0\sim
1/\o$ and energy and spins given by

\be E^3=8\ell N ~s^2\ ,\quad E,s<<\ell N\ .\la{e2}\ee

In flat eleven-dimensional spacetime an analogous relation holds
between mass and spin for all values of the spin (in flat space
this relation follows from dimensional analysis). Thus, in flat
spacetime the mass is minimized for given spin by sending $\ell\ra
0$ and $\o\ra 0$ (keeping $s$ fixed). The flat space spectrum
therefore contains massless states arbitrary spin, which can be
thought of as infinitely long, thin string-like membranes which
are virtually at rest.

In fact, long ago bosonic open membrane (a disk) rotating
simultaneously about two axis was considered in \cite{rot1} where
the relation a relation like \eq{e2} was derived. Such solutions
are possible for $D\ge 5$. Later, this solution was generalized in
\cite{rot2} for the $D=11$ supermembrane \cite{bst}, by gluing two
copies of the open membrane of \cite{rot1} along their edges to
obtain a `pancake' membrane. The zero-point energy of this
membrane was studied by these authors and later in \cite{rot3}. It
was conjectured in \cite{rot2} that the (semi-classical)
energy-angular momentum relation of the kind \eq{e2} would be
modified by an integral or half integral number due to the  fact
that the fermionic coordinates of the supermembrane also carry
intrinsic angular momentum. See \cite{pkt2} for a review of this
fascinating subject.

Going back to $AdS_7\times S^4$, for slow rotation, $\o\sim 1$,
$\o>1$ and finite width $\ell$, the solution \eq{m2c} describes
long membranes whose energy and spin now obeys

\be E-s={3\pi^{2/3}\over 2^{1/3}}(\ell N)^{2/3}s^{1/3}\ ,\quad
E,s>>\ell N\ .\la{e3}\ee

For $\o\ra 1$ the energy and spin diverges and the rotating
membrane develops a boundary given by a folded closed string of
length $\ell$ which trace out a Wilson surface in the stack of
five-branes. Thus, the long membranes of width $\ell$ with finite
energy describe operators in the the $A_{N-1}(2,0)$ theory which
arise in the operator product expansion of the Wilson surface. The
shape of the Wilson surface together with \eq{e} suggest that its
expansion contains bilinear higher spin operators which have
asymptotically small anomalous dimensions, $(E-s)/s<<1$
 for high spin, $s>>\ell
N>>1$. In the limit $s\ra \infty$ their interactions should be
equivalent to those described by the singletons.

Suppose there is no boundary condition which fixes $\ell$ to a
finite value. The prescription is then to vary $\ell$ keeping $s$
fixed as to minimize $E$. The minimal energy configuration for
given spin $s$ is obtained by taking $\ell\ra 0$, $\o\ra 1$ which
results in an infinitely long string-like membrane with energy
$E=s$ (the ratio $E/s$ is larger for short wide membranes than for
long thin ones). Note that this geometry is assumed for any value
of $s$, unlike in the case of the Type IIB closed string which
became infinitely long only as $s/\sqrt{\l}\ra \infty$. As
$\ell\ra 0$ the dual Wilson surface collapses and the higher
derivative corrections to the $A_{N-1}(2,0)$ theory becomes
suppressed, resulting in a flow down to the free tensor theory
describing the unbroken phase with $hs(8^*|4)$ gauge symmetry.

Let us examine the supersymmetry of this solution. The condition
for worldvolume supersymmetry is \cite{bdps1}

\be \C\e=\e\ ,\quad \C={1\over \sqrt{-\det \c}}{1\over 3!}
\e^{\a\b\c}\partial_\a X^M\partial_\b X^N\partial_\c X^N\
,\la{Gamm}\ee

and that $\e$ is the Killing spinor of the $AdS_7\times S^4$
background. An important property of these Killing spinors is that
as we approach the boundary of $AdS_7$, i.e. as $\rho \ra \infty$,
they become an eigenstate of a constant $\C$-matrix as follows
\cite{bdps1}

\be \tilde{\C}\e =\e\ ,\quad \tilde{\C}=\C_{012345}\ , \ee

where $\C_a$ are flat Dirac matrices and $a=0,...,5$ are the
indices tangent to the boundary of $AdS_7$. We have relabeled the
coordinates of $AdS_7$ as

\be (t,\phi,\theta, \psi',\theta',\phi',\rho) \ra
(x_0,x_1,...,x_5,\rho)\ , \ee

where $(\psi',\theta',\phi')$ are the $S^3$ angles. Now, inserting
the solution \eq{m2c} into the definition of $\C$ in \eq{Gamm}
gives

\be \C= {\Big( c\C_0+(\o s)\sin\th\C_1\Big)\C_{62}\over
\sqrt{c^2-\o^2s^2\sin^2\th}}\ . \la{Gammahat} \ee

where $c=\cosh \r$ and $s=\sinh \r$. Next, we find that
$[\C,\tilde \C]=0$. Therefore the worldvolume supersymmetries can
be written as

\be \e=(1+\C)(1+\tilde \C)\eta\ , \la{susye} \ee

for arbitrary $\eta$. We conclude that in the limit $\ell \ra 0,
 \o \ra 1, \rho \ra \rho_0\ra \infty$ keeping $s$ fixed, the
solution \eq{m2c} preserves the $8$ supersymmetries described by
\eq{susye}. The remaining $8$ supersymmetries are broken, which
means that the limiting solutions belong to semi-short multiplets
which we identify as the massless HS multiplets arising in the
tensor product of two tensor singletons listed in Table \ref{tss}.
Hence the corresponding anomaly free operators in the dual SCFT
must fall into the same semi-short multiplet. This suggests that
the dual operators are the Konishi-like superfields \eq{kon6} and
\eq{kon7} containing the conserved HS currents described Section
3.

In the above limit the energy and the spin of the membrane
accumulate at its ends which in turn move along light cones at the
the boundary of $AdS_7$. The condition \eq{susye} has a natural
interpretation as the supersymmetry condition for an intersection
between the five-brane and the boundary of an open membrane. This
suggests that the relevant part of the membrane dynamics are the
fluctuations in this asymptotic region. It will be interesting to
study more carefully the fluctuation spectrum about these
worldvolume singletons, and in particular to examine whether they
exhibit features such as fixed critical tension and discrete
spectrum.

From the fact that the membrane interactions are concentrated at
the boundary of the AdS spacetime we conclude that the $hs(8^*|4)$
gauge theory is a high energy limit of M theory on $AdS_7\times
S^4$.

We remark that the rotating membrane limit is a Lorentzian analog
of the $pp$-wave limit on $AdS_7\times S^4$ \cite{ppwave} which
can be thought of as a collapsed membrane rotating around the
equator of $S^4$. A difference that might be important is that the
collapsed membrane has spherical topology while the rotating
membrane has cylindrical topology.

One important test of the free $CFT_6/7D$ HS gauge theory
correspondence is the matching of the holographic Weyl anomaly
\cite{hs}. It was shown in \cite{frolov} that the 6d trace-anomaly
of (free) tensor singletons does not match the holographic anomaly
computed in gauged supergravity in $D=7$ \cite{hs}. To be more
precise, the relative strength of the Euler invariant and the
remaining invariant differs in the two cases by a factor of $4/7$.
It was argued in \cite{frolov}, however, that the trace anomaly of
any CFT$_6$ picks up contributions from four-point stress-energy
correlators. This makes the 6d trace anomaly sensitive to the
actual interactions of the bulk theory, unlike in e.g. $d=4$,
where the trace anomaly can be computed at weak coupling. Thus the
6d free field trace anomaly should be compared with the
holographic anomaly of the corresponding 7D massless $hs(8^*|4)$
gauge theory (which does not admit any consistent truncation to
gravity). Note that the corrections from massless HS exchange in
the bulk are higher order in derivatives but of the same order in
$1/N$. The matching of the overall strength is a consequence of
the normalization \eq{lpl}, though the interactions should correct
the factor of $4/7$ \cite{frolov}.

Finally we comment on the breaking of HS gauge symmetries in
$AdS_7$. The situation is much less clear here than in $D=4,5$,
basically due to the fact that we do not know how to deform the
boundary theory. As was discussed in Section 3.3, we do not have
any candidates for the Higgs fields, which may have to do with the
fact that we are writing local expressions whereas a more drastic,
perhaps nonlocal construction, is what is actually required to
break the HS symmetries in $D=7$. Another important difference
between the massless spectra in $D=7$ and in $D=4,5$ is the fact
that the latter saturate the unitarity bound for  UIRs belonging
to certain continuous series of the corresponding AdS supergroups,
while the former belongs to an isolated series (see Section 3). In
fact, this can be used to show that there can be no continuous
(marginal) deformations taking the free SCFT to the strongly
coupled fixed point \cite{as}. Another curious fact is that the
massless HS theory described in Section 2.3 does not make use of
the `massless' states which saturate the unitarity bound for UIRs
belonging to the continuous series A. These operators are
described by cubic tensor singletons, and it will be interesting
to attempt to incorporate these into the master field formulation
for the 7D HS theory described in Section 2.3.


\section{Summary and Discussion}


We have proposed that Type IIB string theory with $N$ units of
D3-brane charge and M theory with $N$ units of M2-brane or
M5-brane charge have unbroken phases described by HS gauge
theories which admit consistent truncations to massless HS gauge
theories in $D=4,5,7$ with holographic duals given by $SU(N)$
valued scalar singleton theories in $d=3,4,6$ with $16$
supersymmetries. The corresponding HS algebras are

\bea && hs(8|4) \supset OSp(8|4)\ ,\w2 && hs(2,2|4)\supset
PSU(2,2|4)\ ,\w2  && hs(8^*|4)\supset OSp(8^*|4)\ ,\eea

which are described in Section 2 together with the corresponding
massless HS gauge  theories. These theories also contain massive
fields, some of which are Higgs fields that can be eaten by the
massless fields. Both massless and massive fields also have KK
towers which can be used to re-construct the spectrum of the Type
IIB string and M theory in appropriate limits as discussed in more
detail in Section 5.

In the  case of Type IIB string on $AdS_5\times S^5$, we have
conjectured that the $hs(2,2|4)$ gauge theory arises in a critical
limit of the Type IIB theory in which

\be g_s\sim 1/N\ ,\quad\quad l_s\sim R\quad\quad {\rm fixed}\ R\ ,
\quad N>>1\ , \ee

and that this limit corresponds to the free $4d, {\cN}=4, SU(N)$
SYM in which $g_{YM}=0$ . This means that the relations between
the closed string parameters in $AdS_5\times S^5$ and the gauge
theory parameters for $\l>>1$, which are read off from the
D3-brane solution obtained in the supergravity approximation, are
renormalized, as discussed in Section 5, and summarized in
\eq{eq11} and \eq{eq12}.

In the case of M theory on $AdS_{4/7}\times S^{7/4}$, we have
conjectured the holographic boundary theories to be a $SU(N)$
valued $OSp(8|4)$ singleton field theory which resides at a UV
fixed point in $3d$, and a free $SU(N)$ valued $(2,0)$ tensor
singleton field theory residing at a IR fixed point in $6d$.

The spectrum of massless states in all the HS gauge theories
discussed here have the universal property that they all arise in
the symmetric product of two singletons. This motivates  a
worldsheet sigma model description of these theories based on an
affine extension of  $AdS$ superalgebras in $D=4,5,7$ with
critical KM levels leading to left-moving and right-moving
singleton Verma modules with a maximal number of null-states. In
this respect, the existence of a singleton-like representations of
affine $SO(3,2)$ with level $k_{\rm crit}=5/2$ found in \cite{ds1}
is encouraging.

The idea of obtaining the massless states of a $D=4, {\cN}=8$ HS
theory starting from the free $OSp(8|4)$ singleton theory, which
in turn was obtained from the eleven dimensional supermembrane on
$AdS_4\times S^7$, already appeared long ago \cite{bsst}. We
recall that all the massless fields in this theory, with the
exception of a pseudoscalar, satisfy the energy-spin relation
$E_0=s+1$. More recently, long rotating strings that extend to the
boundary of $AdS_5$ and couple to operators which are
asymptotically anomaly free, i.e. $(E-s)/s\ra 0$ as $E,s\ra
\infty$, have been studied \cite{gkp2}.

Motivated by above the considerations, we have found rotating long
membrane solutions \eq{m2c} to the equations which describe the
M2-brane in $AdS_7\times S^4$ background. These membranes have
width $\ell$ and the geometry of infinitely stretched strings with
energy and spin density concentrated at the end points. They
satisfy the semi-classical energy-spin relation $E=s$. A feature
not present in the string case is that the energy is minimized for
fixed spin by sending the angular velocity $\o\ra 1$ and the width
$\ell\ra 0$ keeping $s$ fixed, resulting in infinitely long
membranes with string-like geometry and semi-classical energy
$E=s$. In Section 7, we have interpreted these as the lowest
weight states of the massless supermultiplets of the 7D HS gauge
theory discussed in Section 2 (see Table \ref{tss}). Further
aspects of this picture, especially the quantization issue, remain
to be studied.

It would also be interesting to study  the spherical membrane in
$AdS_4$ and examine whether it admits `breathing' and `rotation'
modes similar to those of strings in $AdS_3$ with NS-fluxes
\cite{ooguri}.

As there is effectively no separation in AdS energy between the
massless HS fields and the massive HS fields, we have proposed
that the massless HS theories (based on HS extension of the  $32$
supercharge $AdS_{d+2}$ superalgebras in $d=3,4,6$) arise as a
result of consistent truncation of the full HS theories. This
proposal can be tested explicitly since for large $N$, the
singleton theory and the HS gauge theory can be compared order by
order in the $1/N$ expansion: consistent truncation implies that
the massless HS theory action is the generating functional of
correlators of bilinear operators. Indeed, a correlation function
of four bilinear operators in a singleton theory can be written in
a manifestly $s$-$t$-$u$ symmetric form in terms of two- and
three-point functions involving only bilinear operators, as
discussed in Section 4.

We have also examined mechanisms for spontaneous breaking of HS
gauge symmetry down to the symmetries underlying ordinary
supergravity. In $D=4,5$ the `order parameter' for breaking of HS
gauge symmetry is the holographic Yang-Mills coupling. In $4d$
this is a marginal deformation which corresponds to a finite
dilaton VEV in the bulk. The broken theory has an AdS vacuum in
which the broken gauge fields have non-critical masses
$m^2-m^2_{\rm crit}\sim N g_{\rm YM}^2/R^2$. Using the
non-intersection principle we argue these cross over into the
leading Regge trajectory as $N g_{\rm YM}^2$. We have also
identified the Higgs multiplets at arbitrary level in the HS
spectrum, and the realization of the level-one Higgs multiplet in
terms of composite operators (i.e. anomaly multiplets) in the free
singleton SCFT.

Also in $3d$, where the Yang-Mills coupling is a relevant
perturbation, we have identified the Higgs multiplets at arbitrary
level in the HS spectrum, and the realization of the level-one
Higgs multiplet in terms of composite operators (i.e. anomaly
multiplets) in the free $OSp(8|4)$ singleton field theory.

In $D=7$ we do not know what is the order parameter for breaking
HS gauge symmetry, nor have we identified the Higgs multiplets.
This is presumably related to the fact that the massless gauge
fields in $D=7$ belong to the discrete B series (see \eq{6d2} in
Appendix B). We believe this issue should have a simple resolution
in a framework where the nature of the mysterious interactions on
the fivebrane is well understood. We stress that the Higgsing of
the 7D HS gauge theory is dual to weak irrelevant perturbations of
the tensor theory in the IR, which should be describable using a
field theoretic, perhaps non-local, construction in $6d$. One may
also speculate that the continuous A series (see \eq{6d1}) could
play a role in this, since the corresponding fields can be
Higgsed, which signals the existence of the corresponding anomaly
multiplets. This, in turn, would provide valuable data on the
details of the interactions in $6d$.

An interesting open problem is to use the HS gauging techniques
described in Section 2 and 6 to construct interactions between
massless HS fields and Higgs fields. Clearly, the issue of
consistent truncation becomes moot once we include (massive) Higgs
fields. It is therefore a challenge to examine whether some
generalized truncation scheme, perhaps of the type described in
\cite{pioline}, may temper the fluctuations in the massive sector.

In testing various aspects of the AdS/HS gauge theory
correspondences discussed in this paper, it will be very useful to
develop a deeper understanding of the geometrical nature of HS
interactions, possibly formulating them in a generalized
superembedding approach. This would provide a universal tool for
studying the HS dynamics \cite{johan} which would not only
simplify the task of coupling Higgs master fields to HS gauge
theories but also yield a superfield formulation \cite{johan} that
would simplify the treatment of the bulk interaction and the
computations of the attendant Witten diagrams. On the boundary
side, the existing literature on the OPE computations involving
free fields should be extended to cases where subleading in $1/N$
contributions will arise \cite{su1}. We have described few
examples of such correlators in Section 4.

In this paper we have focused our attention on HS gauge theories
in $D=4,5,7$. No doubt these results can be extended to $AdS_6$ as
well. In $D=3$ the HS gauge fields do not propagate physical
degrees of freedom. Nonetheless, physical matter fields of spin
$s=0,1/2$ can be coupled to massless HS gauge theory
\cite{3dv1,3dv2}. The advantage here is that an action principle
is known and the mathematics is much simpler than in higher
dimensions. It would be interesting to study this model in the
context of massless higher spins and holography.

At the algebraic level there is in principle no bound on the
number of supersymmetries in HS gauge theories and we expect
consistent massless interactions for any $\cN$ in $D\leq 7$,
though certain restrictions follow from the requirement of an R
symmetry neutral vierbein \cite{johan}. As discussed in Section 4,
the restrictions on the spacetime superdimension are instead
expected to be related to the consistency of the full HS quantum
theory, including both massless and massive states, which requires
the full generating functional \eq{genfunc} of the free singleton
SCFT with finite sources for composite operators. Effectively, the
condition that this quantity exists is expected to be as
restrictive in the free singleton SCFT as in the (strongly)
interacting singleton SCFT. This may lead to the restriction that
the  holographic dual cannot have more than $16$ supersymmetries
in $d\leq 6$. Similar restrictions should follow from the quantum
consistency of the yet to be constructed dual bulk sigma models.
In Section 4, similar effects were argued to arise in the
holographic theory due to insertions of sewing operators in the
free singleton field theory required for unitarity.

Another particularly interesting class of singleton CFTs, which we
have not considered here, are the free $4d$ conformal HS theories
constructed in \cite{5dv1}. Here the singleton field is a master
field comprising an infinite set of ordinary singletons which
together form an irreducible representation of a HS extension of
the $d$-dimensional conformal group. In that case the relevant HS
symmetry algebra is an infinite dimensional extension of $Sp(8,R)$
which contains the $AdS$ group in $5D$.

To conclude, we believe that the remarkable algebraic and
geometric structures underlying HS gauge symmetry are natural
extensions of supergravity and will be important guides towards
the true foundations of string and M theory. In particular, the
simplicity of their holographic duals together with the fact that
the bulk physics can still be phrased in a relatively simple
language is both gratifying and compelling. Clearly much remains
to be done in this subject which may be viewed as being still in
its infancy.

\vspace{2cm}


\noindent{\Large \bf Acknowledgements}\\


We thank D. Berman, U. Danielsson, S. Frolov, I. Klebanov, F.
Kristiansen, J. Minahan, P. Rajan, D. Roest, K. Skenderis, M.
Staudacher, I.Y. Park, A. Tseytlin, and K. Zarembo for stimulating
discussions. In particular, we would like to thank L. Rastelli for
valuable discussions on the issue of consistent truncation
discussed in Section 4 and J. Fjelstad for discussions and ideas
on critical singleton closed string theory. This research project
has been supported in part by NSF Grant PHY-0070964.

\pagebreak

\begin{appendix}


\section{Spectra of Massless Higher Spin Gauge Theories in $D=5,7$}


In this Appendix, we tabulate the spectra of singletons and the
generators of the super HS groups and the field content of the
master scalar fields in $AdS_5$ and $AdS_7$. The case of $D=4$ is
relatively simpler and has been presented in Section 2.1. The
spectrum of physical states described by the master gauge fields
in $D=4,5,7$ are also given in Section 2.\\

\tfa

\tfb

\pagebreak

\tfid

\tfc

\pagebreak

\tsa

\tsb

\pagebreak

\tsd

\tsc

\end{appendix}

\pagebreak


\section{UIRs of AdS Superalgebras in $D=4,5,7$}


In this  Appendix, we define notation for the irreps of AdS
superalgebras in $D=4,5,7$, and list these irreps as well as the
BPS short supermultiplets. These results are especially useful for
the  discussions of Section 2 and 3.


\subsection{The UIRs of $OSp(8|4)$ and BPS Multiplets}


Recall that a UIR of $OSp(8|4)$ is a multiplet of $SO(3,2)\times
SO(8)$ UIR's denoted by

\be D(E_0,s;a_1,a_2,a_3,a_4)\ ,\la{3dnot}\ee

where $E_0$ is the minimum eigenvalue of the AdS energy generator
$M_{05}$, $s$ denotes $SO(3)\subset SO(3,2)$ spin and
$(a_1,a_2,a_3,a_4)$ are the Dynkin labels of the $SO(8)$ irrep
carried by the lowest energy state. There exist two series of
supermultiplets \cite{min}:

\bea && {\rm A)}\quad\quad\quad\quad E_0 \ge 1+s+a_1+a_2
+\ft12(a_3+a_4)\ , \la{ub1}\w2 && {\rm B)}\quad\quad\quad\quad
E_0=a_1+a_2 +\ft12(a_3+a_4)\ , \quad\quad s=0\ . \la{ub2} \eea

These are the irreps carried by the lowest components of the
supermultiplets, and the entire $OSp(8|4)$ supermultiplets are
obtained by acting with supercharges.

The lowest components of the massless supermultiplets shown in
Table \ref{tfos} saturate the unitarity bound of series A as
$E_0=s+1$, except in level $\ell=0$ supergravity multiplet, in
which case $D(1,0;0,0,2,0)$ belongs to series $B$. The discrete
series B contains the BPS multiplets. In particular, the singleton
multiplet is characterized by the irrep $D(1/2,0;0,0,0,1)$ carried
by its lowest component and it belongs to series B. It can be
described by a suitably constrained superfield. Taking a suitably
symmetrized and constrained product of  $2E_0$ singleton
superfields one can construct BPS superfields whose lowest
components carry the following irreps \cite{fs1}

\bea {\rm BPS\ 1/2}&:& \ \ \ D\left(p/2,0;0,0,p,0\right)\ ,
\la{3dbps1}\w2 {\rm BPS\ 1/4} &: & \ \ \
D\left((p+2q)/2,0;0,q,p,0\right)\ , \la{3dbps2}\w2 {\rm BPS\ 1/8}
&:& \ \ \ D\left((p+2q+3r+4s)/2,0;r+2\ell,q,p,r\right)\ .
\la{3dbps3} \eea

All of these belong to series B. In particular, the lowest
components of the KK towers of the level $\ell=0$ supergravity
multiplet carry the irrep $D(k/2,0;0,0,k,0)$ for $k=3,4,...$
\cite{af1}.


\subsection{The UIRs of $PSU(2,2|4)$ and BPS Multiplets}


A UIR of $SU(2,2|4)$ consists of UIRs of $SO(4,2)\times SO(6)$
denoted by

\be D(E_0,j_L,j_R;a_1,a_2,a_3)_Y\ , \la{4dnot}\ee

where $E_0$ is the eigenvalue of the AdS energy generator $M_{06}$
and $(j_L,j_R)$ label the $SO(4)\subset SO(4,2)$ irrep,
$(a_1,a_2,a_3)$ denote the Dynkin labels of the $SO(6)\simeq
SU(4)$ R-symmetry irrep carried by the minimum energy states and
$Y$ denotes the outer $U(1)_Y$ automorphism charge, which will
often be suppressed when it is vanishing. There exist three series
of supermultiplets \cite{dp}:

\bea && {\rm A)}\quad\  E_0 \ge 2+J_L+J_R+a_1+a_2+a_3\ , \quad
J_L-J_R \ge \ft12(a_3-a_1) \la{4ub1} \w4 && {\rm B)} \quad\
E_0=\ft12(a_1+2a_2+3a_3)\ge 2+2J_L +\ft12(3a_1+2a_2+a_3)\ ,\ J_R=0
\la{4ub2}\w4 && \quad\quad\ {\rm which\ implies} \nn\w2 &&
\quad\quad\ E_0\ge 1+J_L+a_1+a_2+a_3\ , \quad 1+J_L \le
\ft12(a_3-a_1) \nn\w4 && {\rm C)}\quad\  E_0= 2a_1+a_2\ , \quad
a_3=a_1\ ,\ J_L=J_R=0 \eea

In the case of series B,  irreps with $(J_L \leftrightarrow J_R,
a_1 \leftrightarrow a_3)$ must also be included. The irreps listed
above are carried by the lowest components of the supermultiplets,
and the entire $PSU(2,2|4)$ supermultiplets are obtained by acting
with supercharges.

The lowest components of the massless supermultiplets shown in
Table \ref{tfis} saturate the unitarity bound of series A as
$E_0=s+2$, with $J_L=J_R=s/2$, except in level $\ell=0$
supergravity multiplet in which case $D(2,0,0;0,2,0)$ belongs to
series $C$. The discrete series C contains the BPS multiplets. In
particular the Maxwell singleton multiplet is characterized by
$D(1,0,0;0,1,0)$ carried by its lowest component and it belongs to
series C. It can be described by a suitably constrained
superfield. Taking a properly symmetrized and constrained product
of $E_0$ singletons superfields one can construct BPS superfields
whose lowest components carry the following irreps \cite{fs1}

\bea {\rm BPS\ 1/2} &: & \ \ \ D(p,0,0;0,p,0)\ , \la{4dbps1}\w2
{\rm BPS\ 1/4}  &: & \ \ \ D\left(p+2q,0,0;q,p,q\right)\ ,
\la{4dbps2}\w2 {\rm BPS\ 1/8} & : & \ \ \
D\left(p+2q+3r,0,0;q,p,q+2r\right)\ . \la{4dbps3} \eea

The BPS 1/2 and BPS 1/4 multiplets belong to series C, and the BPS
1/8 multiplets belong to series B. The KK towers of the level
$\ell=0$ supergravity are the BPS 1/2 multiplets given by
$D(k,0,0;0,k,0)$ with $k=3,4,...$ \cite{witten,ffz,af3}.

There exists an extensive literature on the OPEs of various BPS
1/2 operators. The UIRs which can appear in these OPEs belong to
series A with $J_L=J_R=s/2$, and series C \cite{fs2}.


\subsection{The UIRs of $OSp(8^*|4)$ and BPS Multiplets }


A UIR of $OSp(8^*|4)$ consists of UIRs of $SO(6,2)\times USp(4)$
denoted by

\be D(E_0,J_1,J_2,J_3; a_1,a_2)_Y\ ,\ee

where $E_0$ is the eigenvalue of the AdS energy generator
$M_{08}$, $(J_1,J_2,J_3)$ denote the Dynkin labels of the $SU(4)
\simeq SO(6) \subset SO(6,2)$ irrep, $(a_1,a_2)$ denote the Dynkin
labels of the $USp(4)$ irrep carried by the minimum energy states
and $Y$ denotes the outer $U(1)_Y$ automorphism charge, which will
often be suppressed when it is vanishing. There exist four series
of supermultiplets \cite{min}:

\bea && {\rm A)}\quad\ E_0 \ge 6+\ft12(J_1+2J_2+3J_3) +
2(a_1+a_2)\ , \la{6d1}\w4 && {\rm B)}\quad\ E_0 =
4+\ft12(J_1+2J_2)+2(a_1+a_2)\ ,\quad J_3=0\ , \la{6d2}\w4 && {\rm
C)}\quad\ E = 2+\ft12 J_1 +2(a_1+a_2)\ ,\quad J_3=J_2=0\ ,
\la{6d3}\w4 && {\rm D)}\quad\ E_0=2(a_1+a_2)\ ,\quad
J_3=J_2=J_1=0\ . \la{6d4} \eea

These are the irreps carried by the lowest components of the
supermultiplets, and the entire $OSp(8^*|4)$ supermultiplets are
obtained by acting with supercharges.

The lowest components of the massless supermultiplets shown in
Table \ref{tss} have $E_0=s+4$ and belong to series B, while the
level $\ell=0$ supergravity multiplet carries the irrep.
$D(4,0,0,0;0,2)$ which belongs to series $D$. The discrete series
D contains the BPS multiplets. The singletons are contained in
series C and D. The superfields in terms of which they are
realized, and the UIRs carried by their lowest components are as
follows \cite{min,g7,fs1}:

\be \tabcolsep=5mm
\begin{tabular}{lp{2cm}p{4cm}}
\medskip
{\rm D)} & $W^{ij}$ & $D(2,0,0,0;0,1)$  \\
\medskip
{\rm D)} & $W^i$ & $D(2,0,0,0;1,0)$ \\
\medskip
{\rm C)} & $W$      & $D(2,0,0,0;0,0)$\\
\medskip
{\rm C)} & $\omega_{\a_1...\a_{\ell-2}}$ &
$D(\ft{\ell}2+1,\ell-2,0,0;0,0)$ \\
\end{tabular}
\ee

The index $i=1,...,4$ labels the $4$-plet of $USp(4)$, the index
$\a=1,...,4$ labels the chiral spinor of $SO(6)$, $W^{ij}=-W^{ji}$
and symplectic traceless, $\O^{ij}W_{ij}=0$, and
$\omega_{\a_1...\a_{\ell-2}}$ is totally symmetric in its indices
(see Table \ref{tsd} for further details). The superspace
constraints imposed on these superfields can be found in
\cite{fs1}. The superfield $W_{ij}$ represents the well known
$(2,0)$ tensor singleton and it is singlet under an $SU(2)_Z$
group defined in Section 2.3. The singleton superfields
$(W^i,W,\omega_{\a_1...\a_{\ell-2}})$, on the other hand, carry
$SU(2)_Z$ spins $(1/2,1,\ell/2)$, respectively. These are the
level $\ell=1,2$ and $\ell\ge 3$ singletons shown in Table
\ref{tsd}.

Taking suitably symmetrized and constrained products of singleton
superfields, one can construct BPS superfields whose lowest
components carry the following UIRs \cite{fs1}:

\bea {\rm BPS\ 1/2}: & D(2p,0,0,0;0,p)\ , \la{6dbps1}\w2 {\rm
BPS\ 1/4} : & \quad\quad\ \  D\left(2p+4q,0,0,0;2q,p\right)\ .
\la{6dbps2} \eea

Both of these belong to series D. The KK towers of the level
$\ell=0$ supergravity are the BPS 1/2 multiplets given by
$D(2k,0,0,0;0,k)$ with $k=3,4,...$ \cite{fs1}.

The OPEs of BPS 1/2 operators have been studied \cite{fs2,as}. The
supermultiplets that can appear in the OPE of two BPS 1/2
operators belong to series A with $(J_1,J_2,J_3)=(0,s,0)$; series
B with $(J_1,J_2)=(0,s)$ and $E_0=4+s+2(a_1+a_2)$; series C with
$J_1=0$ and $E_0=2+2(a_1+a_2)$, and series D \cite{fs2}.


\section{ Labeling of $USp(8)$, $SU(4)$, $USp(4)$ and $SO(8)$ Irreps}


\subsection{\bf $USp(8)$}

The highest weight state (HWS) labels $(n_1,n_2,n_3,n_4)$ of
$USp(8)$ satisfy $n_1\ge n_2\ge n_3\ge n_4$ and are related to the
Dynkin labels $[a_1,a_2,a_3,a_4]$ as follows:

$$ n_1=a_1+a_2+a_3+a_4\ , \quad\quad n_2=a_2+a_3+a_4\ ,\quad\quad
n_3=a_3+a_4\ ,\quad\quad n_4=a_4\ .$$

\subsection{$SU(4)\sim SO(6)$}

The HWS labels of $SU(4)$ irreps are $(n_1,n_2,n_3)$. They satisfy
$n_1\ge n_2\ge n_3$ and they are related to the Dynkin labels
$[a_1,a_2,a_3]$ as follows:

$$ n_1= a_1+a_2+a_3\ ,\quad\quad    n_2= a_2+a_3\ , \quad\quad n_3=a_3\ . $$

The $SO(6)$ HW labels by $(m_1,m_2,m_3)$ obey $m_1\ge m_2\ge
|m_3|$ and they are related to the $SO(6)$ Dynkin labels
$[b_1,b_2,b_3]$ as

$$
m_1= b_1+\ft12( b_2+b_3)\ ,\quad\quad m_2= \ft12(b_2+b_3)\ ,
\quad\quad m_3=\ft12(-b_2+b_3)\ .
$$

These are related to the $SU(4)$ HW labels $(n_1,n_2,n_3)$ and
$SU(4)$ Dynkin labels $[a_1,a_2,a_3]$ as

$$\tabcolsep=5mm
\begin{tabular}{lp{4cm}p{4cm}}
$m_1=\ft12(n_1+n_2-n_3)$ & $m_2=\ft12(n_1-n_2+n_3)$ &
$m_3=\ft12(-n_1+n_2+n_3)$ \\
$b_1=a_2$                & $b_2=a_1               $ & $b_3=a_3
     $ \\
\end{tabular}
$$

\subsection{$USp(4)\sim SO(5)$}

The $USp(4)$ irreps have the HWS labels $(n_1,n_2)$ which satisfy
$n_1\ge n_2\ge $ and they are related to the Dynkin labels
$[a_1,a_2]$ as

$$ n_1= a_1+a_2\ ,\quad\quad    n_2= a_2$$

The irreps of $SO(5)$ have HW labels $(m_1,m_2)$ which satisfy
$m_1\ge m_2\ge 0$ and they are related to the $SO(5)$ Dynkin
labels $[b_1,b_2]$ as

$$m_1=b_1+\ft12\,b_2\ ,\quad\quad m_2=\ft12\,b_2$$

These are related to the $USp(4)$ HW labels $(n_1,n_2)$ and
$USp(4)$ Dynkin labels $[a_1,a_2]$ as

$$\tabcolsep=5mm
\begin{tabular}{lp{4cm}}
$m_1=\ft12(n_1+n_2)$ & $m_2=\ft12(n_1-n_2)$  \\
$b_1\  =a_2$            & $b_2\  =a_1           $  \\
\end{tabular}
$$

\subsection{$SO(8)$}

The irreps of $SO(8)$ have HWS labels $(n_1,n_2,n_3,n_4)$ which
satisfy $n_1\ge n_2\ge 0\ge n_3\ge |n_4|$ and they are related to
the $SO(8)$ Dynkin labels $[a_1,a_2,a_3,a_4]$ as

$$n_1=a_1+a_2+\ft12(a_3+a_4)\ ,\quad\quad n_2=a_2+\ft12(a_3+a_4)\ ,\quad\quad
n_3=\ft12(a_3+a_4)\ ,\quad\quad n_4= \ft12(-a_3+a_4)\ .$$\\


\section{Compact and Non-compact Bases for $SO(d,2)$}


We write the $SO(d,2)$ algebra in canonical form as
($A=0,\dots,d,d+2$):

\be [M_{AB},M_{CD}]=i\eta_{BC}M_{AD}+\mbox{$3$ more}\ ,\ee

where $\eta={\rm diag}(-,+,...,+,-)$. The compact basis, which is
suitable for describing physical AdS fields, consists of the AdS
energy $E=-M_{0,d+2}$, the $SO(d)$ generators $M_{ij}$
($i=1,...,d$) and the spin-boosts $L^{\pm}_i=M_{i,d+2}\mp i
M_{0i}$, which shift the AdS energy by $\pm 1$. In compact basis
the $SO(d,2)$ weight spaces $D(E_0;m_1,\dots m_{[d/2]})$ are
obtained by acting with $L_i^+$ on lowest weight states, which
have minimal energy $E=E_0$ and carry $SO(d)$ highest weights
$(m_1,\dots m_{[d/2]})$. Note that the label $m_1$ is the
$SO(3)\subset SO(3,2)$ spin in the case of $AdS_4$ and the sum
$j_L+j_R$ of $SU(2)_L\times SU(2)_R \simeq SO(4)\subset SO(4,2)$
spins in the case of $AdS_5$. The non-compact basis, which is
suitable for describing conformal fields, consists of the
dilatation generator $D=M_{d,d+2}$, the $SO(d-1,1)$ generators
$M_{\m\n}$ ($\m=0,1,\dots,d-1$), and the $d$-dimensional momentum
$P_\m=M_{\m d}+M_{\m,d+2}$ and generator of special conformal
transformations $K_\m=M_{\m d}-M_{\m,d+2}$. The compact basis
$(E,M_{ij}, L_i^{\pm})$ and non-compact basis
$(D,M_{\m\n},K_\m,P_\m)$ are related \cite{gnc} by a similarity
transformation executed by the (non-unitary) operator

\be S=\exp iL^+_d\ ,\ee

with the following properties

\bea S D S^{-1}&=&-iE+\frac12 L^-_{d}\ ,\\
SM_{0a} S^{-1}&
=&-iM_{a,d}-\frac{i}2L^-_a\ ,\qquad S M_{ab} S^{-1}= M_{ab}\ ,\\
S K_0 S^{-1}&=& -\frac{i}2 L^-_d\ ,\qquad \quad\qquad SK_a
S^{-1}=-\frac12 L^-_a\ ,\eea

where we have split the indices as follows

\be i=\overbrace {1,2,...,d-1}^a,d\ ,\qquad\qquad \m=0,
\overbrace{1,2,...,d-1}^a\ .\ee

Hence $(d+1)$-dimensional time-evolution and spatial rotation are
equivalent to $d$-dimensional dilatation and Lorentz rotation.
Thus $$S^{-1}D(E_0;m_1,\dots,m_{[d/2]})=\cO_{\D}(0)\ket{0}$$ where
$\ket{0}$ is the vacuum of the CFT$_d$ and $\cO_{\D}(x)=e^{i x^\m
P_\m}\cO_{\D}(0)e^{-i x^\m P_\m}$ is a conformal tensor with
scaling dimension
$$\D=E_0$$ and Lorentz spin given by $(m_1,\dots,m_{[d/2]})$.

\pagebreak


\end{document}